\documentclass[final,3p,times]{elsarticle}
\pdfoutput=1

\usepackage{amssymb}
\usepackage{subfigure}
\usepackage{subfigmat}
\usepackage{makecell}
\usepackage{multirow}
\usepackage{amsmath}
\usepackage{mathtools}
\usepackage{float}

\begin{document}

\begin{frontmatter}

\title{On Secondary Tones Arising in Trailing-Edge Noise at \\ Moderate Reynolds Numbers}

 \author{T\'ulio R. Ricciardi}
 \ead{tulioricci@fem.unicamp.br}
 \author{Walter Arias-Ramirez}
 \ead{walter.arias.ramirez@gmail.com}
 \author{William R. Wolf}
 \ead{wolf@fem.unicamp.br}
 \address{University of Campinas, Campinas, SP, 13083-860, Brazil}

\begin{abstract}
Direct numerical simulations are carried out to investigate the flow features responsible for secondary tones arising in trailing-edge noise at moderate Reynolds numbers. Simulations are performed for a NACA 0012 airfoil at freestream Mach numbers 0.1, 0.2 and 0.3 for angle of incidence 0 deg. and for Mach number 0.3 at 3 deg. angle of incidence. The Reynolds number based on the airfoil chord is fixed at $Re_c=10^5$. Flow configurations are investigated where noise generation arises from the scattering of boundary layer instabilities at the trailing edge. Results show that noise emission has a main tone with equidistant secondary tones, as discussed in literature. An interesting feature of the present flows at zero incidence is shown; despite the geometric symmetry, the flows become non-symmetric with a separation bubble only on one side of the airfoil. A separation bubble is also observed for the non-zero incidence flow. For both angles of incidence analyzed, it is shown that low-frequency motion of the separation bubbles induce a frequency modulation of the flow instabilities developed along the airfoil boundary layer. When the airfoil is at 0 deg. angle of attack an intense amplitude modulation is also observed in the flow quantities, resulting in a complex vortex interaction mechanism at the trailing edge. Both amplitude and frequency modulations directly affect the velocity and pressure fluctuations that are scattered at the trailing edge, what leads to secondary tones in the acoustic radiation.
\end{abstract}

\begin{keyword}
airfoil tonal noise \sep direct numerical simulation \sep trailing-edge noise scattering \sep flow instabilities
\end{keyword}

\end{frontmatter}

\section{Introduction}

The understanding of trailing-edge noise is an overriding concern for design of noise-efficient aerodynamic shapes including wings and high-lift components, wind turbine and helicopter blades, fans and even car roof racks~\cite{massarotti2018}. For such devices, tonal noise may be an important component of noise spectrum, especially at moderate Reynolds number flows. 
At these conditions, vortex shedding due to laminar boundary layer instabilities and blunt trailing edges are identified by Brooks {\em et al.}~\cite{Brooks1989} as potential sources of airfoil tonal noise. Wolf {\em et al.}~\cite{Wolf2012, Wolf2013, Wolf:DU96, Ricciardi2019} performed numerical investigations of turbulent flows past NACA0012 and DU96 airfoils showing that tonal noise may also appear in far-field acoustic predictions for blunt trailing edges even in the presence of fully turbulent boundary layers.

Several pioneering studies of airfoil noise were conducted in the 1970s in order to examine airfoil tonal noise generation. These investigations showed that discrete tones are emitted from isolated airfoils at specific flow conditions \cite{Smith1970,Clark1971,Hersh1971,Longhouse1977}. Such findings triggered some of the first systematic and detailed studies of airfoil tonal noise~\cite{Paterson1973,Tam1974,Fink1975, Fink1976,Arbey1983}. Paterson {\em et al.}~\cite{Paterson1973} performed noise measurements from symmetric NACA airfoils for a Reynolds number range between $10^{5}$ and $10^{6}$ at various angles of attack; their results showed the existence of multiple tones in a ladder-like structure pattern in terms of frequency and freestream velocity. Furthermore, they measured spanwise surface pressure fluctuations on the airfoils and found a strong correlation along the airfoil surface. This indicated that the flow phenomenon associated with airfoil tonal noise generation could be modeled as two-dimensional.

Tam~\cite{Tam1974} suggested that the ladder-like structure of frequency as function of flow speed was due to a self-excited feedback loop between the trailing edge and near wake. Fink~\cite{Fink1975} assumed that the discrete tonal frequencies were linked to the laminar boundary layer developing along the pressure side of the airfoil.
Arbey and Bataille~\cite{Arbey1983} repeated the experimental studies from Paterson in an open wind tunnel for three different NACA airfoils and showed that the noise spectrum was composed of a broadband contribution with a main tonal peak plus a set of equidistant secondary tones. The broadband component was assumed to appear due to scattering of Tollmien-Schlichting (TS) instabilities on the boundary layer. Lowson {\em et al.}~\cite{Lowson1994} found that the presence of tones was related to a separation bubble developed on the airfoil pressure side. In this case, another explanation of the feedback loop mechanism was provided where TS instabilities developed along the laminar boundary layer would lead to acoustic scattering on the trailing edge. Acoustic waves would then propagate upstream closing the loop, and the separation bubble would act as an amplifier of acoustic disturbances.

Later, supporting this model, Desquenes {\em et al.}~\cite{Desquesnes2007} showed that the main acoustic tone frequency radiated to the far-field was close to that most amplified along the pressure side boundary layer. These authors performed 2D direct numerical simulations (DNS) of flows past a NACA0012 airfoil for Reynolds numbers $1\times10^{5}$ and $2\times10^{5}$ with angles of attack of 2 and 5 degs. Their results presented  multiple tonal peaks consistent with experimental observations from Arbey and Bataille~\cite{Arbey1983}. The same authors also proposed a new model of feedback loop mechanism to explain the presence of secondary tones by considering a secondary feedback loop due to boundary layer instabilities forming on the airfoil suction side.

Nash {\em et al.}~\cite{Nash1999} performed experimental studies of airfoil tonal noise for a NACA0012 profile with a Reynolds number of $1.45\times10^{6}$ and several angles of attack. A closed-section wind tunnel, with and without acoustic-absorbing lining on its walls, was used in the experiments and results from the hard-wall tunnel revealed multiple frequency peaks. However, the authors argued that these tonal peaks were correlated to resonant frequencies of the wind tunnel. Thus, they carried out measurements with lined walls simulating anechoic conditions and, under such conditions, a single dominant tone was observed instead of several peaks. Furthermore, no ladder-like structure of tonal frequency was observed, in disagreement with previous studies of Paterson {\em et al.}~\cite{Paterson1973}, Fink,~\cite{Fink1975} and Arbey and Bataille~\cite{Arbey1983}. It is important to mention that secondary tones were often observed in experiments conducted in open-jet facilities.

Kurotaki {\em et al.}~\cite{Kurotaki2008}, Chong and Joseph~\cite{Chong2013a} and Plogmann {\em et al.}~\cite{Plogmann2013} also found multiple tones in their experiments and the so-called ladder-like structure pattern. The latter authors demonstrated that tripping the pressure side boundary layer would transition the flow to turbulent, eliminating the separation bubble and tones. Meanwhile, Tam and Ju~\cite{Tam2011} conducted 2D DNS of a NACA0012 airfoil with three different trailing edge configurations in the Reynolds and Mach number ranges of $2\times10^{5} < Re_c < 5\times10^{5}$ and $0.08 < M_{\infty} < 0.2$ at zero angle of attack. Under these conditions, their numerical results showed a single tone for each simulation, in agreement with measurements of Nash {\em et al.}~\cite{Nash1999}. 

Following the results of \cite{Desquesnes2007}, Pr\"obsting {\em et al.}~\cite{Probsting2014} employed particle image velocimetry to investigate the mechanisms associated with tonal noise generation. These authors discussed about an amplitude modulation of velocity fluctuations measured near the trailing edge. Fosas de Pando {\em et al.}~\cite{FosasdePando2014} employed both 2D DNS and global stability analysis to study the dynamics of hydrodynamic and acoustic wavepackets driving the feedback loop  mechanism. Another recent work that makes the connection between tones and TS waves is presented by Sanjose {\em et al.}~\cite{Sanjose2018} for a controlled-diffusion airfoil at $Re_c < 5\times10^{5}$. These authors employed a suit of modal analysis techniques to investigate the tonal noise generation problem and found that intermittency plays a significant role in the flow dynamics and noise emission of the airfoil configuration investigated. 

\begin{figure}[H]
\centering
{\includegraphics[width=0.75\textwidth,trim={1mm 1mm 1mm 1mm },clip]
	{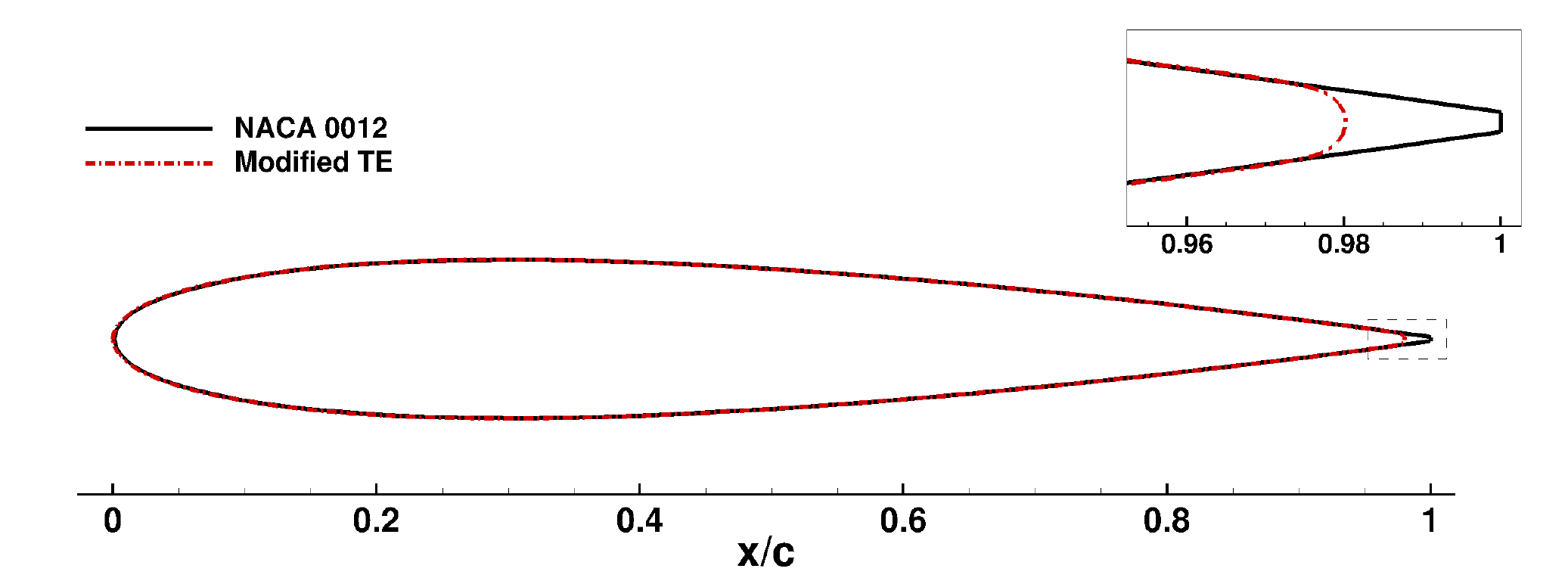}}
\caption{Modified NACA0012 trailing edge profile.}
\label{fig1}
\end{figure}

It is clear that since the 1970s, great efforts have been carried out to improve the understanding of the airfoil tonal noise phenomenon and we suggest the review paper by Arcondoulis {\em et al.}~\cite{Arcondoulis2010} for a general discussion on the topic. As one can see, there are still several open questions and disagreements in literature with respect to trailing edge tonal noise. The current work presents a numerical study of airfoil tonal noise generation at moderate Reynolds number flows. Direct numerical simulations are conducted for 2D flows past a NACA0012 airfoil with a rounded trailing edge as shown in Fig.~\ref{fig1}. The current effective chord is 98\% of that from the original NACA0012 profile and the trailing-edge radius has 0.4\% of the chord length. We investigate effects of compressibility and angle of incidence on both the acoustic signature and flow dynamics near the trailing edge, and particular attention is given to the appearance of secondary tones in spectra.
The flow configurations analyzed comprise angles of incidence of $0$ and $3$ deg. and freestream Mach numbers $M_{\infty}=0.1$ to $0.3$, for a fixed Reynolds number based on the airfoil chord of $Re_{c}=10^5$. The features of secondary tones are interpreted by comparison with a simple theoretical model constructed based on the acoustic signature of the simulated flows.

\section{Numerical Methodology}

Direct numerical simulations are performed solving the two-dimensional compressible Navier-Stokes equations in general curvilinear coordinates. Length, velocity components, density and pressure are non-dimensionalized by the airfoil chord, $L$, freestream speed of sound, $a_{\infty}$, freestream density, ${\rho}_{\infty}$ and ${\rho}_{\infty}a_{\infty}^{2}$, respectively. For comparison, results in terms of time and frequency are presented non-dimensionalized by freestream velocity as $t=t^* \, U_{\infty} / L$ and $f=f^* \, L / U_{\infty}$. In these definitions, $t^*$, $f^*$ and $f$ are time (dimensional), frequency (dimensional) and Strouhal number, respectively.
A staggered grid sixth-order accurate compact scheme \cite{Nagarajan2003} is employed for the spatial discretization of the governing equations. A high wavenumber compact filter \cite{Lele1992} is applied to the computed solution to control numerical instabilities arising from mesh non-uniformities and interpolation at grid interfaces. This filter is only applied in flow regions away from solid boundaries.

The time integration of the fluid equations is carried out by the implicit second-order scheme of Beam and Warming \cite{Beam1978} in the near-wall region in order to overcome the time step restriction due to the usual near-wall fine-grid numerical stiffness. A third-order Runge-Kutta scheme is used for time advancement of the equations in flow regions far away from solid boundaries. No-slip adiabatic wall boundary conditions are applied along the solid surfaces and characteristic plus sponge boundary conditions are applied in the far-field locations to minimize wave reflections. The numerical tool has been previously validated for several 2D and 3D simulations of compressible flows involving airfoil sound generation \cite{Wolf2012, Wolf:DU96, AriasJBSMSE2015}.

Three O--type grids are used in the simulations and a mesh refinement study is performed (see more details in the results section). The computational domain extends 14 chords from the airfoil in each direction and the grids are symmetric with respect to the x-axis. Based on previous studies \cite{AriasAIAA2015}, all grids have 600 points in the wall-normal direction ($N_y=600$), which is sufficient to resolve the flow structures along the boundary layers considering the current wall-normal stretching. In all grids, the distance from the surface to the first mesh point is $\Delta y_{wall}=0.0002$. Along the flow direction, the refinement is applied considering $N_x = 400$, 800 and 1200 points for grids 1, 2 and 3, respectively. In this refinement study, the resolution along the flow direction is exactly doubled from grid 1 to grid 2 and tripled from grid 1 to 3. Values of mesh resolution at the leading and trailing edges are displayed in Table \ref{table1}. Here, these values are presented in a similar fashion compared to other simulations available in Refs. \cite{Desquesnes2007,FosasdePando2014}. We also employed other grid configurations with localized refinement along the leading and trailing edges to verify the sensitivity of these regions with respect to resolution and observed that an overall grid refinement was sufficient for converged solutions. A detail view of all 3 grids used in the simulations is provided in Fig. \ref{fig_mesh} and a summary of the grids is presented in Table \ref{table1}.
\begin{table}[H]
\begin{center}
	\caption{Description of the different mesh setups employed.}
	\label{table1}
	\begin{tabular}{c|ccccc}
		Grid  & $N_{x}$ & $N_{y}$ & $\Delta y_{wall}$ & $\Delta x_{LE}$ & $\Delta x_{TE}$  \\ \hline
		\#1   & 400     & 600        & 0.0002      & 0.002500   & 0.000510   \\
		\#2   & 800     & 600        & 0.0002      & 0.001250   & 0.000255   \\
		\#3   & 1200    & 600        & 0.0002      & 0.000833   & 0.000170   \\ \hline
	\end{tabular}
\end{center}
\end{table}

A description of the different flows investigated is presented in Table \ref{table2} in terms of freestream Mach number ($M_{\infty}$), grid configuration, angle of attack (AoA), time step ($\Delta t$) and total period of simulation ($T$) for which data is recorded after discarding transients. For all cases, simulations are run for at least 10 flow through times before statistics are collected. This number is a little more conservative than that used by Desquenes {\em et al.} \cite{Desquesnes2007} who collected data after approximately 8 flow through times. Flow snapshots are saved with a time step of 0.018 and 0.009 for the zero and three deg. angle of incidence, respectively. Unless specified in the text, computations of pressure and velocity spectra employ bins with 2048 snapshots and overlap of 67\% using a Hanning window function.
\noindent
\begin{minipage}{0.4\textwidth}
	\begin{table}[H]
	\begin{center}
		\caption{Description of flow configurations investigated.}
		\label{table2}
		\begin{tabular}{cc|ccc}
	$M_{\infty}$         & Grid & AoA & $\Delta t \times 10^{5}$ &  $T$ \\ \hline
	\multirow{4}{*}{0.1} & \#1  &  0  &      6   &     160.0       \\
	                	 & \#2  &  0  &      6   &     100.0       \\
	                 	 & \#2  &  3  &      6   &      70.0       \\
	                     & \#3  &  0  &      4   &      95.0       \\ \hline 
	\multirow{4}{*}{0.2} & \#1  &  0  &     12   &     150.0       \\
	                 	 & \#2  &  0  &     12   &     335.0       \\
	                 	 & \#2  &  3  &     12   &     300.0       \\
	                 	 & \#3  &  0  &      8   &     165.0       \\ \hline 
	\multirow{4}{*}{0.3} & \#1  &  0  &     18   &     415.0       \\
						 & \#2  &  0  &     18   &     240.0       \\
						 & \#2  &  3  &     18   &     315.0       \\
						 & \#3  &  0  &     12   &     150.0       \\ \hline
		\end{tabular}
	\end{center}
	\end{table}
\end{minipage}
\begin{minipage}{0.56\textwidth}
\begin{figure}[H]
\centering
	\subfigure[$N_x = 400$]
	{\includegraphics[width=0.99\textwidth,trim={1mm 1mm 1mm 1mm },clip]
		{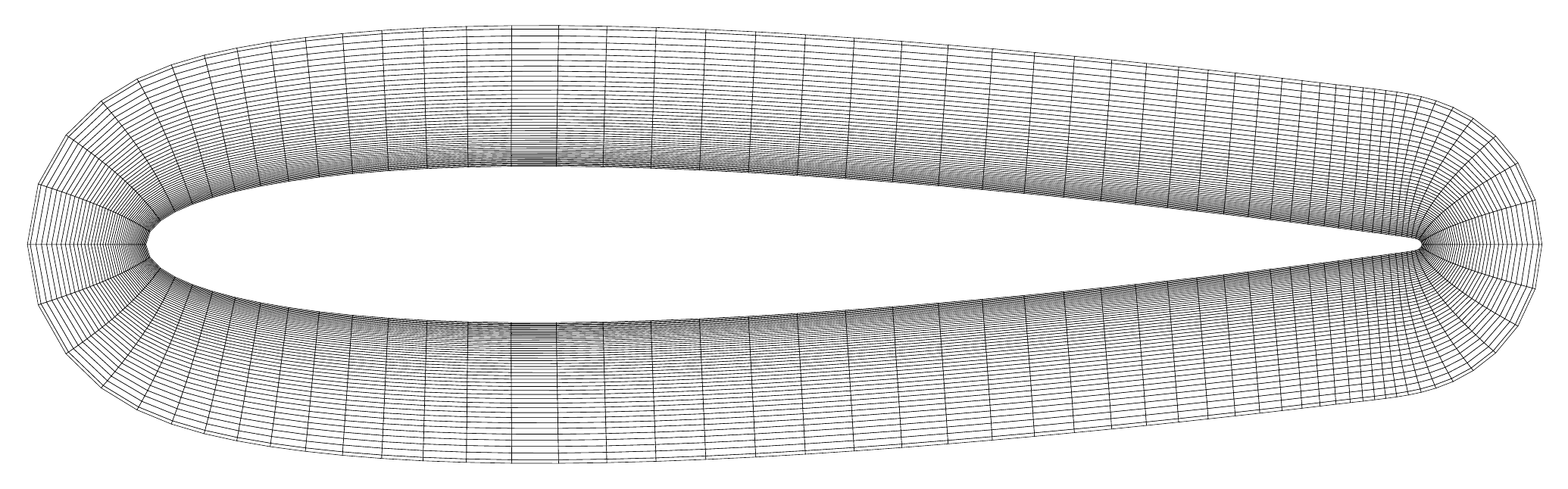}}
	\subfigure[$N_x = 800$]
	{\includegraphics[width=0.99\textwidth,trim={1mm 1mm 1mm 1mm },clip]
		{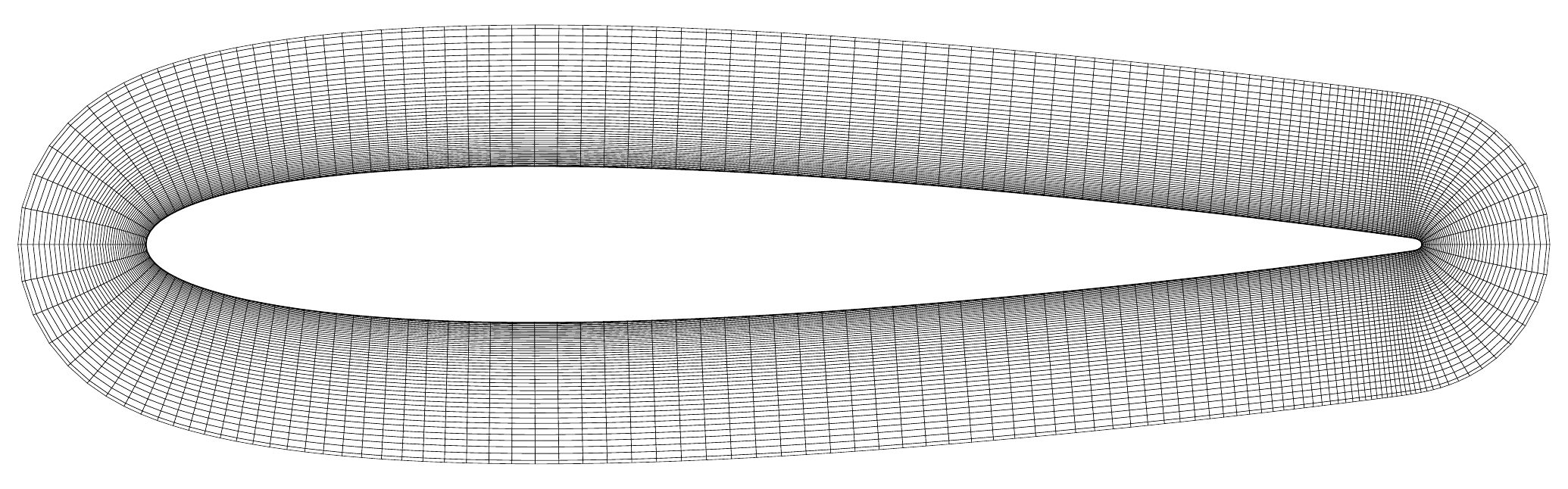}}
	\subfigure[$N_x = 1200$]
	{\includegraphics[width=0.99\textwidth,trim={1mm 1mm 1mm 1mm },clip]
		{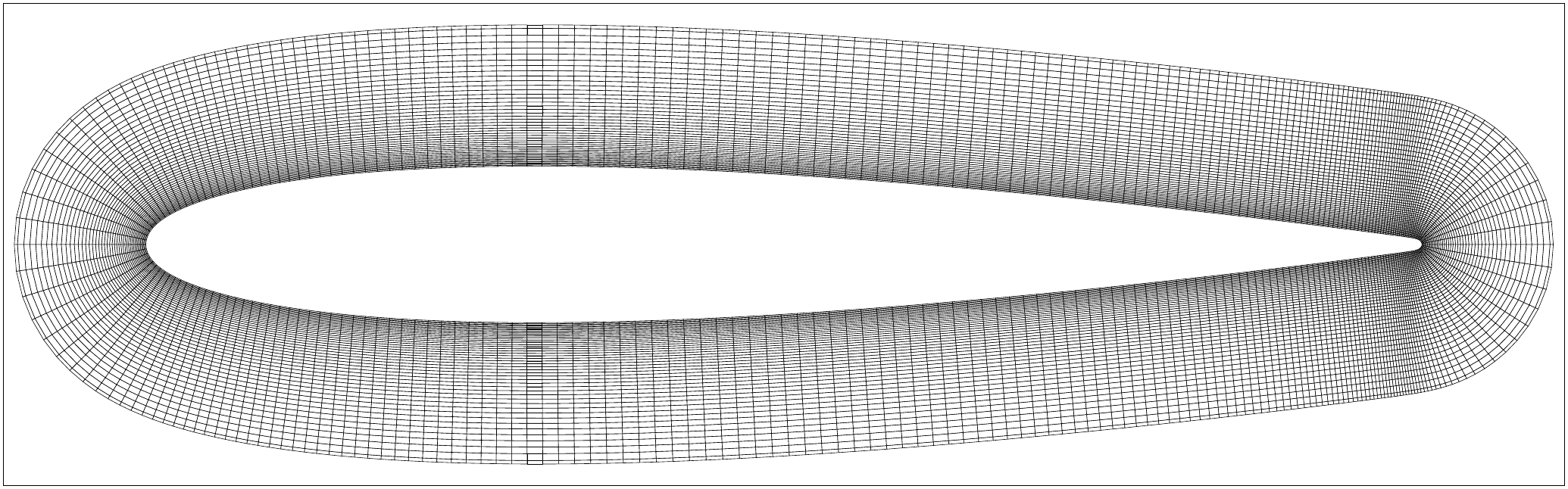}}
	\caption{Detail view of O--grids employed in the mesh refinement study (every 4th grid point shown).}
	\label{fig_mesh}
\end{figure}
\end{minipage}

\section{Results and Discussion}

An airfoil immersed in a low Reynolds number flow emits a single tone due to vortex shedding. At moderate Reynolds numbers the mechanisms of noise generation are modified by the presence of Tollmien-Schlichting (TS) instabilities developed along boundary layers. These flow instabilities are spanwise coherent, being efficient noise sources that lead to acoustic scattering at the airfoil trailing edge and causing the appearance of multiple tones on pressure spectrum.

In this section, we present results of direct numerical simulations for moderate Reynolds number flows past a NACA0012 airfoil. The effects of compressibility and angle of incidence on tonal noise generation are evaluated for $Re_{c}= 10^5$. Simulations are performed for freestream Mach numbers $M_{\infty}=0.1$, $0.2$ and $0.3$ for $0$ deg. and for $M_{\infty}=0.3$ at $3$ deg. angle of incidence.

\subsection{Airfoil at 0 deg. angle of incidence}

\begin{figure}[H]
		\subfigure[$M_\infty=0.1$]
		{\includegraphics[width=0.33\textwidth,trim={1.5mm 5mm 15mm 20mm},clip]
			{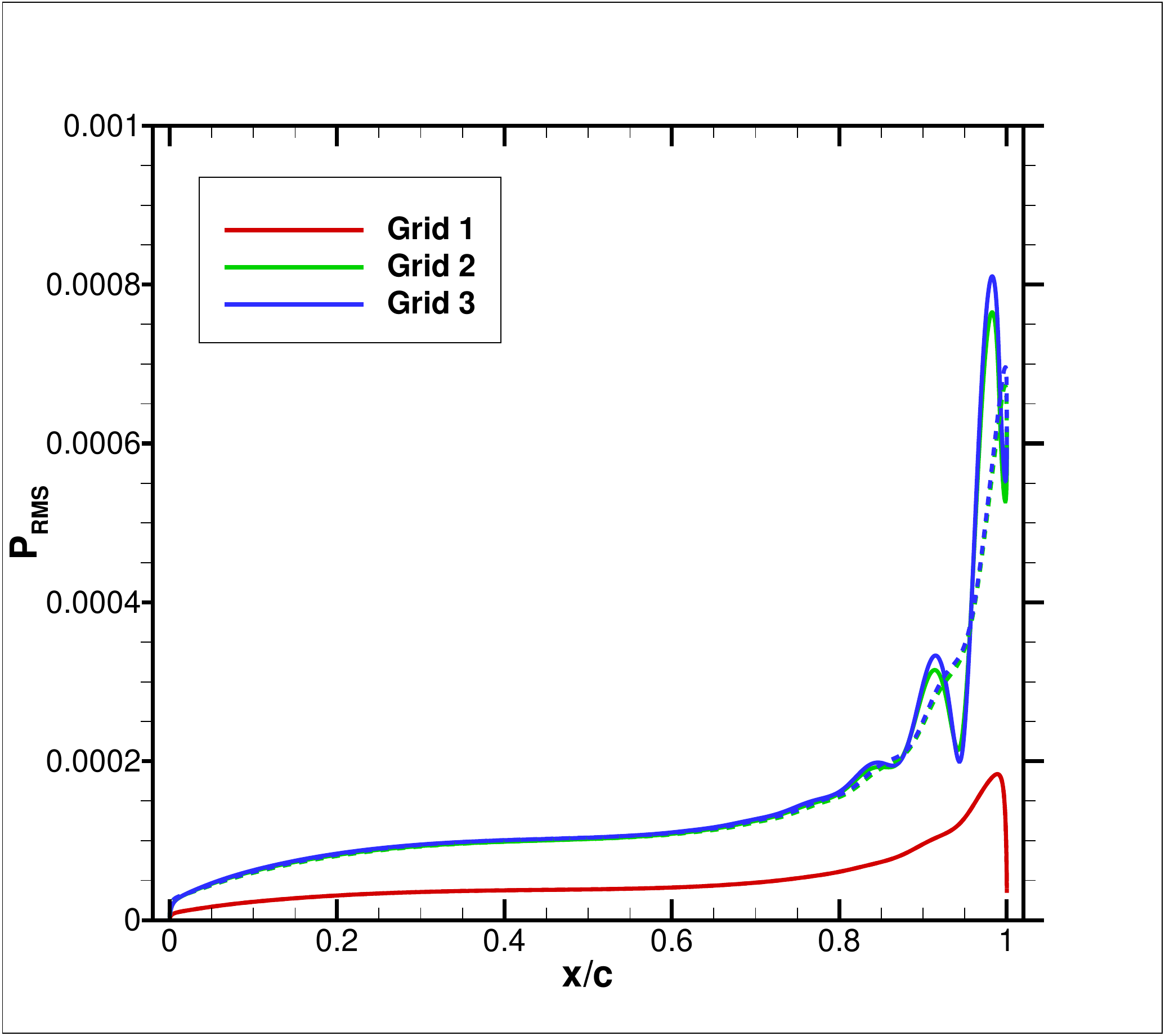}}
		\subfigure[$M_\infty=0.2$]
		{\includegraphics[width=0.33\textwidth,trim={1.5mm 5mm 15mm 20mm},clip]
			{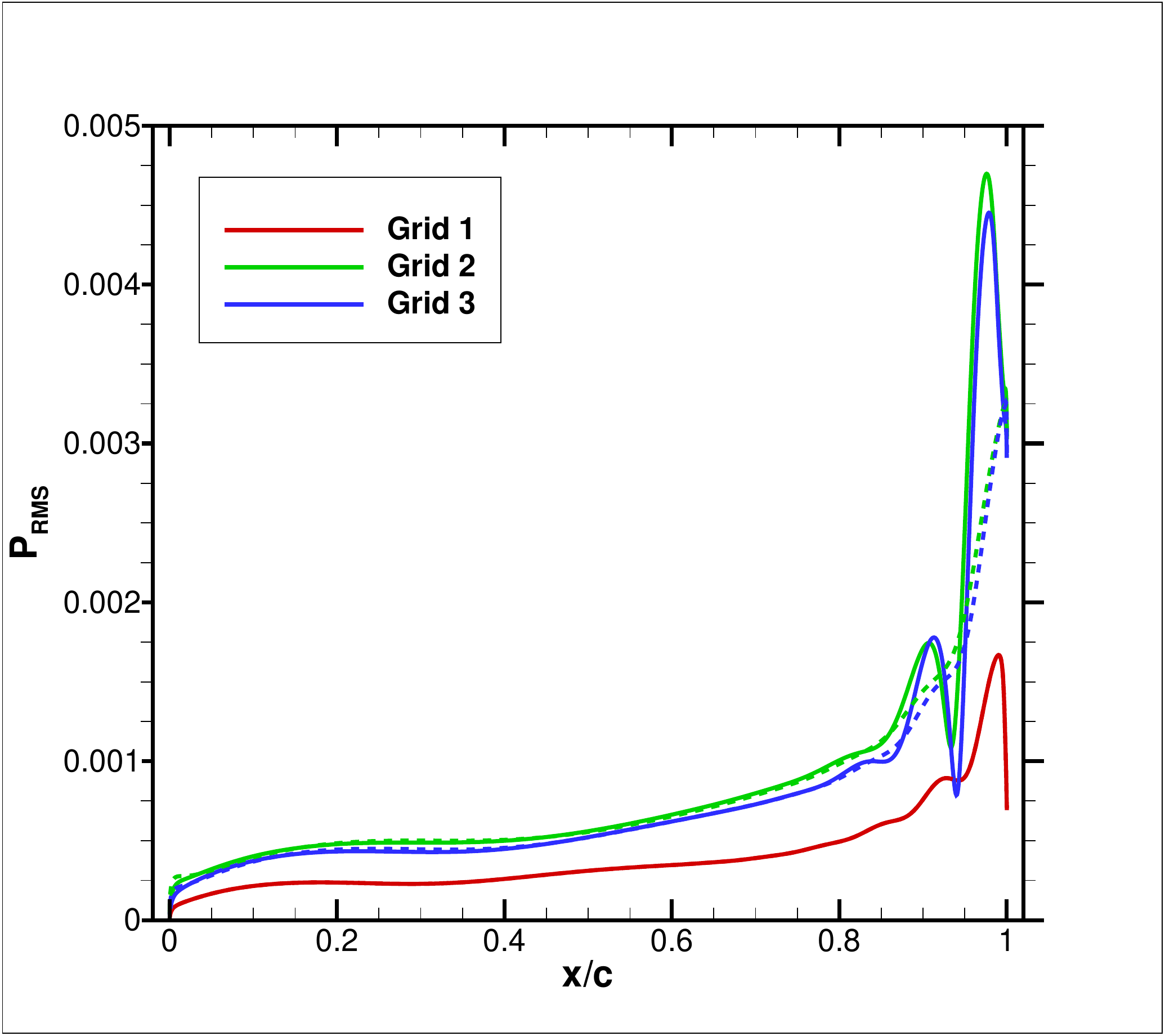}}
		\subfigure[$M_\infty=0.3$]
		{\includegraphics[width=0.33\textwidth,trim={1.5mm 5mm 15mm 20mm},clip]
			{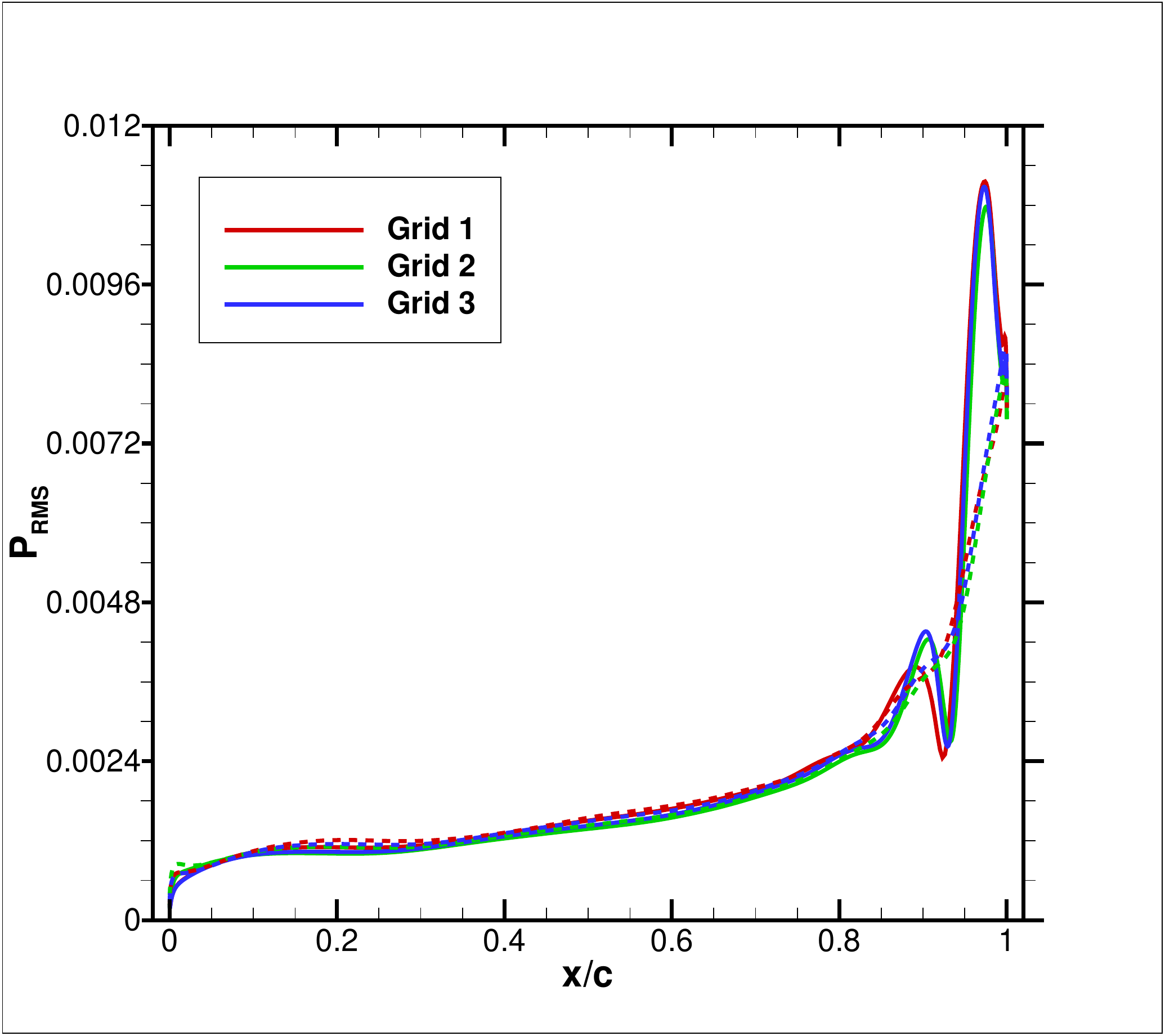}}
	\caption{Distribution of pressure RMS along airfoil surface for different grids. Solid lines indicate pressure values on the side of the separation bubble.}
	\label{fig_Prms_gridref}
\end{figure}

Based on the theories of Curle \cite{Curle1955} and Amiet \cite{Amiet1976}, the noise emitted from an airfoil in a low Mach number flow can be computed using the surface pressure fluctuations. Figure \ref{fig_Prms_gridref} presents the RMS values of pressure for $M_{\infty} = 0.1$, 0.2 and 0.3 for all three grids used in the simulations. As one can see, solutions obtained for grids 2 and 3 present similar pressure distributions indicating convergence for all Mach numbers. A curious feature of the current flows is that pressure distributions are not symmetric despite the fact that freestream flow and grids are symmetric. A movie with vorticity contours around the airfoil trailing edge is provided for $M_{\infty} = 0.2$ and Grid 2. This movie shows the transient portion of the flow when symmetry is broken. A similar behavior is observed for axisymmetric bluff-body wakes by \cite{Bury2012,Rigas2016}.
In the cases studied by these refences, the flows undergo multiple transitions depending on Reynolds number and, for some cases, low frequency beating leads to formation of secondary tones and lack of symmetry. As for the previous cases, it may be that the present flow goes through a bifurcation at the present Reynolds number and becomes asymmetric.
	
\begin{figure}[H]
	%
	\subfigure[Grid 1]
	{\includegraphics[width=0.33\textwidth,trim={1mm 10mm 15mm 1mm},clip]
		{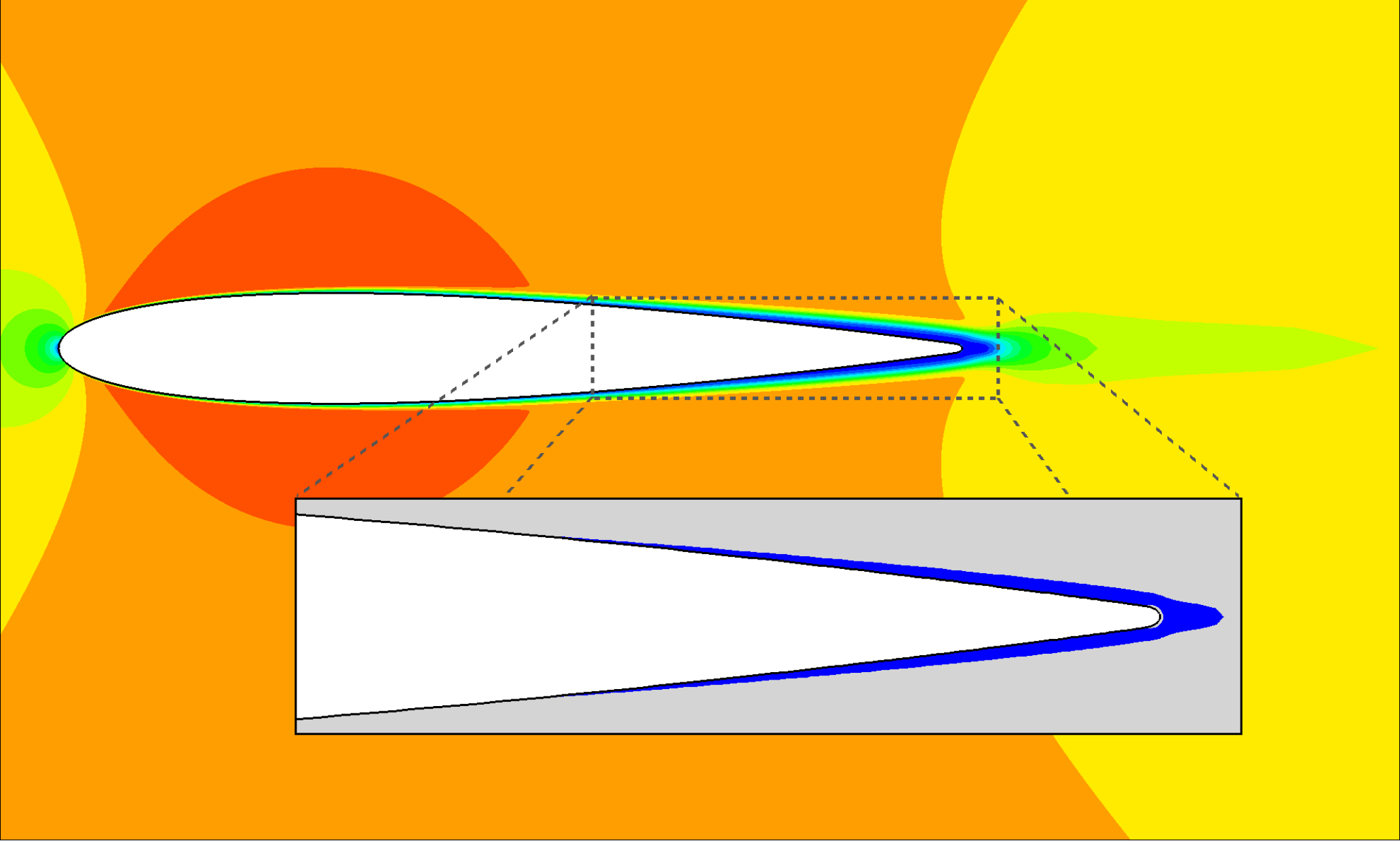}}
	\subfigure[Grid 2]
	{\includegraphics[width=0.33\textwidth,trim={1mm 10mm 15mm 1mm },clip]
		{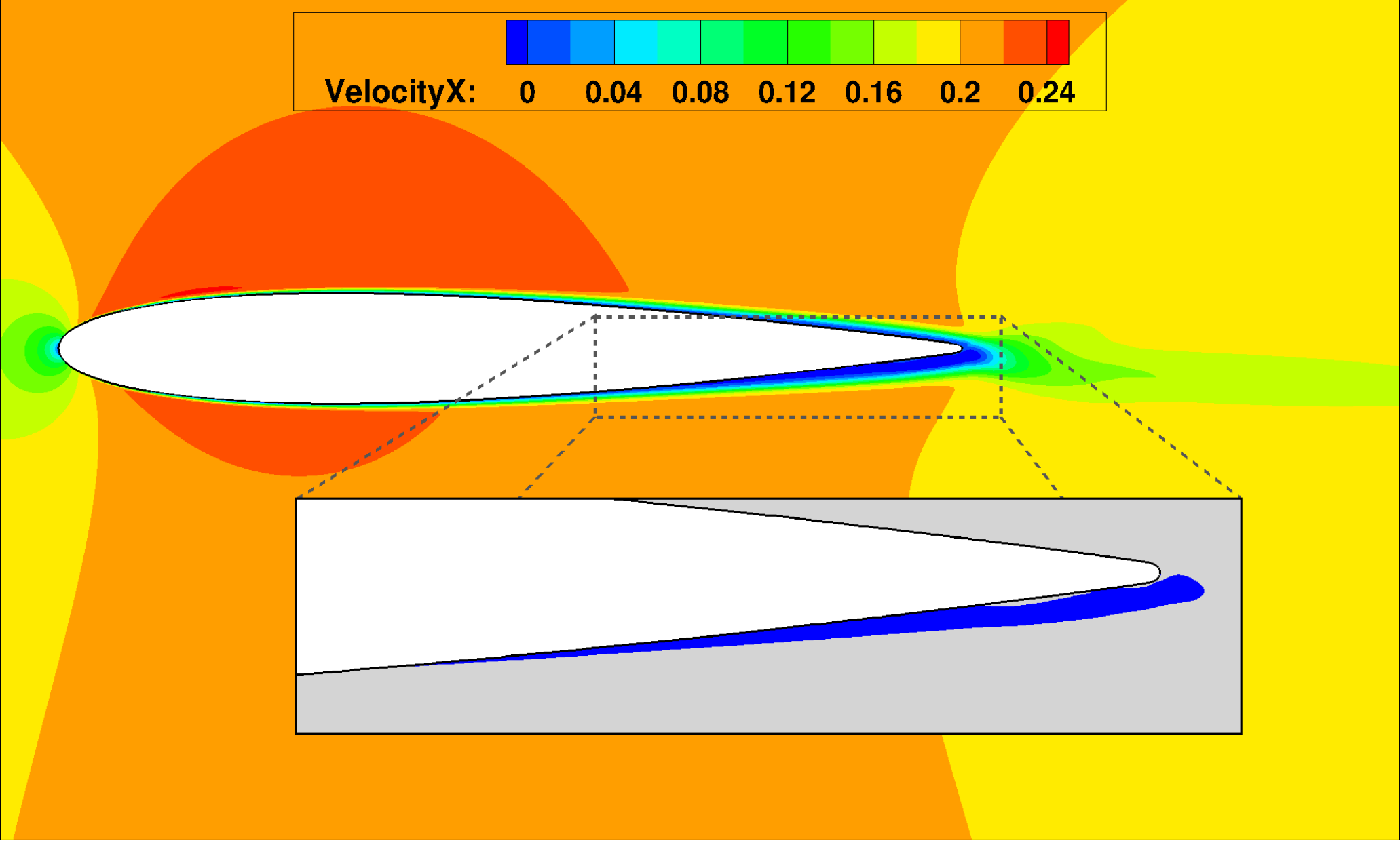}}
	\subfigure[Grid 3]
	{\includegraphics[width=0.33\textwidth,trim={1mm 10mm 15mm 1mm },clip]
		{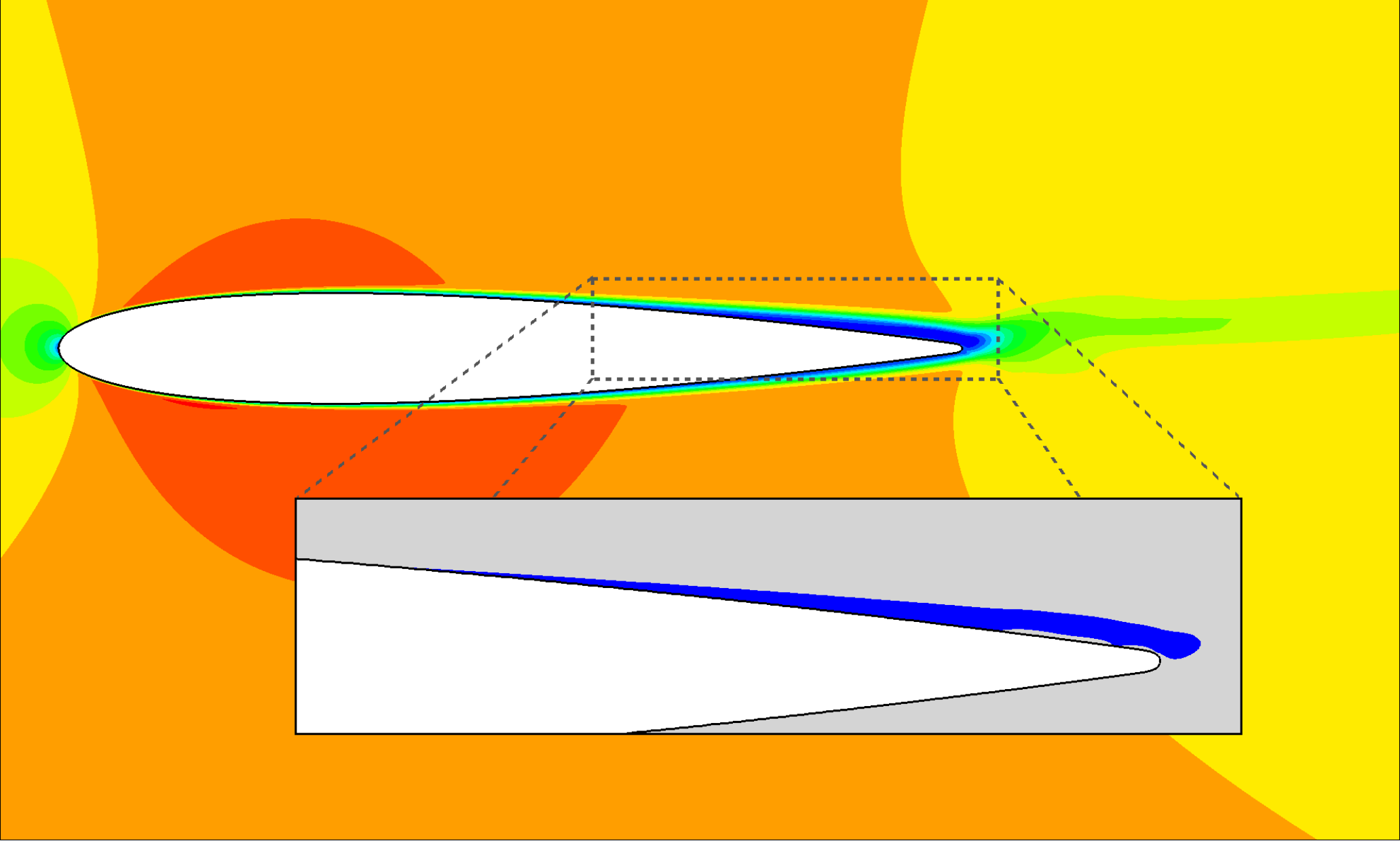}}
	\caption{Flow contours of mean u-velocity for $M_{\infty} = 0.2$ with detail view of separation bubble.}
	\label{fig_bubble_M02}
\end{figure}

In Fig. \ref{fig_Prms_gridref}, solid lines indicate pressure values on the airfoil side where a separation bubble appears while dashed lines indicate values on the opposite side. When the mesh resolution is improved, this asymmetry effect becomes evident for all cases analyzed. For Mach numbers 0.1 and 0.2, results obtained for the coarsest mesh (Grid 1) are still symmetric as shown by red lines in Figs. \ref{fig_Prms_gridref}(a) and (b). 
This effect can be further visualized in the mean flow computed for $M_{\infty}=0.2$ in Fig. \ref{fig_bubble_M02}. The solution obtained for Grid 1 is symmetric, with identical separation bubbles on both sides of the airfoil. When the grid resolution is improved, the separation bubble selects one of the sides of the airfoil, as shown in Figs. \ref{fig_bubble_M02}(b) and (c), and mean lift is generated by pressure variation along the airfoil surface. The mean flows computed for Grids 2 and 3 are almost identical, being only mirrored with respect to the x-axis.
%
%

\begin{figure}[H]
	\begin{subfigmatrix}{3}
		\subfigure[$M_\infty=0.1$]
		{\includegraphics[width=0.3\textwidth,trim={5mm 5mm 20mm 20mm},clip]
			{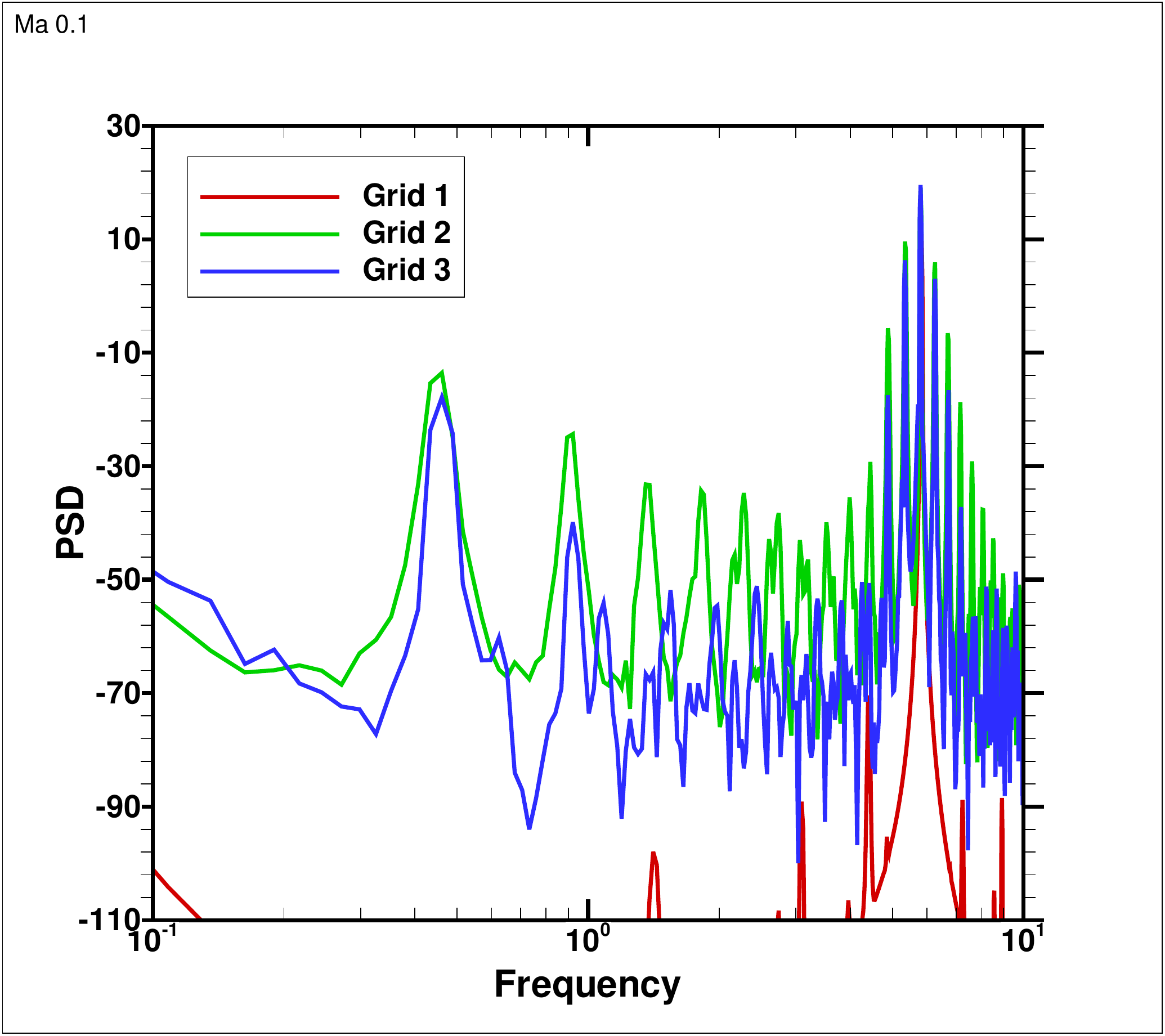}}
		\subfigure[$M_\infty=0.2$]
		{\includegraphics[width=0.3\textwidth,trim={5mm 5mm 20mm 20mm},clip]
			{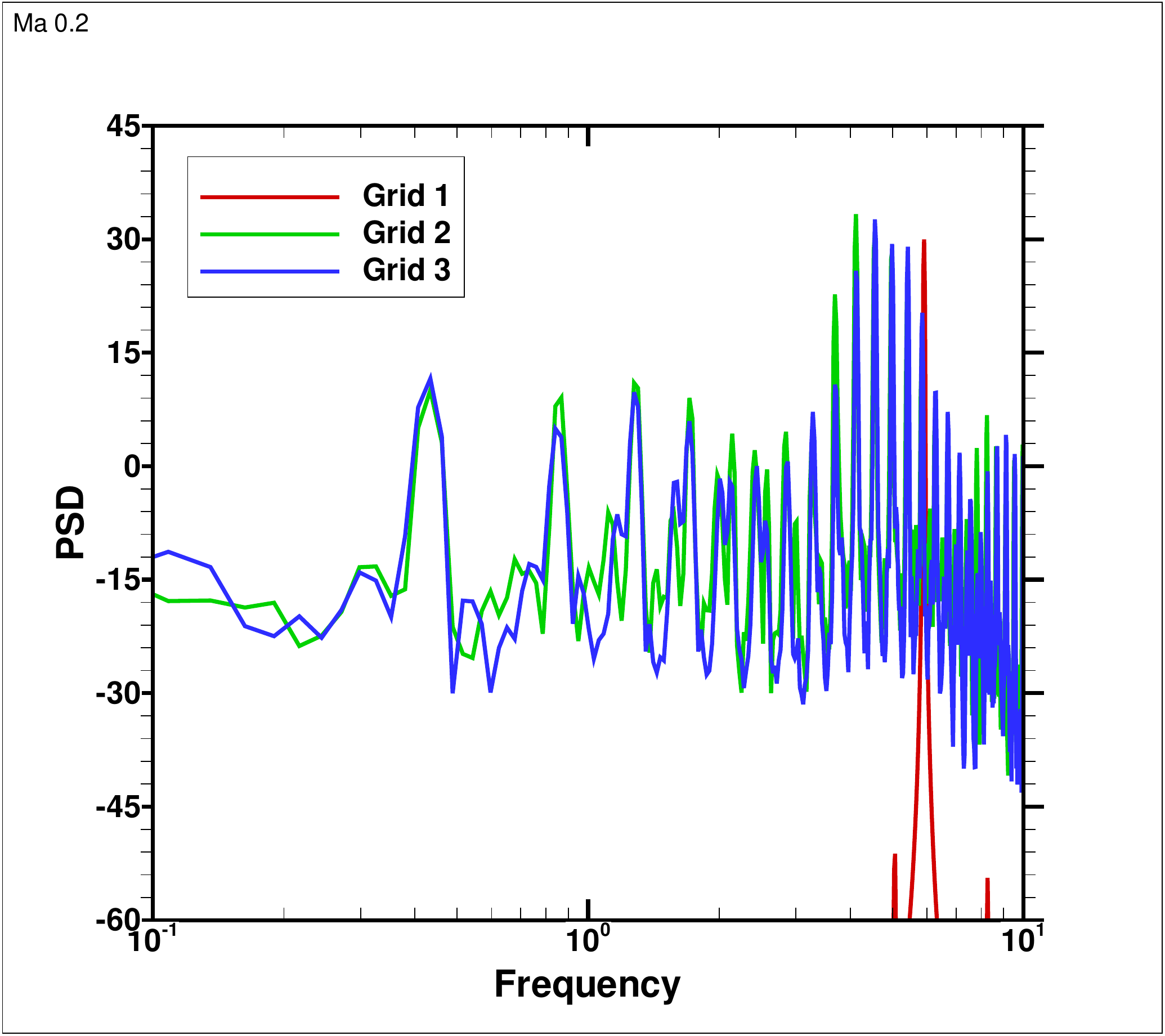}}
		\subfigure[$M_\infty=0.3$]
		{\includegraphics[width=0.3\textwidth,trim={5mm 5mm 20mm 20mm},clip]
			{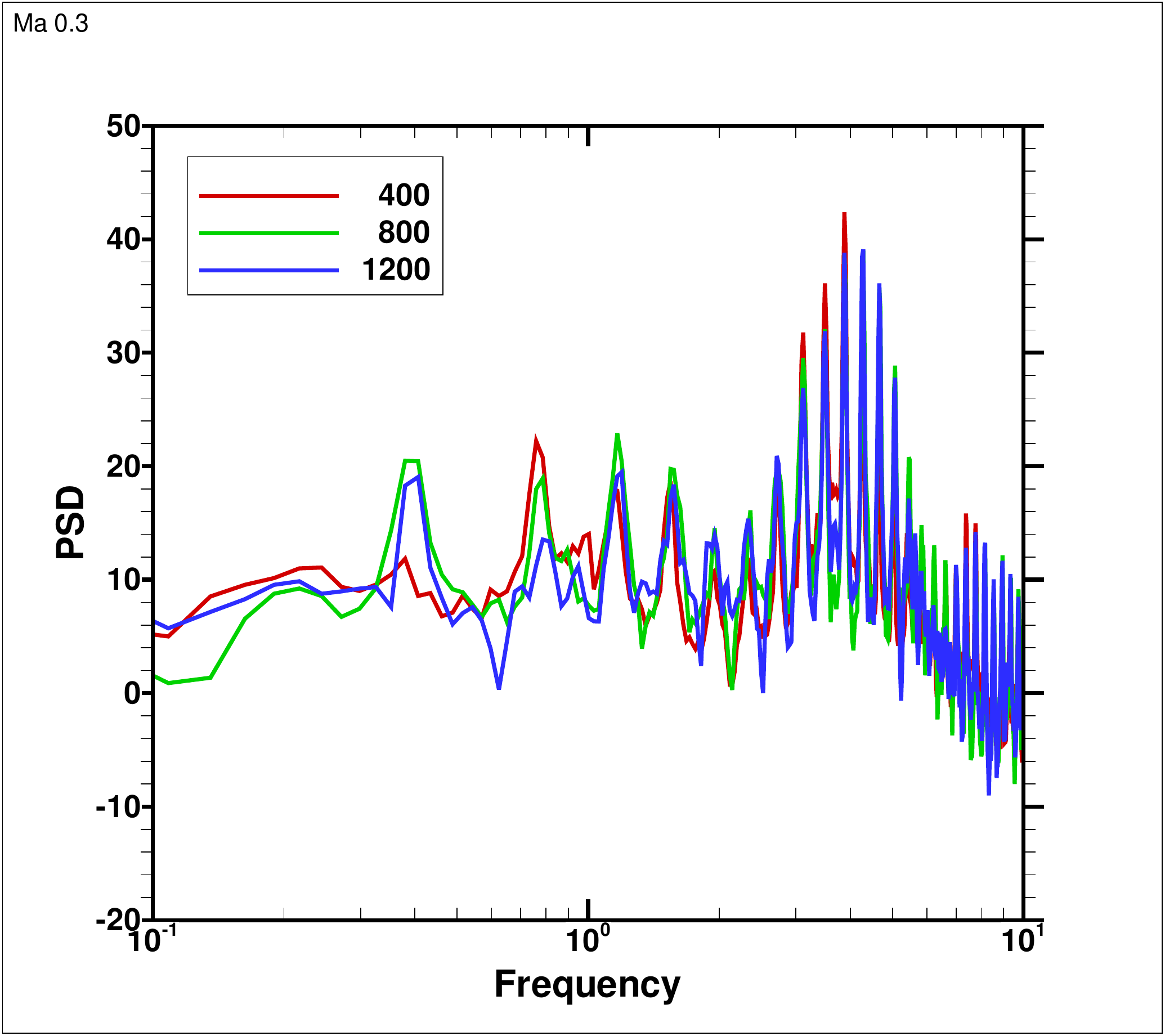}}
	\end{subfigmatrix}
	\caption{Pressure PSDs in dBs obtained for different grids and Mach numbers as a function of non-dimensional frequency $f = f^* \, L /U_{\infty}$. Probe location is at $x=0.98$ and $y=0.98$.}
	\label{fig_spectra_gridref}
\end{figure}

In order to assess the effects of compressibility on noise generation and emission, pressure power spectral densities (PSDs) are presented in dBs in Fig. \ref{fig_spectra_gridref} for different mesh resolutions. The PSDs are computed in the acoustic field, one chord length away from the airfoil trailing edge at $x=0.98$ and $y=0.98$ if the bubble appears on the airfoil top surface, or at $x=0.98$ and $y=-0.98$ when the bubble appears on its bottom surface. Solutions are shown as a function of non-dimensional frequency given by Strouhal number ($f = f^* \, L /U_{\infty}$) and results obtained for Grids 2 and 3 show good convergence, especially for the higher Mach numbers. For the lower Mach numbers, solutions obtained for Grid 1 display different features which are related to flow symmetry as previously discussed. The PSDs have similar characteristics for all Mach numbers including a main tonal peak with secondary tones superposed on a broadband content. Compressibility effects impact mainly the PSD levels and scaling with Strouhal number is questionable since frequency variations are observed for the main tones, an issue discussed later in this section. However, the non-dimensional tonal frequencies are close for the various Mach numbers, what can be more easily visualized for the low frequency tones.
%

\begin{figure}[H]
	\begin{subfigmatrix}{2}
		\subfigure[Temporal window 1]
		{\includegraphics[width=0.495\textwidth,trim={5mm 1mm 2mm 5mm},clip]
			{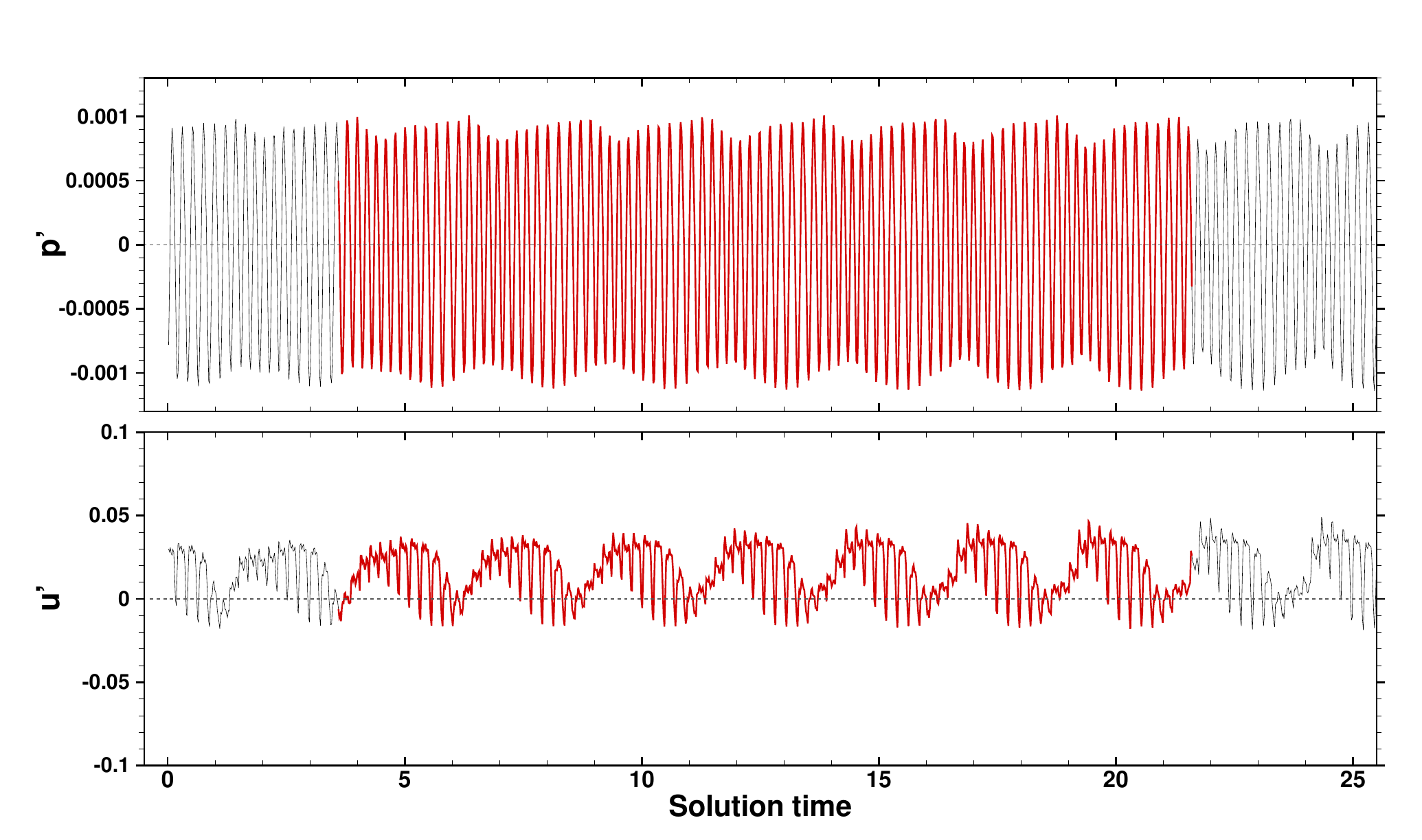}}
		\subfigure[Temporal window 2]
		{\includegraphics[width=0.495\textwidth,trim={5mm 1mm 2mm 5mm},clip]
			{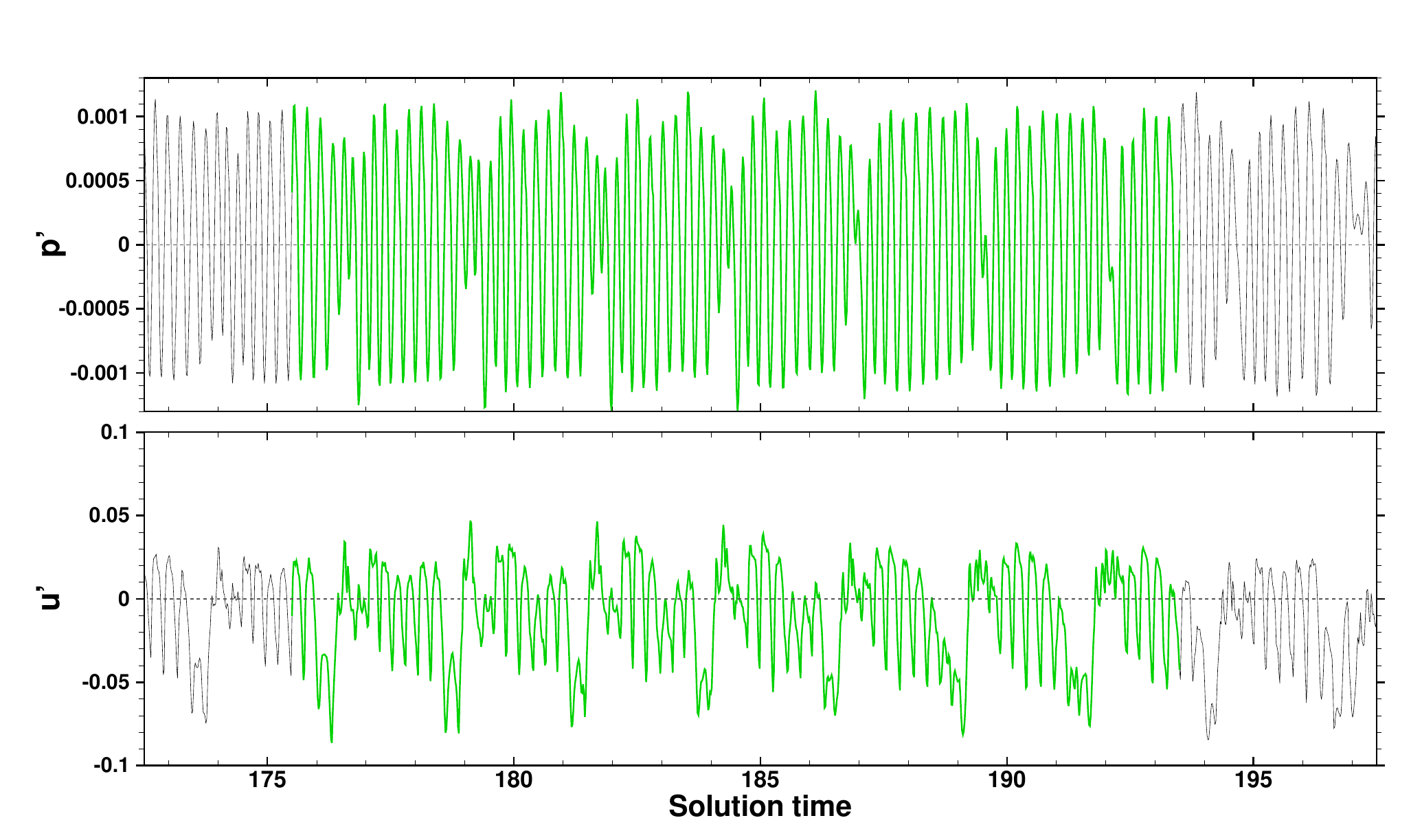}}
	\end{subfigmatrix}
	\caption{Pressure and u-velocity fluctuation signals for $M_{\infty} = 0.3$ and different time windows as function of non-dimensional time $t=t^* \, U_{\infty} / L$.}
	\label{fig_time_M03}
\end{figure}

Similar flow features are observed for all configurations from a visual inspection of the numerical solutions. Hence, from now on, we concentrate our analysis in the $M_{\infty}=0.3$ configuration. For this case, the separation bubble appears on the airfoil top surface for all grids tested. Considering that convergence is achieved for both Grids 2 and 3, further results and analyses will be presented for Grid 2. To provide a better understanding of the spectral characteristics of noise generation and emission, pressure and velocity fluctuation signals are presented in Fig. \ref{fig_time_M03} for different temporal windows of the simulation. Velocity fluctuations are computed near the trailing-edge surface at $x=0.940$ and $y=0.026$ while pressure fluctuations are computed in the acoustic field at $x=0.98$ and $y=0.98$.
Results are presented as function of non-dimensional time $t=t^* \, U_{\infty} / L$. It should be clarified that the simulation is run for a significant amount of time and initial transients are discarded before statistics are collected. Portions of the signals are highlighted in red and green in windows 1 and 2, Figs. \ref{fig_time_M03}(a) and (b), respectively. These windows show that pressure and velocity fluctuations exhibit changing patterns during the flow. Intense amplitude modulations occur for the signals, a feature that was also observed by \cite{Desquesnes2007,Probsting2014} but not discussed with respect to the changing forms of modulation. Further discussion about this issue will be provided later.
%

\begin{figure}[H]
	\begin{subfigmatrix}{2}
		\subfigure[U-velocity (red signal in Fig. \ref{fig_time_M03})]
		{\includegraphics[width=0.4\textwidth,trim={5mm 5mm 5mm 5mm},clip]
			{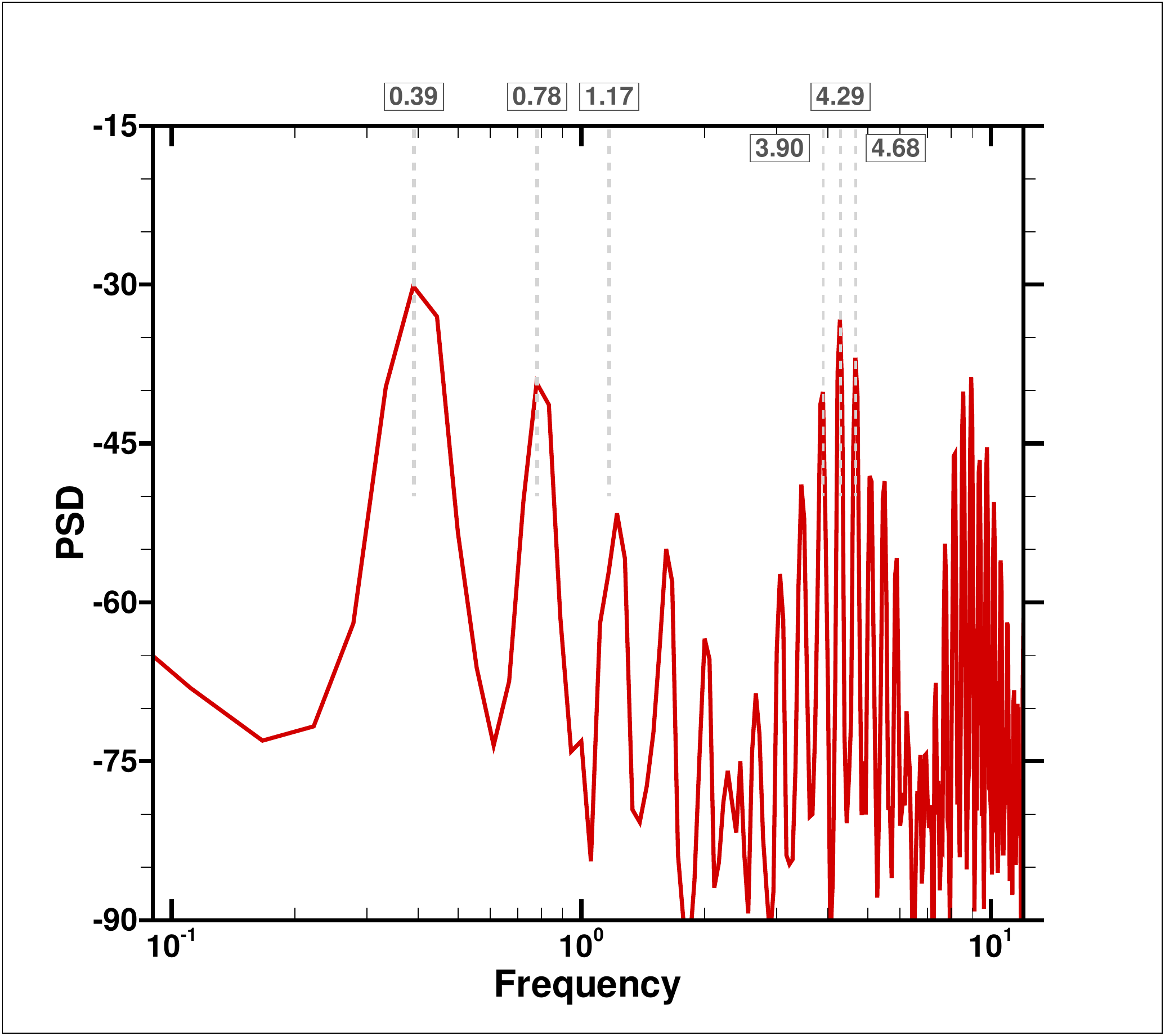}}
		\subfigure[Pressure (red signal in Fig. \ref{fig_time_M03})]
		{\includegraphics[width=0.4\textwidth,trim={5mm 5mm 5mm 5mm},clip]
			{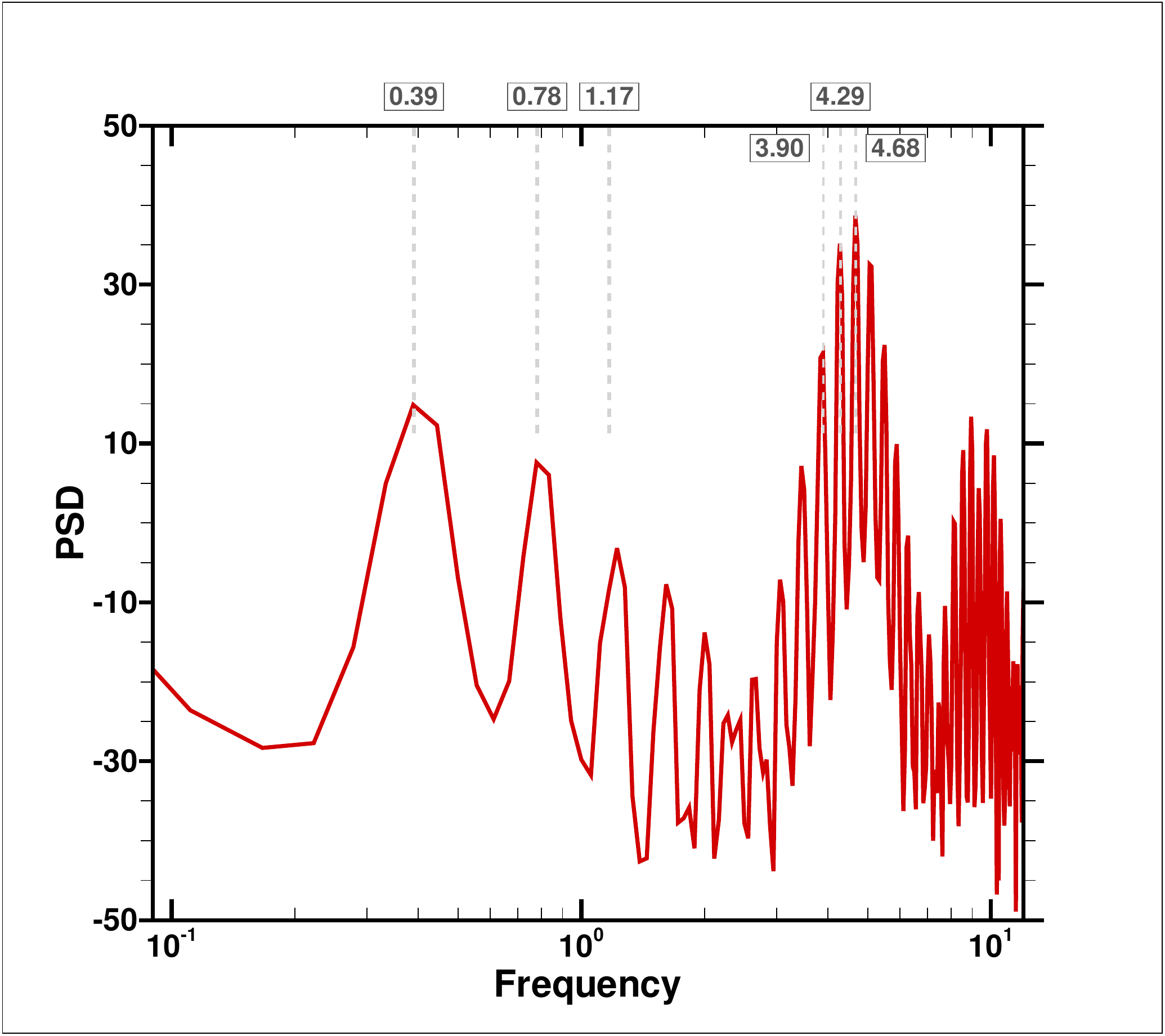}}
		\subfigure[U-velocity (green signal in Fig. \ref{fig_time_M03})]
		{\includegraphics[width=0.4\textwidth,trim={5mm 5mm 5mm 5mm},clip]
			{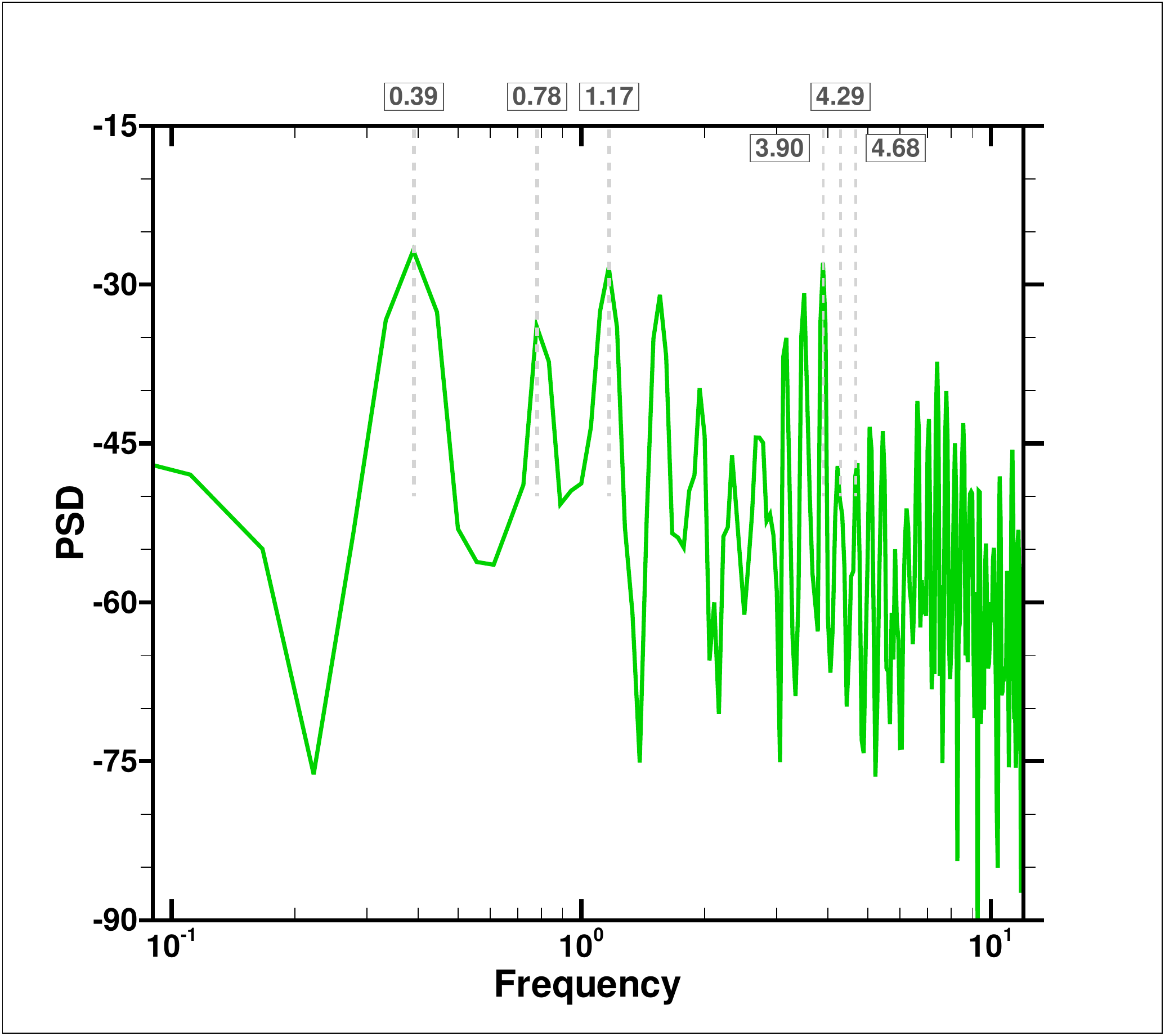}}
		\subfigure[Pressure (green signal in Fig. \ref{fig_time_M03})]
		{\includegraphics[width=0.4\textwidth,trim={5mm 5mm 5mm 5mm},clip]
			{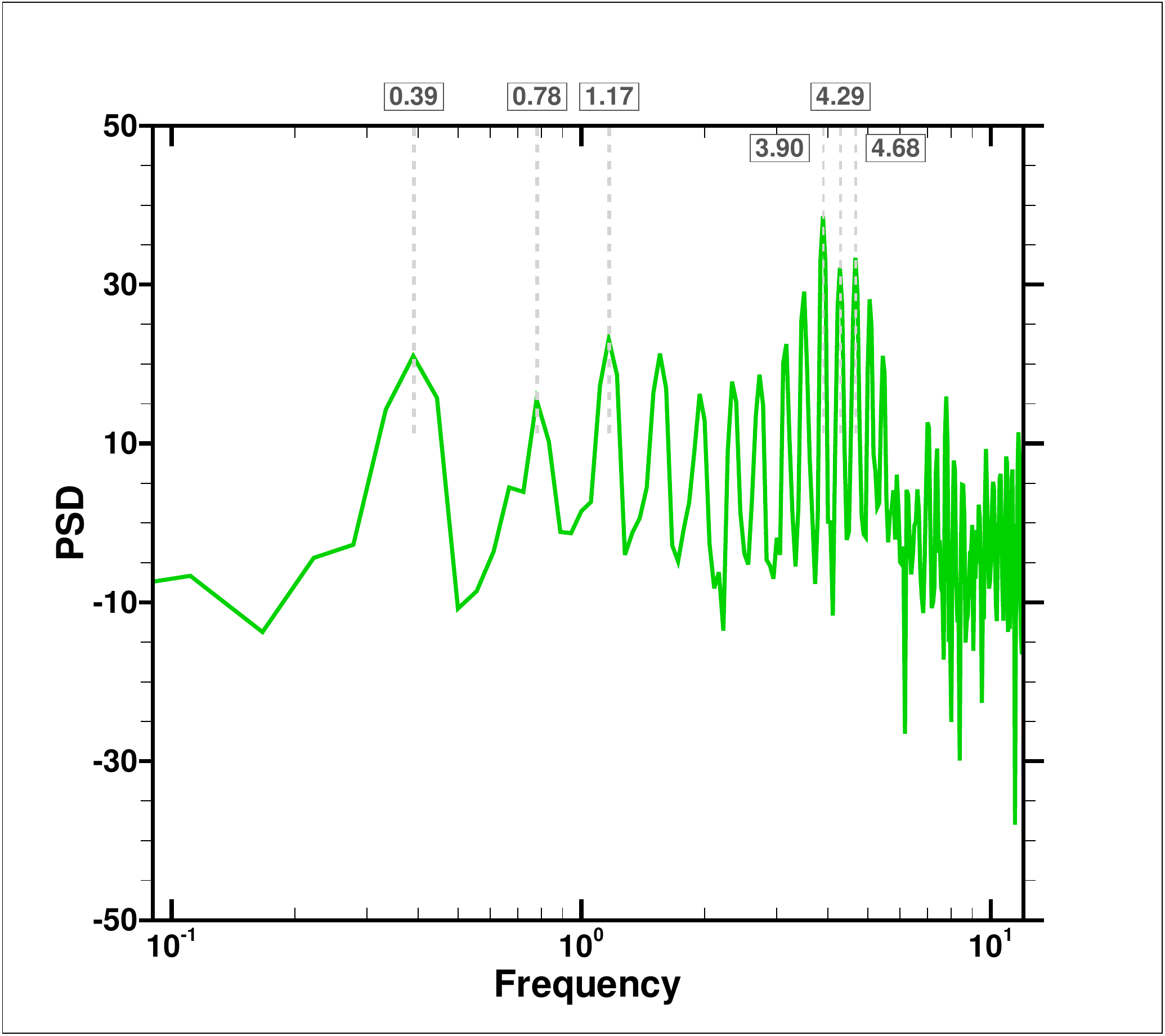}}		
	\end{subfigmatrix}
	\caption{Power spectral densities of different time windows for $M_{\infty} = 0.3$.}
	\label{fig_spectra_M03}
\end{figure}

Figure \ref{fig_spectra_M03} presents PSDs of the individual temporal signals highlighted in red and green in Fig. \ref{fig_time_M03}. Despite the short period of acquisition for these signals, main tone and secondary peaks are clearly visible in both u-velocity and pressure PSDs. For both velocity spectra, the highest amplitudes are observed for a low frequency tone at Strouhal number $f = f^* \, L /U_{\infty}=0.39$.
On the other hand, for the pressure spectra measured in the acoustic field, the most pronounced tones are obtained at $f = 4.68$ and $3.90$ for the red and green spectra, Figs. \ref{fig_spectra_M03}(b) and (d), respectively. Hence, a frequency shift of the main tonal peak is observed and it is caused by the pattern variations observed in Fig. \ref{fig_time_M03}. For both spectra, secondary peaks are still equidistant and displaced by the frequency of the lowest tone $f =0.39$.
%

\begin{figure}[H]
	\begin{subfigmatrix}{2}
		\subfigure[U-velocity]
		{\includegraphics[width=0.4\textwidth,trim={5mm 5mm 5mm 5mm},clip]
			{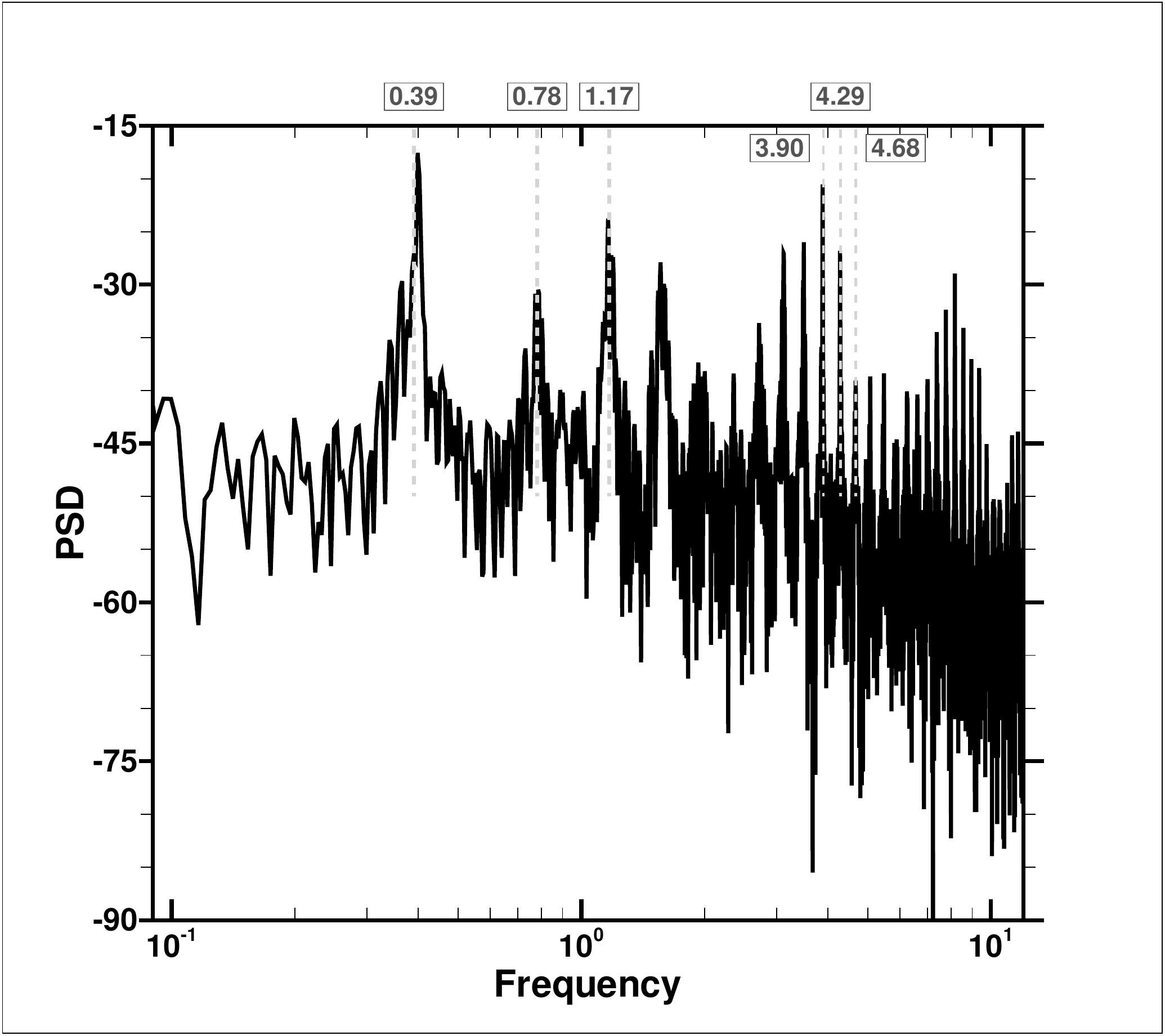}}
		\subfigure[Pressure]
		{\includegraphics[width=0.4\textwidth,trim={5mm 5mm 5mm 5mm},clip]
			{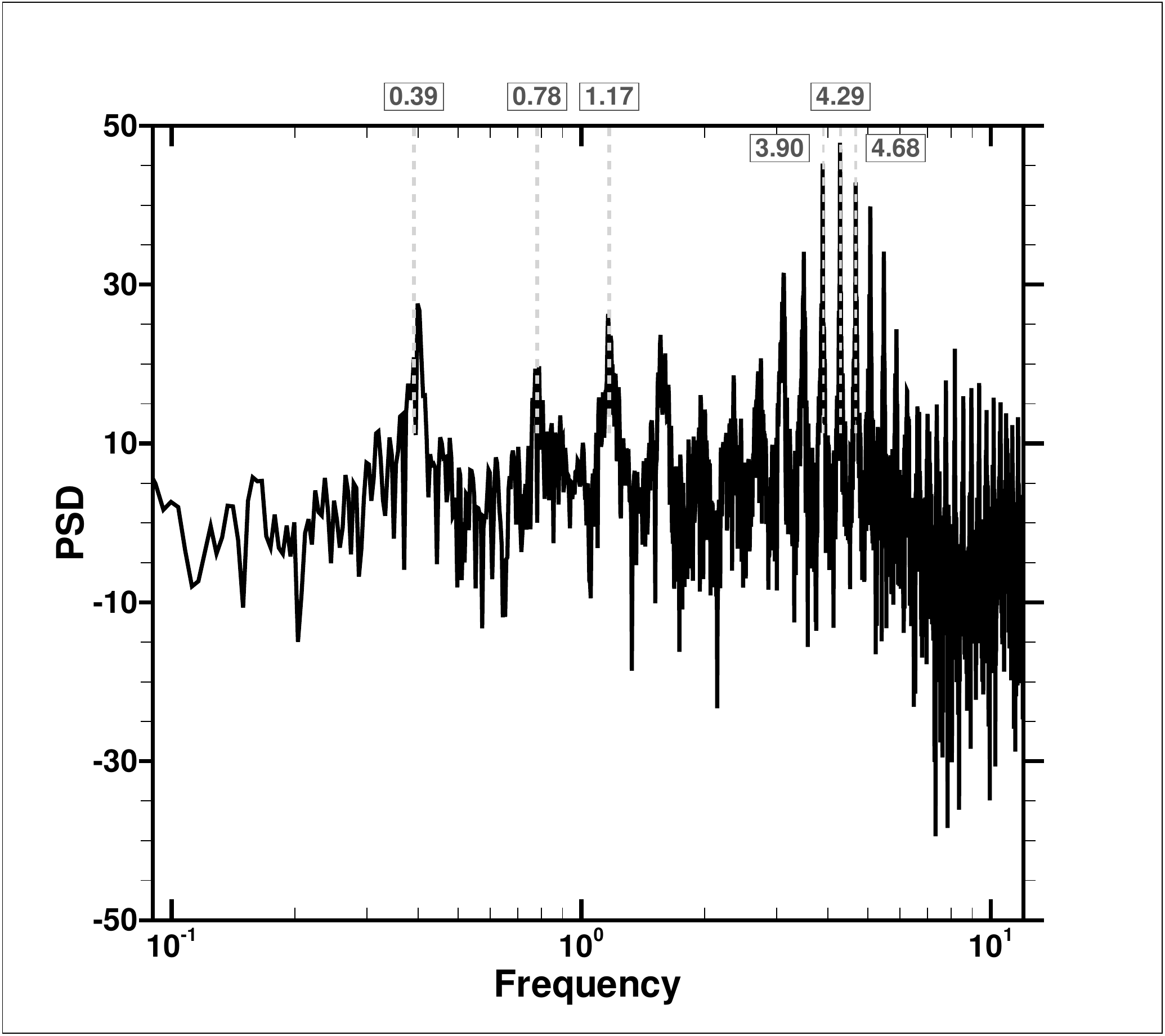}}
	\end{subfigmatrix}
	\caption{Power spectral density of longer temporal signal for $M_{\infty} = 0.3$.}
	\label{fig_spectrafull_M03}
\end{figure}

Figure \ref{fig_spectrafull_M03} shows the u-velocity and pressure PSDs computed for larger temporal windows spanning the entire signal ($0\leq t \leq 250$). These PSDs are calculated with longer bins containing 8192 snapshots with overlap of 67\% using a Hanning window. Results are similar to those observed in Fig. \ref{fig_spectra_M03}, however, the main tone in the pressure spectrum is at $f=4.29$. Again, the amplitude of the lowest frequency tone is more prominent for the u-velocity spectrum. From these results, there is an indication that TS instabilities are generated at frequencies around $f =4.29$ with frequency shifts proportional to the lowest tone at $f =0.39$.
%

\begin{figure}[H]
	\begin{subfigmatrix}{2}
		\subfigure[U-velocity, $f= 0.39$]
		{\includegraphics[width=0.495\textwidth,trim={1mm 1mm 1mm 1mm},clip]{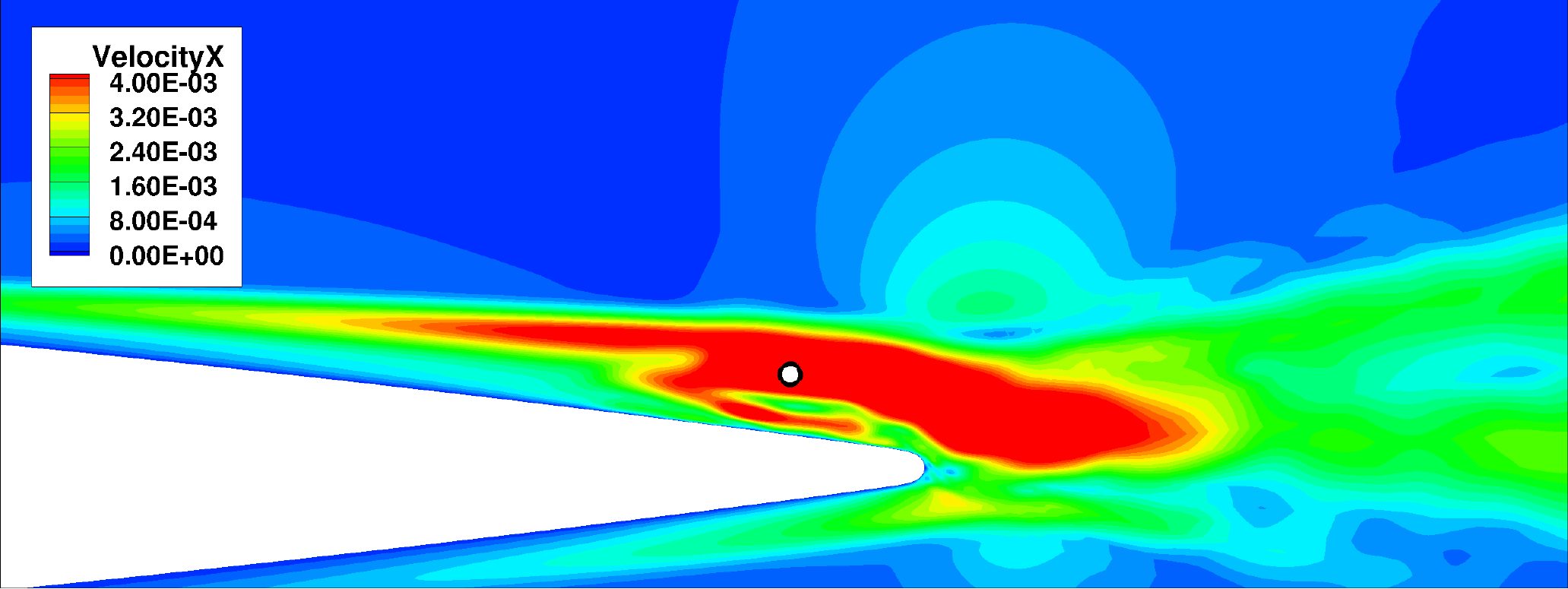}}
		\subfigure[Pressure, $f= 0.39$]
		{\includegraphics[width=0.495\textwidth,trim={1mm 1mm 1mm 1mm},clip]{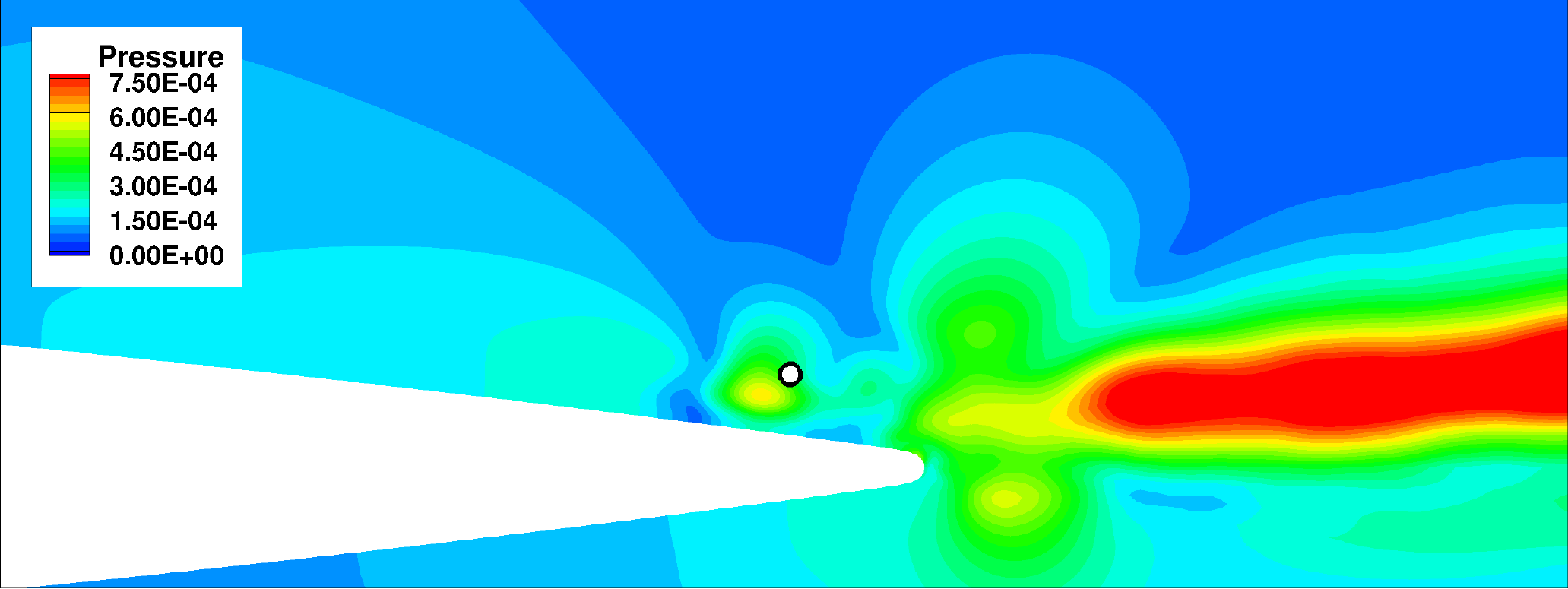}}
		\subfigure[U-velocity, $f= 4.29$]
		{\includegraphics[width=0.495\textwidth,trim={1mm 1mm 1mm 1mm},clip]{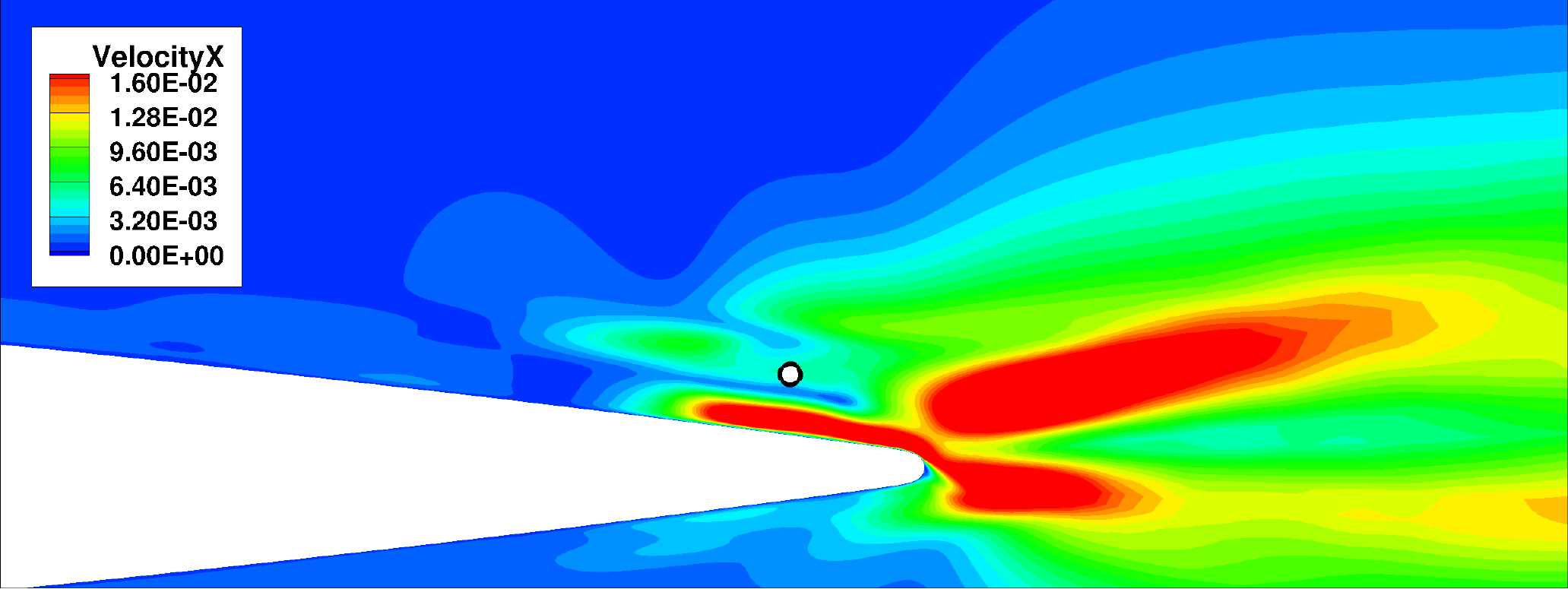}}
		\subfigure[Pressure, $f= 4.29$]
		{\includegraphics[width=0.495\textwidth,trim={1mm 1mm 1mm 1mm},clip]{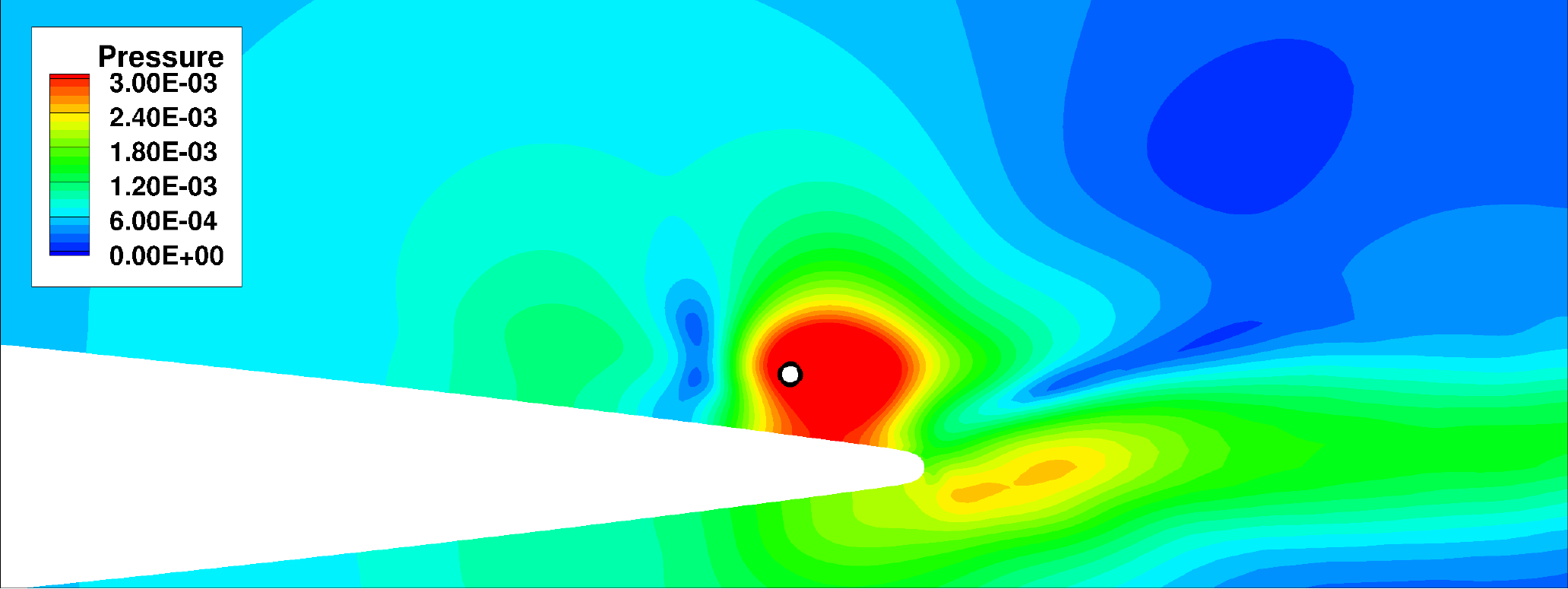}}
	\end{subfigmatrix}
	\caption{Contours of Fourier mode amplitudes of longer temporal signal for $M_{\infty} = 0.3$.}
	\label{fig_fourier_M03}
\end{figure}

In Fig. \ref{fig_fourier_M03}, one can observe the contours of Fourier mode amplitudes for the longer temporal signal considered in Fig. \ref{fig_spectrafull_M03}. The left and right columns display contours of u-velocity and pressure, respectively, while the first and second rows present results for the lowest frequency and main tones at $f=0.39$ and $4.29$, respectively. It is important to notice that contour values for $f=4.29$ are four times higher than those for $f=0.39$. The lower frequency tone appears to be related to motion of the separation bubble as can be seen in Fig. \ref{fig_fourier_M03}(a). At this frequency, pressure fluctuations shown in Fig. \ref{fig_fourier_M03}(b) are mainly visualized along the airfoil near wake. On the other hand, the main tone frequency is related to velocity and pressure fluctuations close to the airfoil trailing edge surface as observed in Figs. \ref{fig_fourier_M03}(c) and (d). This observation endorses the trailing-edge scattering mechanism of TS instabilities. Current results indicate that noise generation dependends on the multiple frequencies of TS instabilities; such frequencies are related both to the main tone and its shifts, which lead to secondary tones. The latter are caused by coupling between motion of the separation bubble, that occurs at the lowest tonal frequency, and TS instabilities which occur at frequencies around that of the main tone. This results in a frequency modulation behavior of the flow quantities. Moreover, acoustic emission is also impacted by amplitude modulation that can be observed in the temporal signal of Fig. \ref{fig_time_M03}.
%

\begin{figure}[H]
		%
		\subfigure[$t=t_a=215.496$]
		{\includegraphics[width=0.33\textwidth,trim={2mm 2mm 2mm 2mm },clip]
			{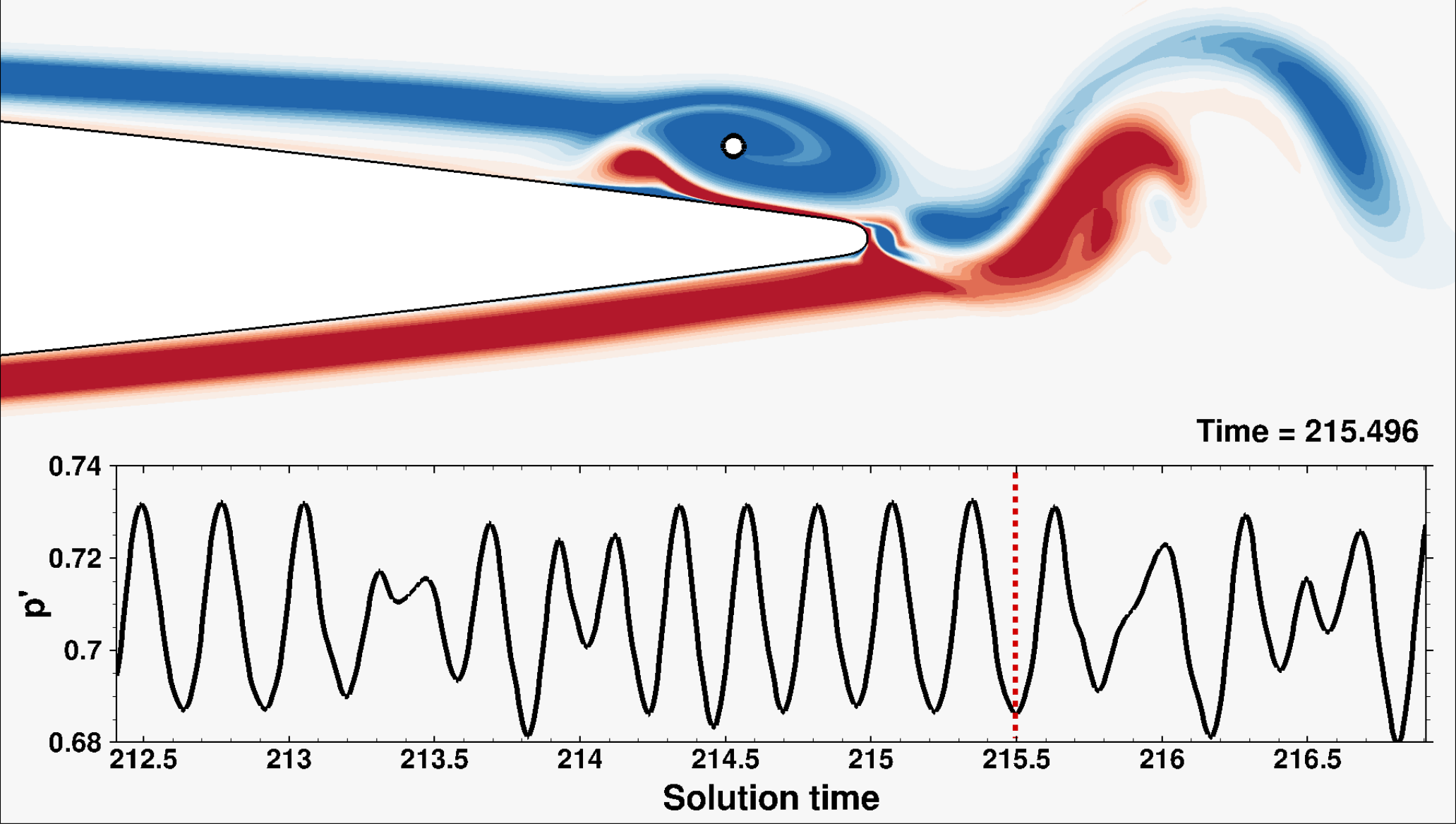}}
		\subfigure[$t_b=t_a+0.079$]
		{\includegraphics[width=0.33\textwidth,trim={2mm 2mm 2mm 2mm },clip]
			{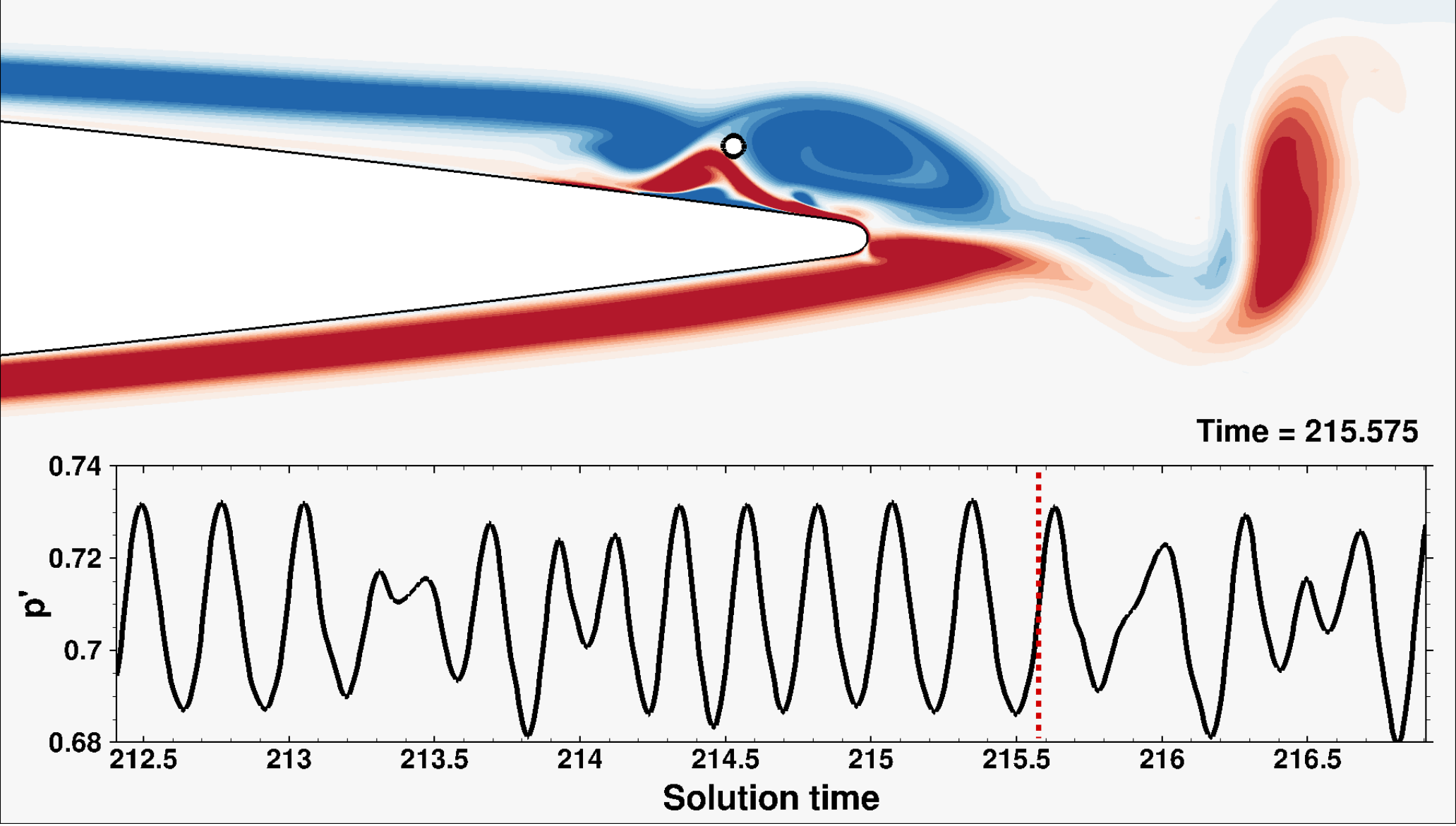}}
		\subfigure[$t_c=t_b+0.065$]
		{\includegraphics[width=0.33\textwidth,trim={2mm 2mm 2mm 2mm },clip]
			{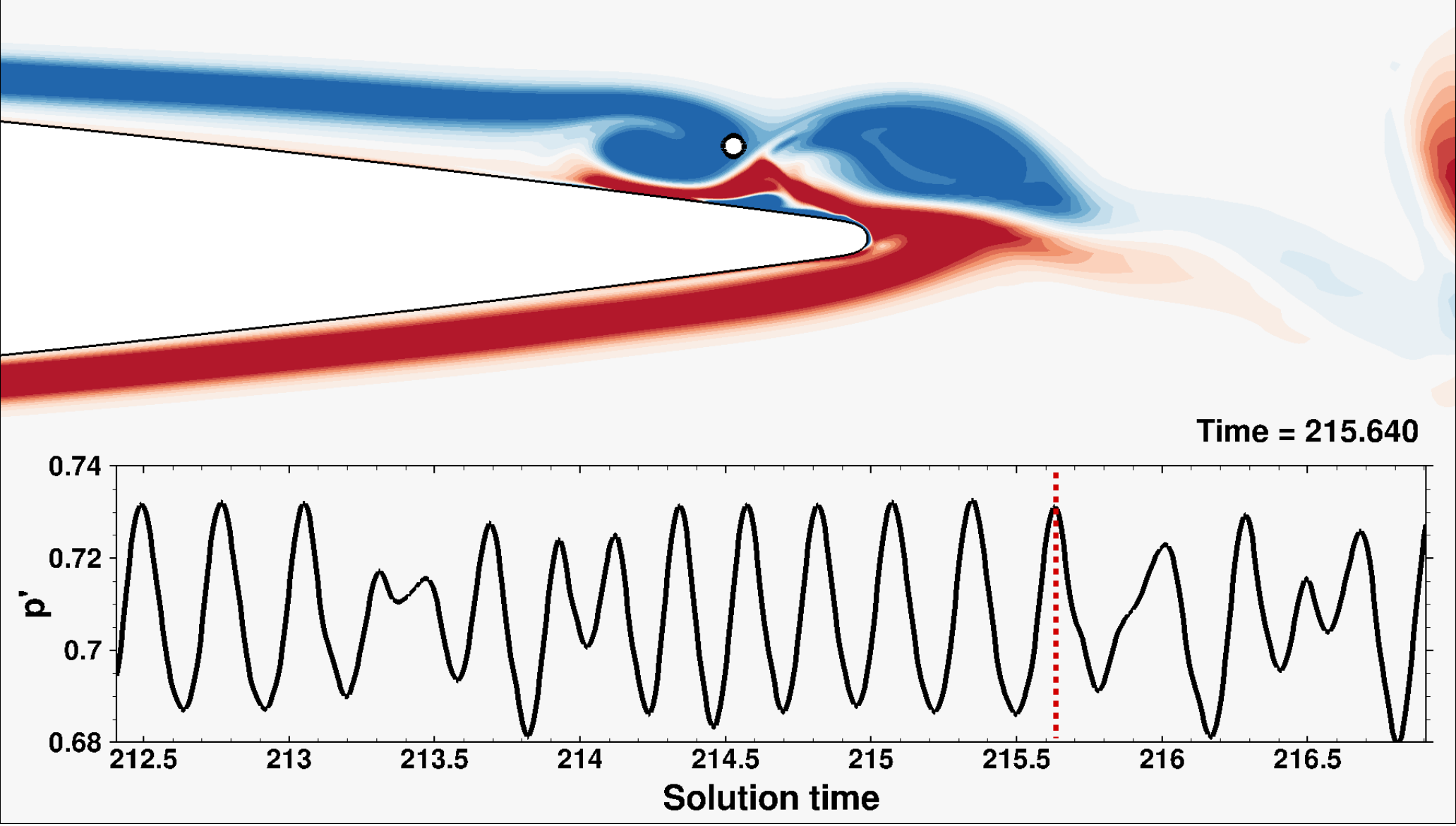}}
		\subfigure[$t_d=t_c+0.058$]
		{\includegraphics[width=0.33\textwidth,trim={2mm 2mm 2mm 2mm },clip]
			{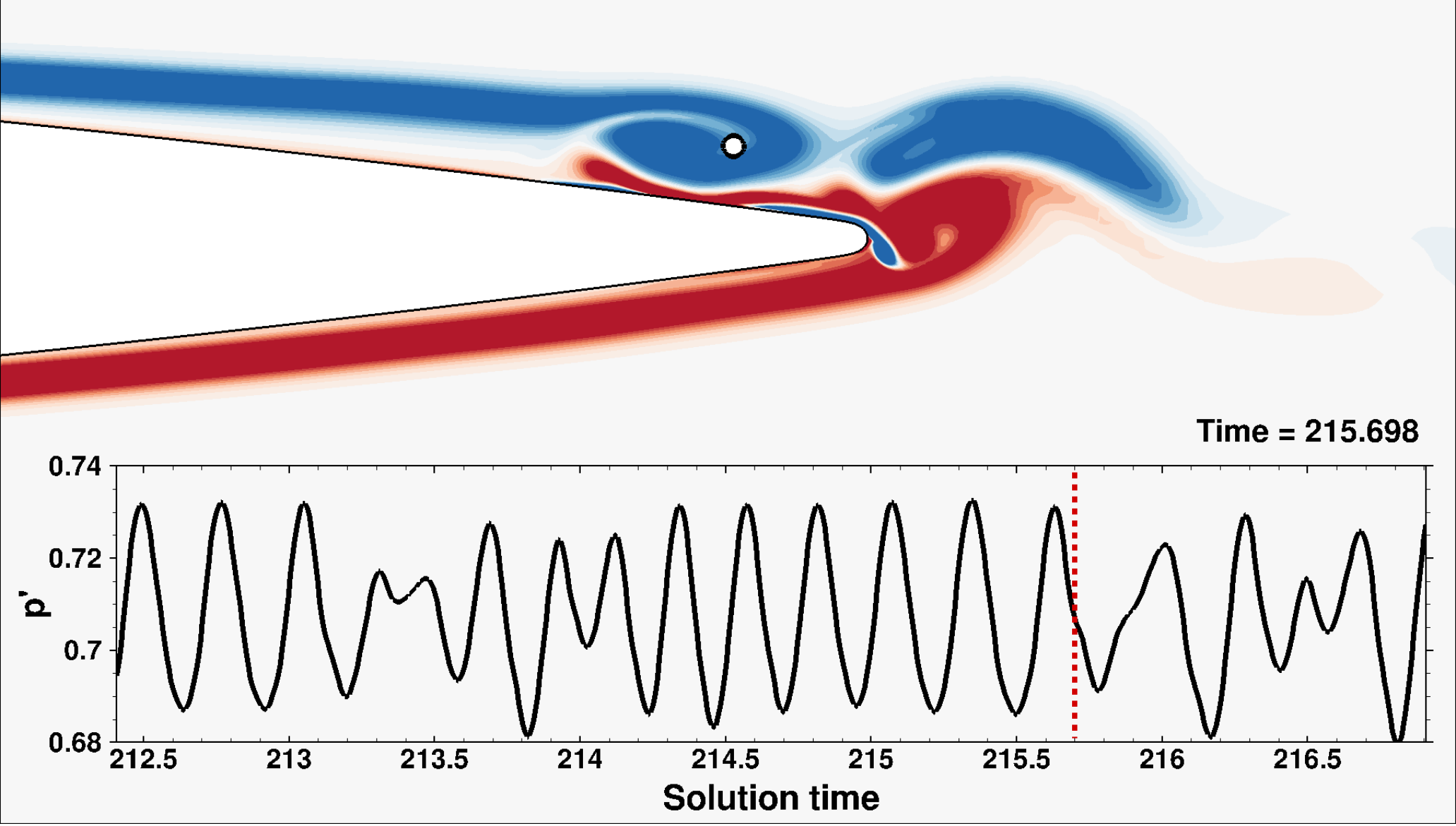}}
		\subfigure[$t_e=t_d+0.086$]
		{\includegraphics[width=0.33\textwidth,trim={2mm 2mm 2mm 2mm },clip]
			{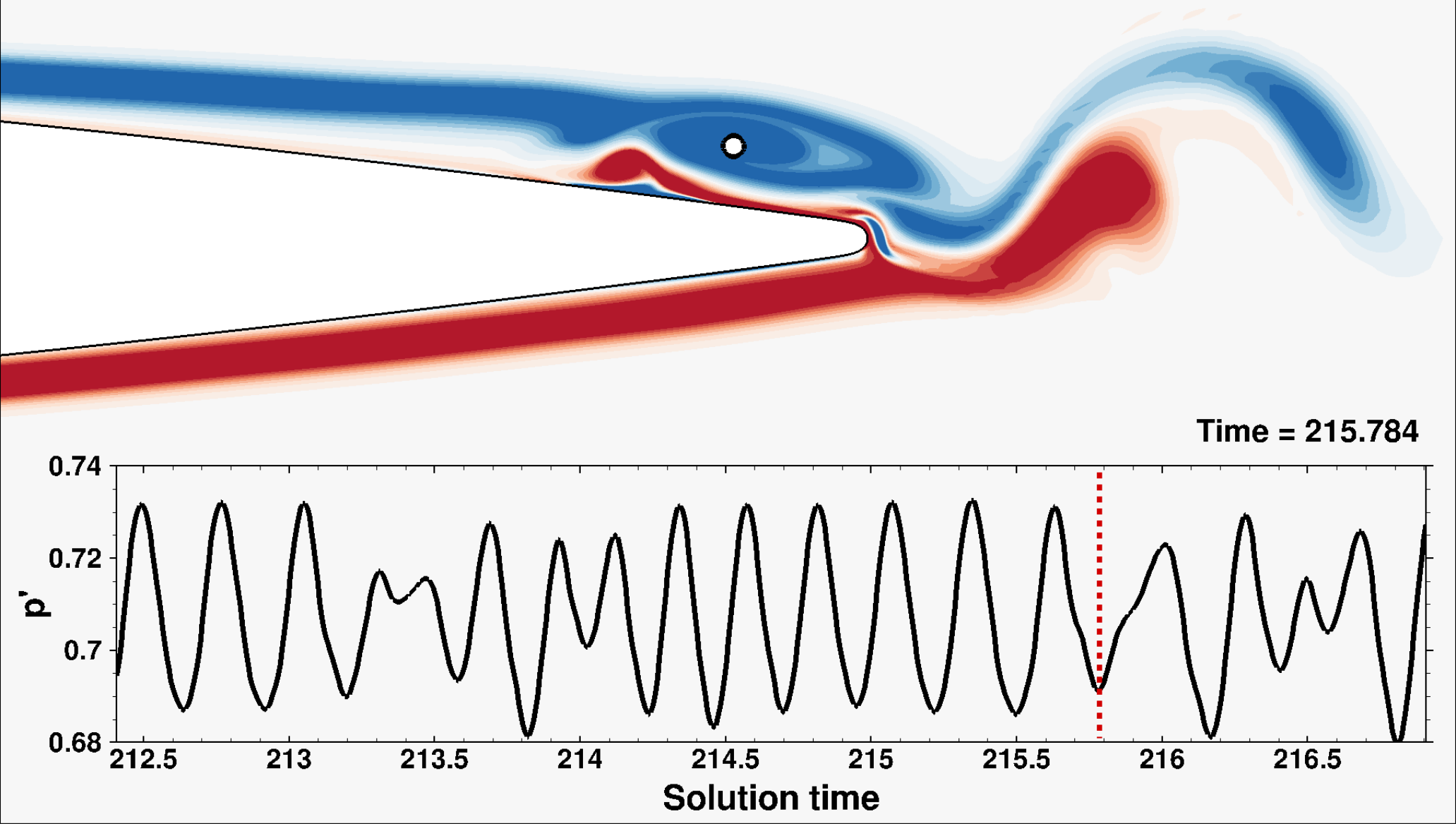}}
		\subfigure[$t_f=t_e+0.108$]
		{\includegraphics[width=0.33\textwidth,trim={2mm 2mm 2mm 2mm },clip]
			{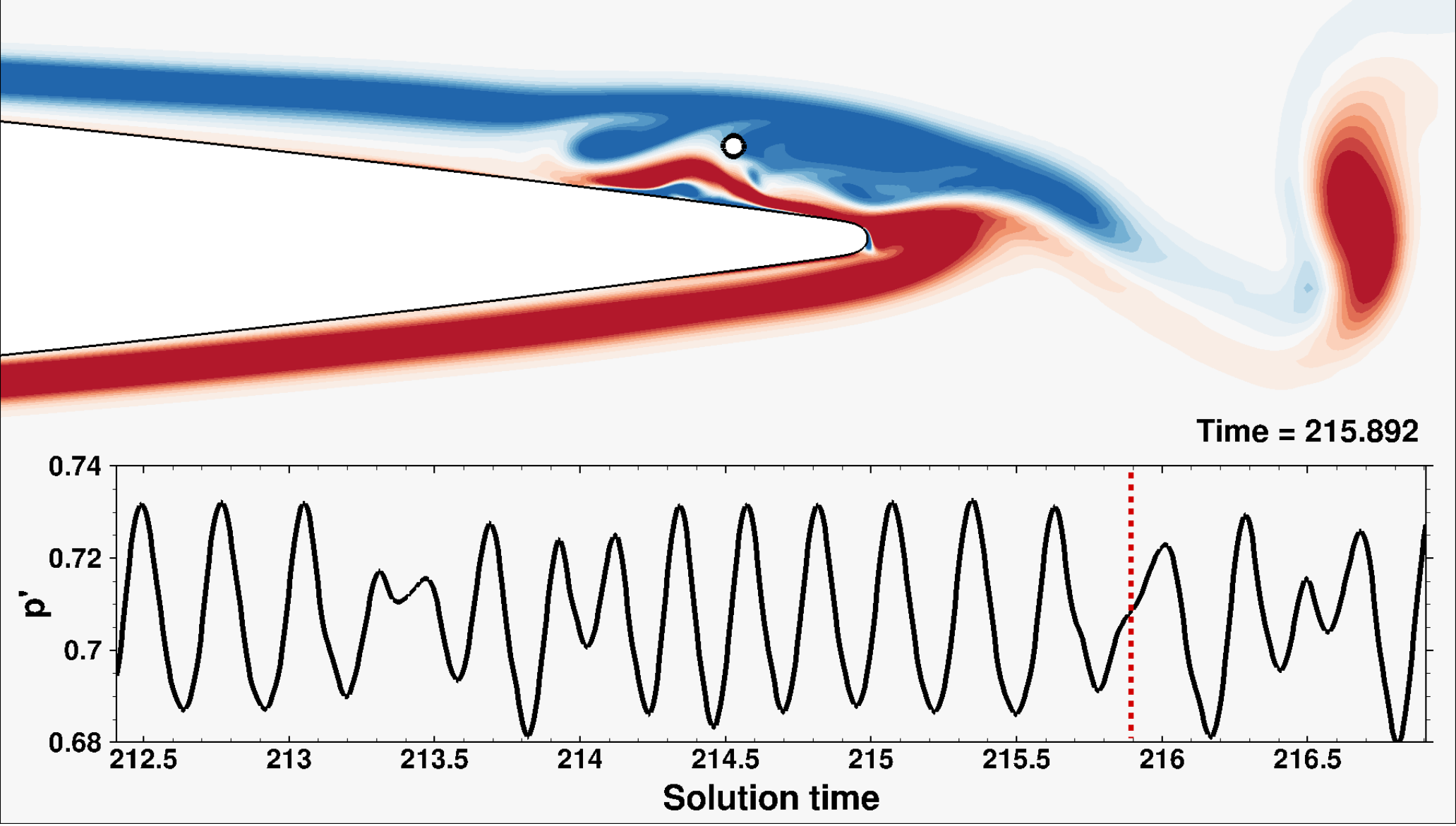}}
		\subfigure[$t_g=t_f+0.115$]
		{\includegraphics[width=0.33\textwidth,trim={2mm 2mm 2mm 2mm },clip]
			{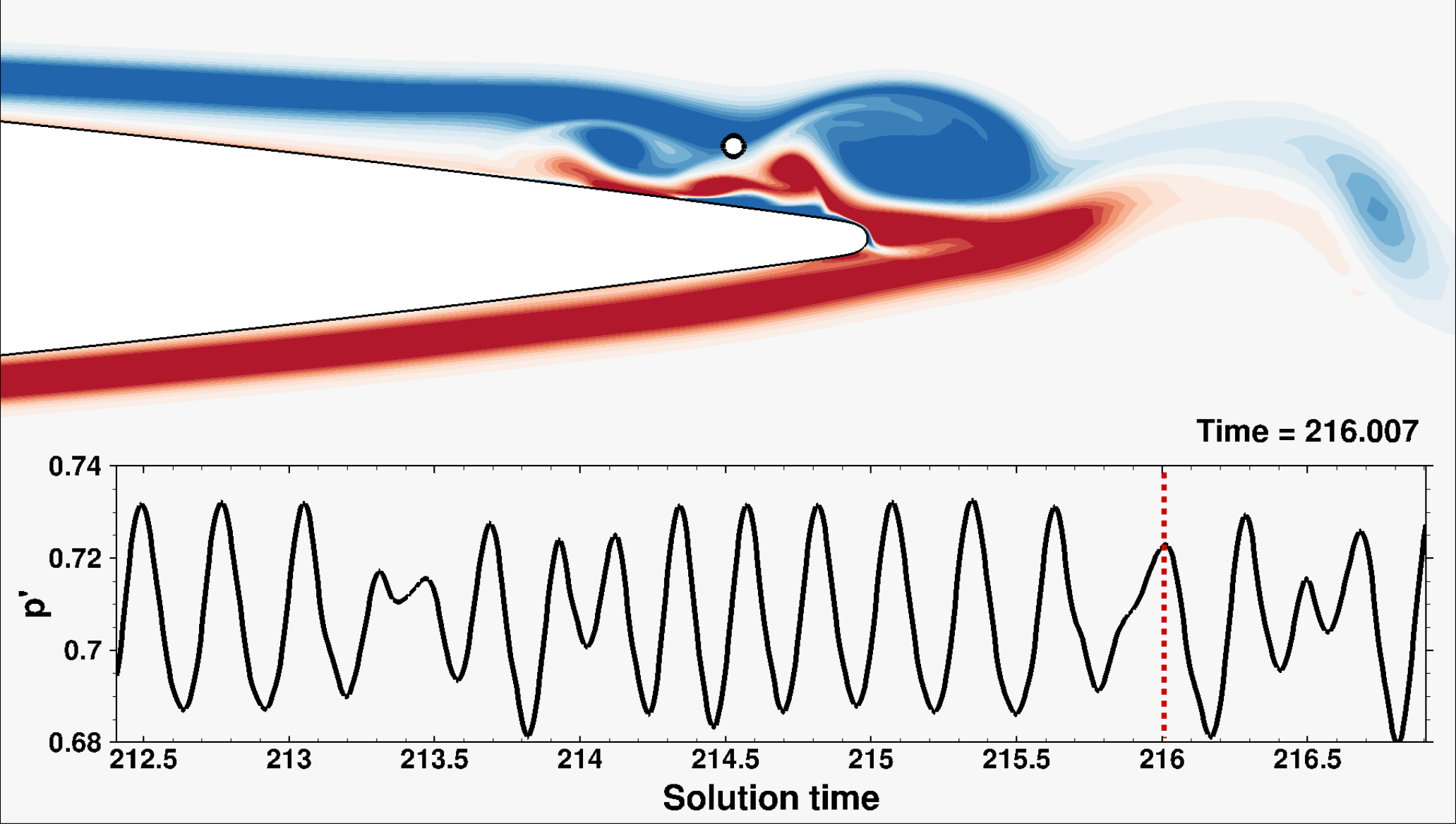}}
		\subfigure[$t_h=t_g+0.072$]
		{\includegraphics[width=0.33\textwidth,trim={2mm 2mm 2mm 2mm },clip]
			{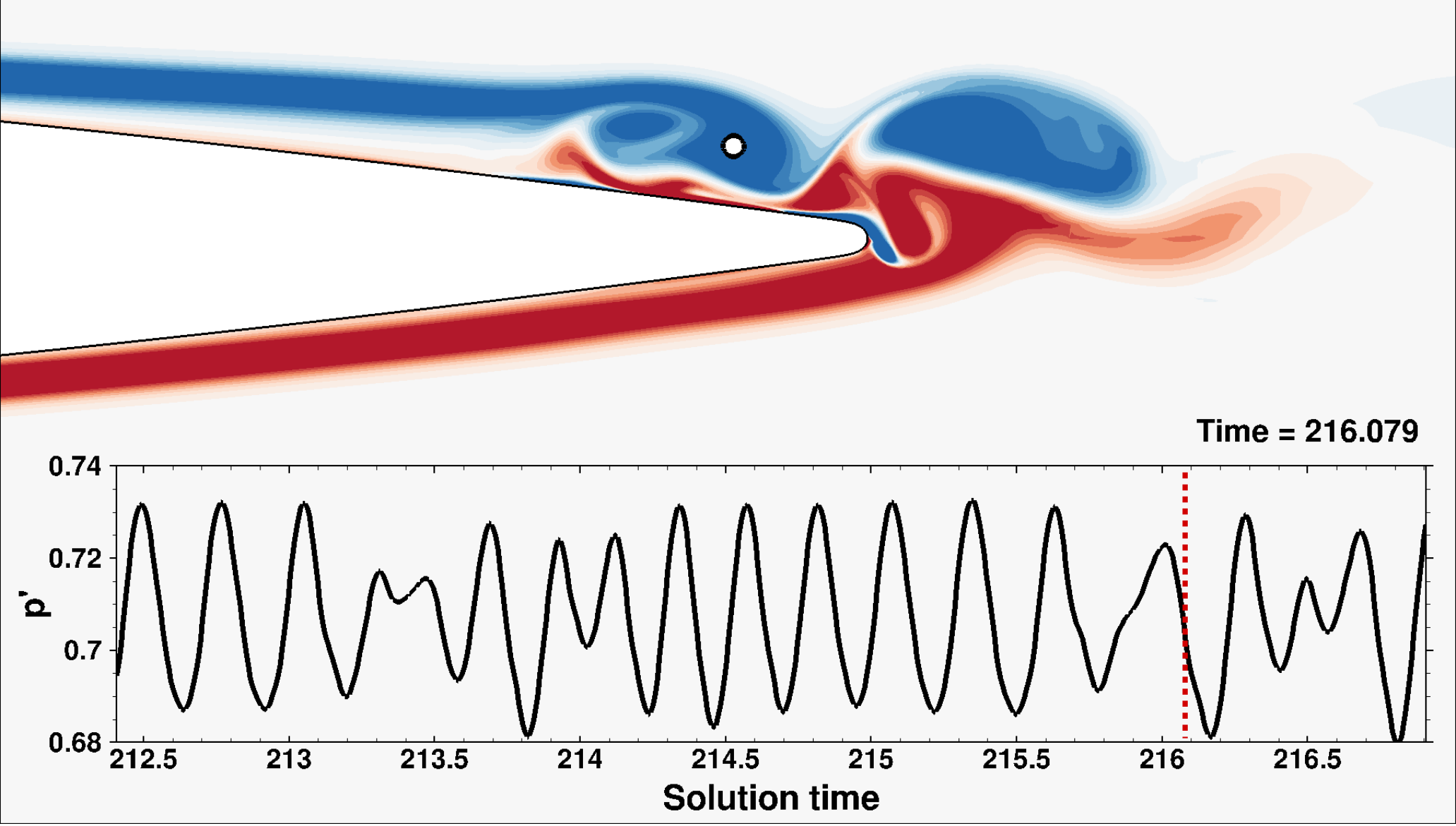}}
		\subfigure[$t_i=t_h+0.087$]
		{\includegraphics[width=0.33\textwidth,trim={2mm 2mm 2mm 2mm },clip]
			{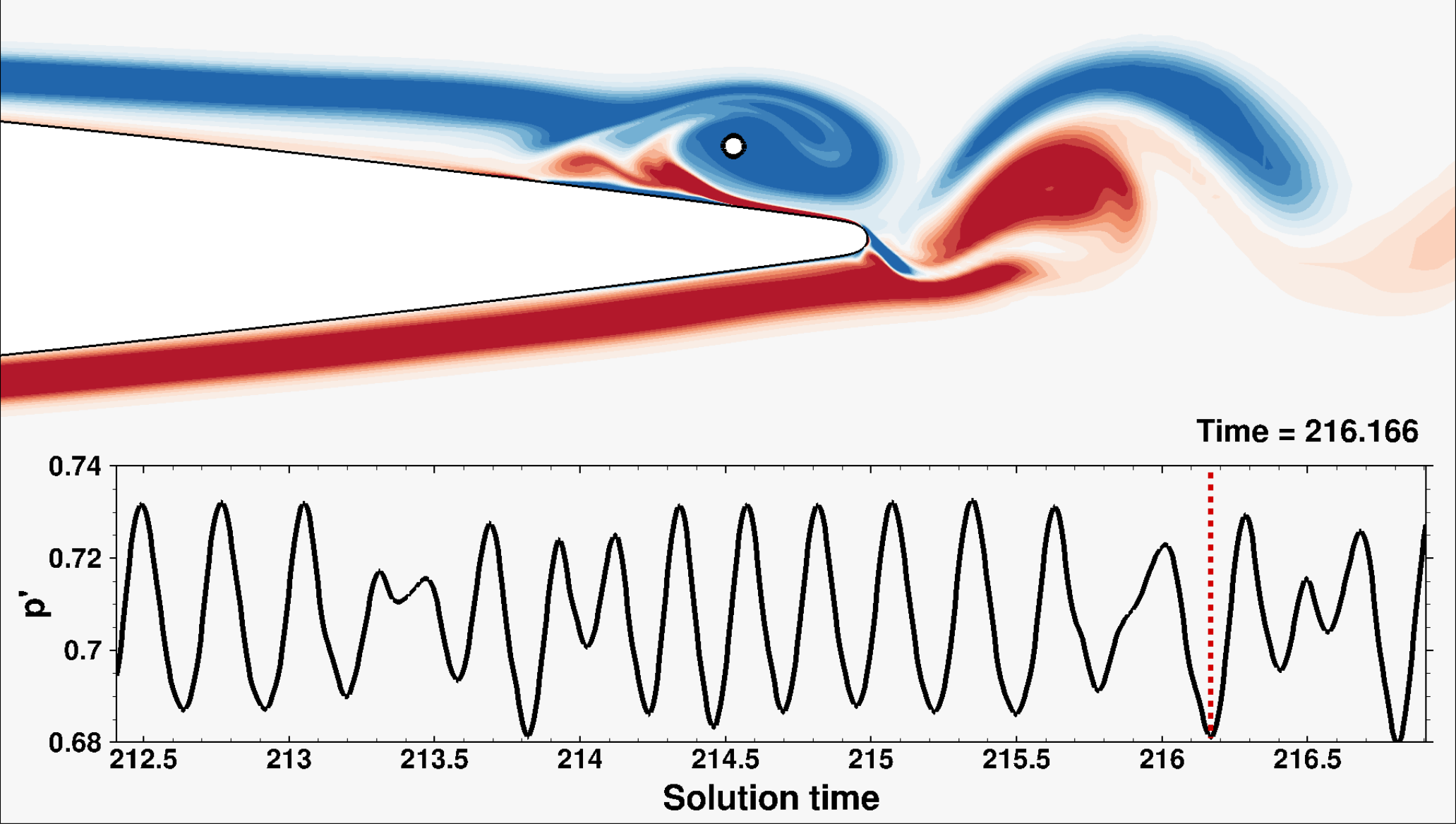}}
	\caption{Flow snapshots of vorticity contours and near-field acoustic pressure signal for $M_{\infty} = 0.3$.}
	\label{fig_flow_M03}
\end{figure}

Snapshots of vorticity contours show the temporal evolution of flow structures in Fig. \ref{fig_flow_M03}. Pressure signals measured near the trailing edge at $x=0.940$ and $y = 0.026$ are also presented to understand the correlation between shedding from vortical structures and near-field pressure fluctuations. The probe position is represented by a black and white dot in the figures and a vertical dashed line shows the pressure value for a particular time instant. The pressure signal contains two major regions: one where pressure variations are close to sinusoidal ($214.5 \leq t \leq 215.5$) and another composed by multiple Fourier modes $t>215.5$. Figures \ref{fig_flow_M03}(a--c) depict the shedding of a vortical structure transported along the trailing edge at the end of the sinusoidal period, represented by a valley, node and peak in the pressure signal. Several small scale structures are also observed in the figures and they are important to the overall dynamics that entrain, push and pull the larger vortex which is shed at the trailing edge. This interaction may delay or anticipate the shedding mechanism.

Figures \ref{fig_flow_M03}(d--i) show the shedding dynamics during the period where pressure variations are composed by multiple Fourier modes. In this sense, Figs. \ref{fig_flow_M03}(e--g) correspond to a valley, node and peak in the pressure signal and could be compared to their counterparts, Figs. \ref{fig_flow_M03}(a--c). One can observe that the main vortical structure in Fig. \ref{fig_flow_M03}(e) is more squeezed compared to that of Fig. \ref{fig_flow_M03}(a). This affects the vortex shedding mechanism and leads to a more stretched structure as seen in Fig. \ref{fig_flow_M03}(f). A more complex entrainment process of the smaller scale structures is then observed near the trailing edge in Figs. \ref{fig_flow_M03}(g) and (h). The overall impact of these flow features, which seem to be related to low frequency flapping of the separation bubble, is the amplitude and frequency modulations observed in the pressure signal. The effect of frequency modulation can be also observed with the cycle-to-cycle temporal variations computed between peaks and valleys and which are described in the individual captions of Fig. \ref{fig_flow_M03}.
%

\begin{figure}[H]
	\begin{subfigmatrix}{2}
		\subfigure[U-velocity]
		{\includegraphics[width=0.495\textwidth,trim={2mm 2mm 20mm 10mm },clip]
			{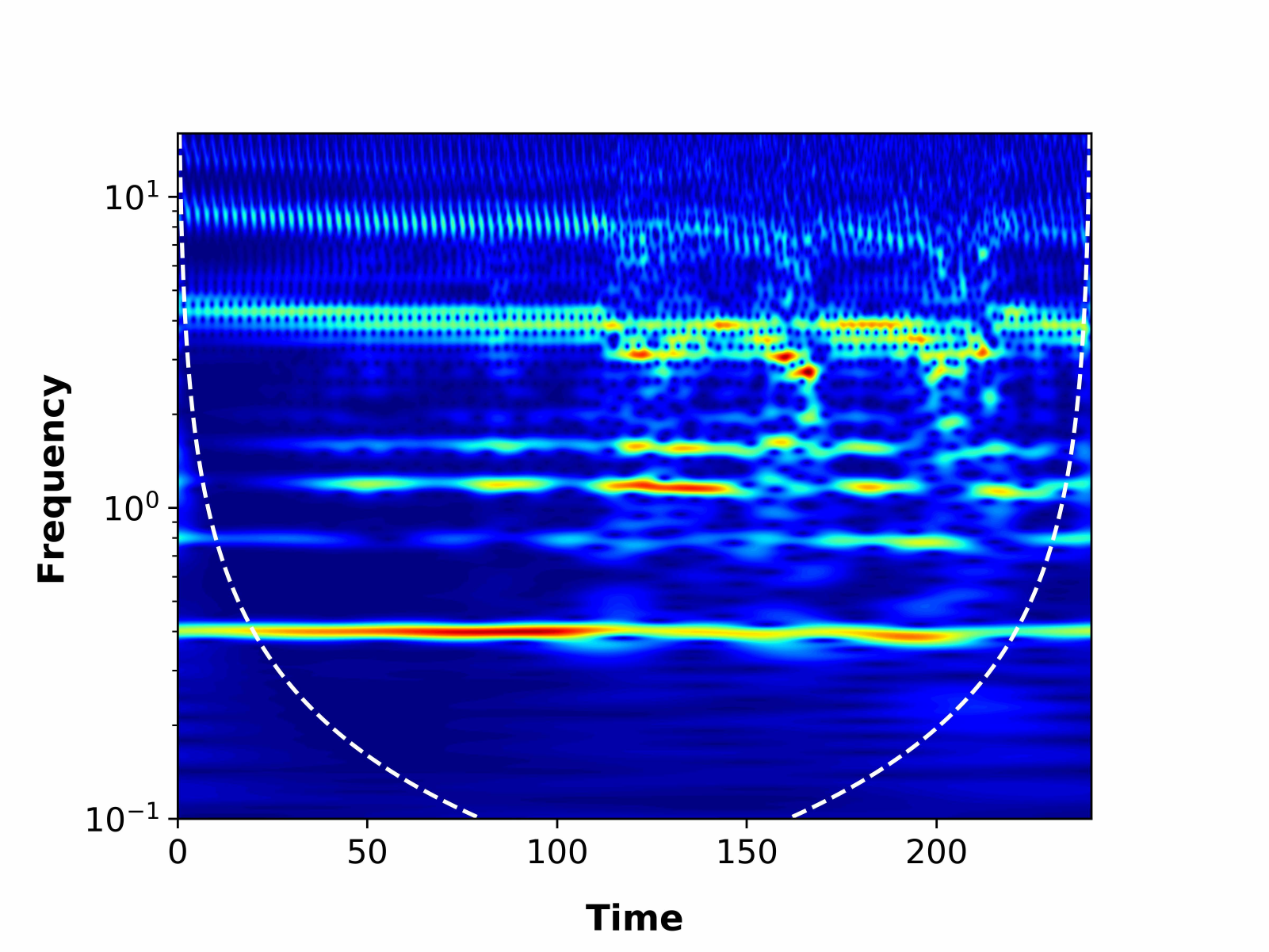}}
		\subfigure[Pressure]
		{\includegraphics[width=0.495\textwidth,trim={2mm 2mm 20mm 10mm },clip]
			{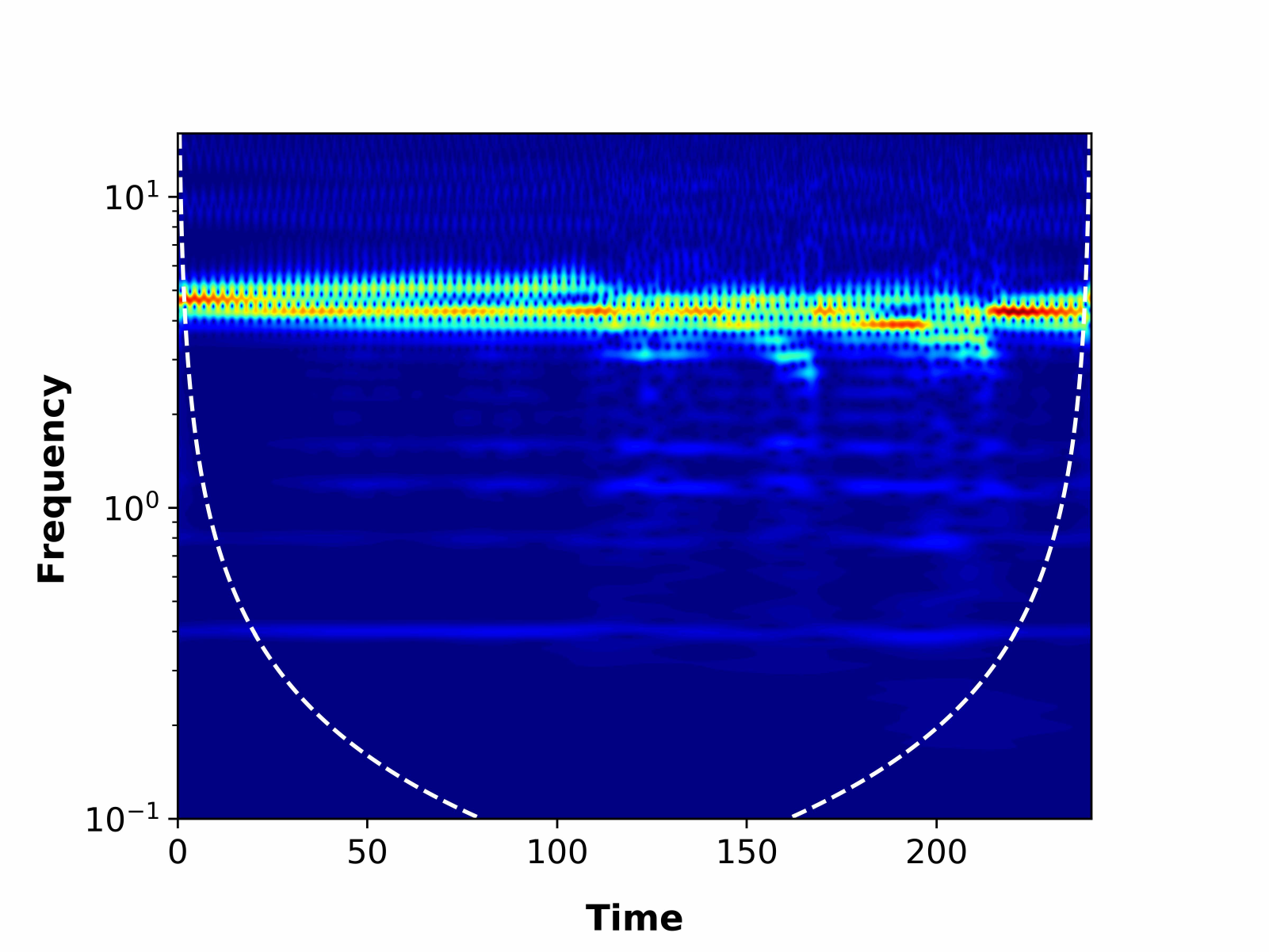}}
	\end{subfigmatrix}
	\caption{Spectrogram of entire temporal signal for $M_{\infty} = 0.3$.}
	\label{fig_wavelet_M03}
\end{figure}

To provide a better visualization of the amplitude and frequency modulation effects that occur in the current flow, near-field u-velocity and acoustic field pressure wavelet spectrograms are shown in Fig. \ref{fig_wavelet_M03}. Non-dimensional time and frequency are presented as $t=t^* \, U_{\infty} / L$ and $f=f^* \, L / U_{\infty}$ in the x and y-axes, respectively. Both spectrograms show the frequency modulation effects that shift frequencies around the main and secondary tones at $f=3.90$, $4.29$ and $4.68$. These tones are displaced by the frequency of the lowest tone $f=0.39$ which can be seen in the u-velocity plot of Fig. \ref{fig_wavelet_M03}(a). Besides the frequency variations, strong intermittency is also observed through amplitude variations in time, an effect also observed by \cite{Sanjose2018}.
%

\subsection{Airfoil at 3 deg. angle of incidence}

\begin{figure}[H]
\centering
	{\includegraphics[width=0.45\textwidth,trim={1mm 10mm 15mm 1mm},clip]
		{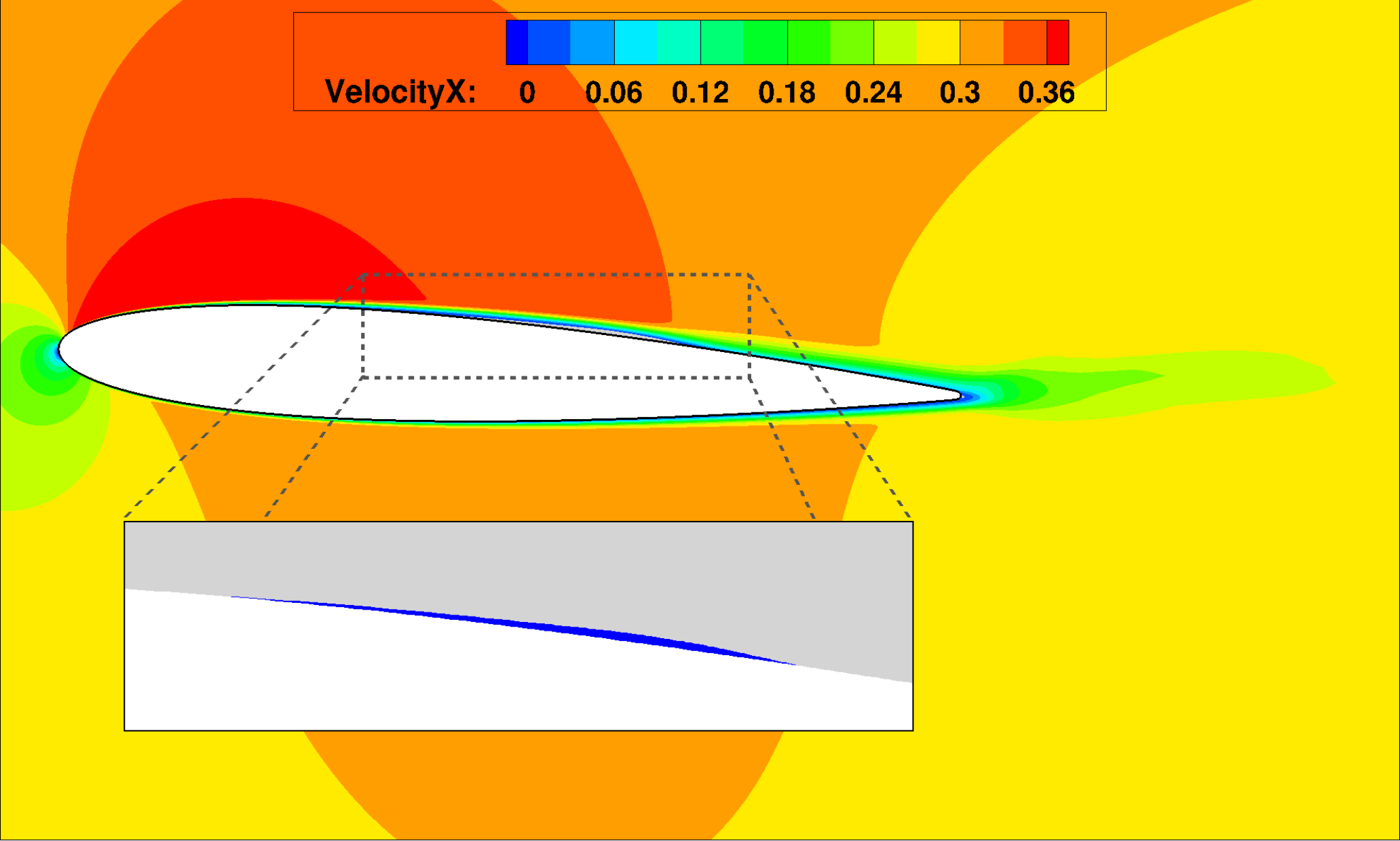}}
	\caption{Mean flow contours of u-velocity with detail view of separation bubble.}
	\label{fig_bubble_M03_3deg}
\end{figure}

The effects of angle of incidence are now investigated for $Re_{c}= 10^5$ and freestream Mach number $M_{\infty}=0.3$, using only Grid 2. Figure \ref{fig_bubble_M03_3deg} presents mean flow contours of u-velocity with a detail view of the separation bubble. Geometry asymmetry induces separation along the airfoil suction side and, differently from the zero deg. angle of attack configuration, the bubble forms away from the trailing edge for the current case.
%

\begin{figure}[H]
	\begin{subfigmatrix}{2}
	\centering
	{\includegraphics[width=0.5\textwidth,trim={5mm 1mm 2mm 5mm},clip]{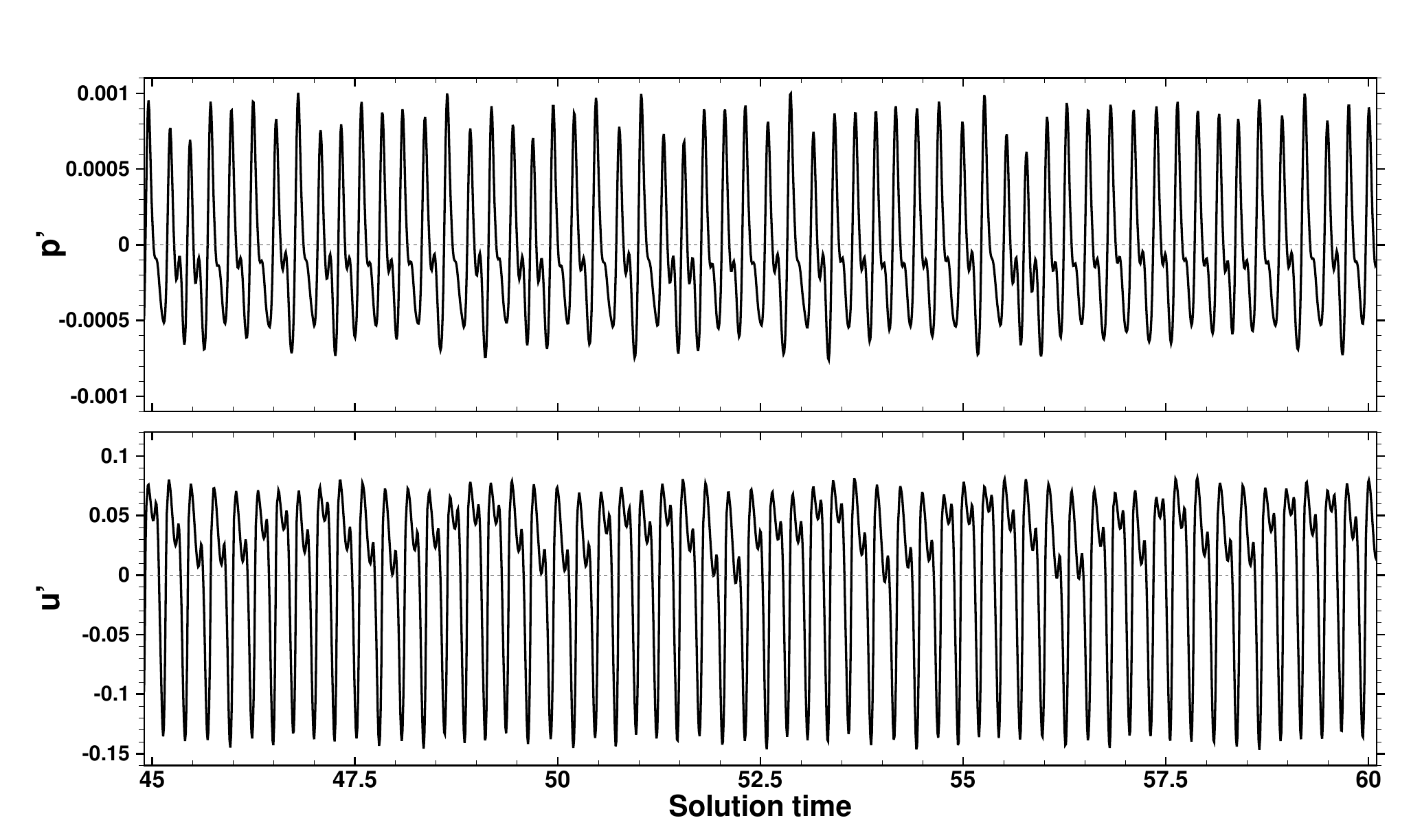}}
	\end{subfigmatrix}
	\caption{Pressure and u-velocity fluctuation signals as function of non-dimensional time $t=t^* \, U_{\infty} / L$.}
	\label{fig_time_M03_3deg}
\end{figure}

\begin{figure}[H]
	\begin{subfigmatrix}{2}
		\subfigure[U-velocity]
		{\includegraphics[width=0.4\textwidth,trim={5mm 5mm 5mm 5mm},clip]
			{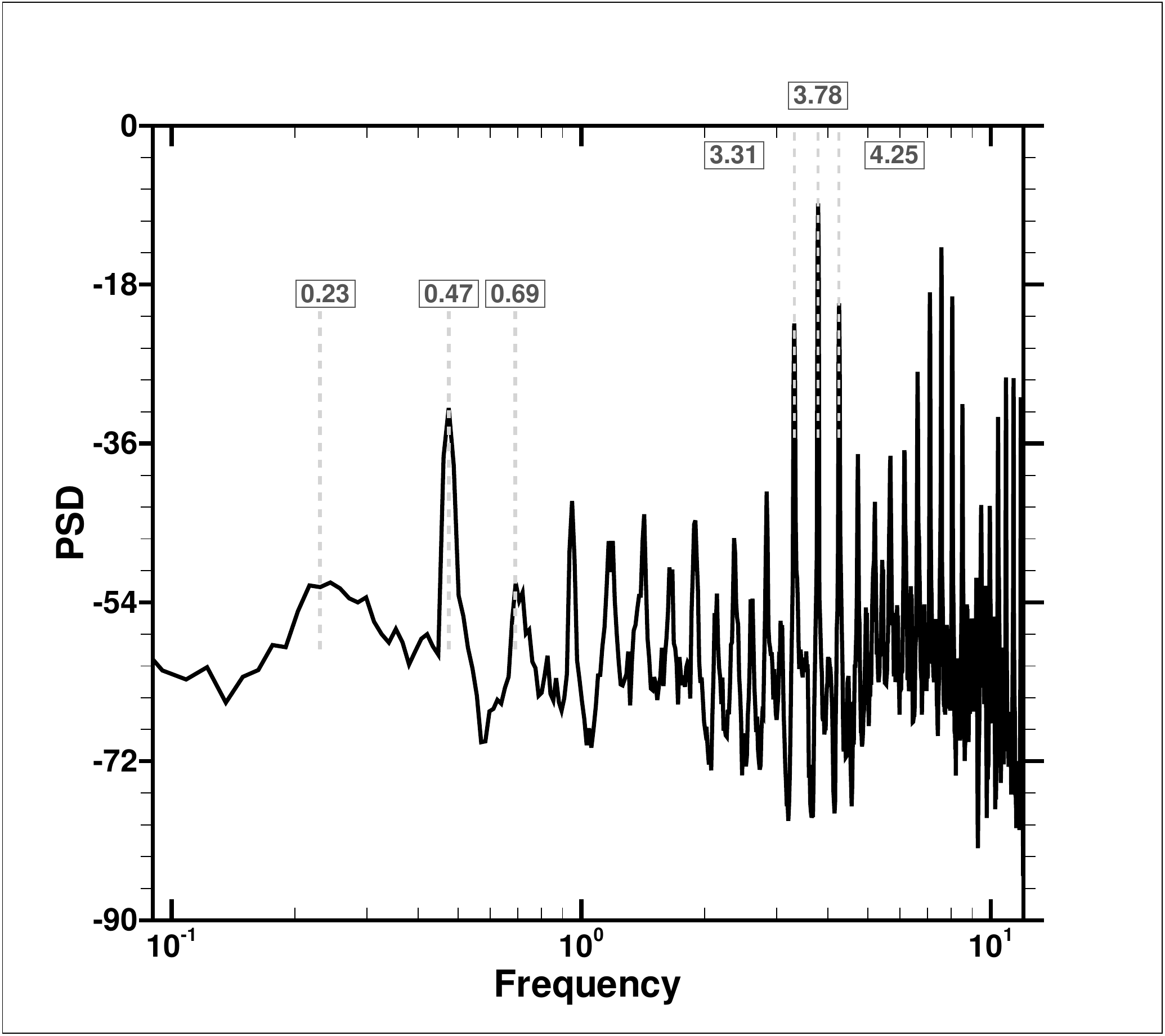}}
		\subfigure[Pressure]
		{\includegraphics[width=0.4\textwidth,trim={5mm 5mm 5mm 5mm},clip]
			{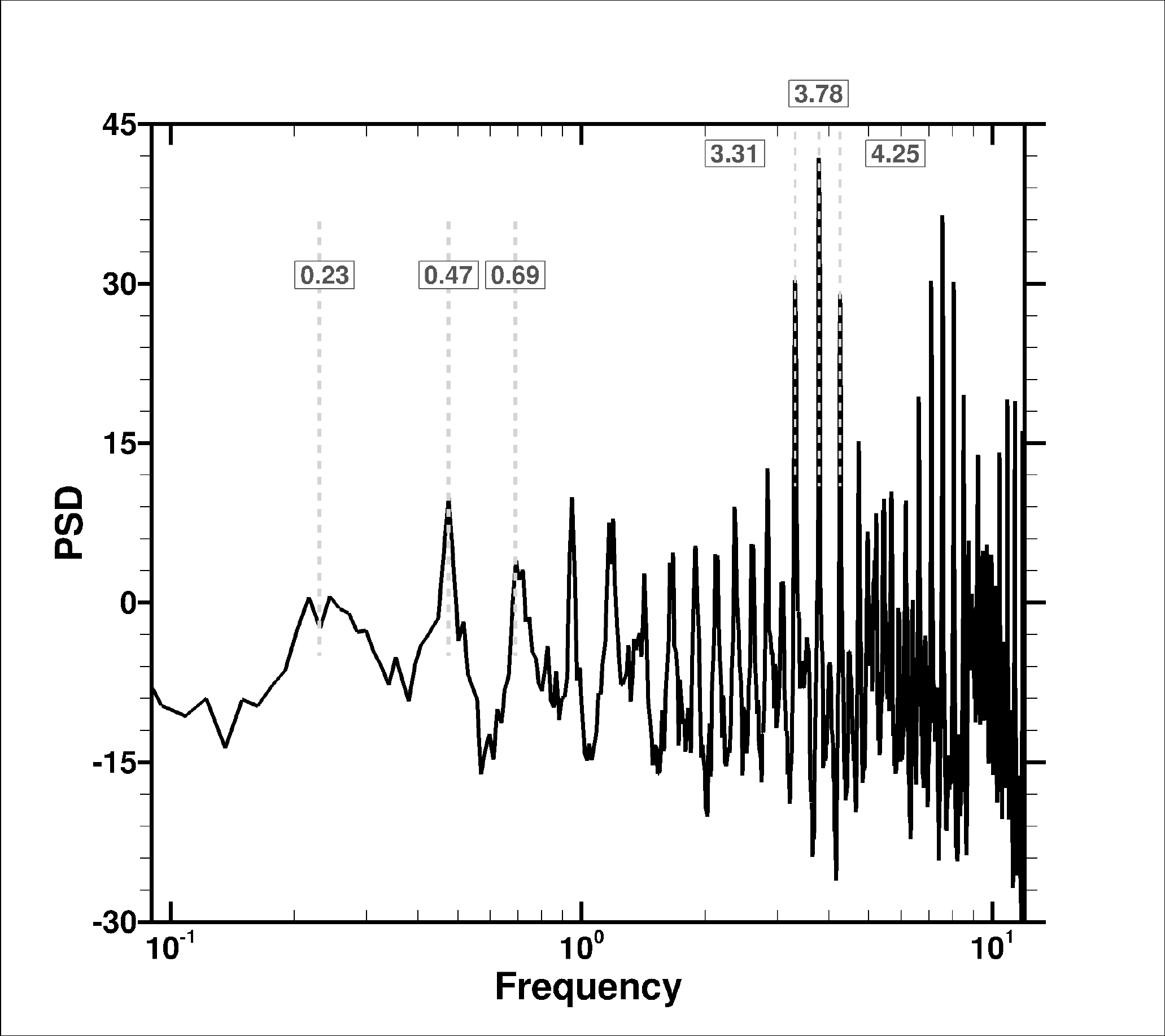}}
	\end{subfigmatrix}
	\caption{Power spectral density of longer temporal signal.}
	\label{fig_spectrafull_M03_3deg}
\end{figure}

In Fig. \ref{fig_time_M03_3deg}, u-velocity and pressure fluctuations are shown for a small period of the simulation. The signals are measured in a near-field position at $x=0.69$ and $y=0.028$ on the airfoil suction side for velocity, and in the acoustic field one chord above the trailing edge for pressure. Amplitude modulation is not as intense as for the previous case with zero incidence. For this case, velocity fluctuations show two patterns: one related to a higher frequency and higher amplitude, associated with more intense negative fluctuations, and another corresponding to a lower frequency sinusoidal pattern observed mainly above the horizontal line marking the zero value. Power spectral densities of these signals are presented in Figs. \ref{fig_spectrafull_M03_3deg}(a) and (b) for u-velocity and pressure, respectively. These PSDs are calculated with longer bins containing 8192 snapshots with overlap of 67\% using a Hanning window. In both spectra, the main and secondary tones are observed as well as lower frequency tones. The overall features of the spectra are similar to those from the zero deg. angle of incidence case (Fig. \ref{fig_spectrafull_M03}). A main tone is observed at $f=3.78$ while a low frequency tone is found at $0.47$. However, for the current flow, intermediate secondary tones are also present in the spectra with a frequency difference of $\Delta f = 0.23$. Such tones are related to a hump in spectra at this frequency, which may be a sub-harmonic of $f=0.47$. These intermediate tones have lower amplitudes than those displaced by $\Delta f= 0.47$. The high frequency pattern observed in the u-velocity fluctuations of Fig. \ref{fig_time_M03_3deg} is related to the frequency of TS instabilities advecting along the boundary layer at $f=3.78$. On the other hand, the low frequency quasi-sinusoinal fluctuations observed in the same figure for $u'>0$ are associated to $f= 0.47$.
%
%

\begin{figure}[H]
\centering
		\subfigure[U-velocity, $f= 0.23$]
		{\includegraphics[width=0.55\textwidth,trim={1mm 1mm 1mm 1mm},clip]
			{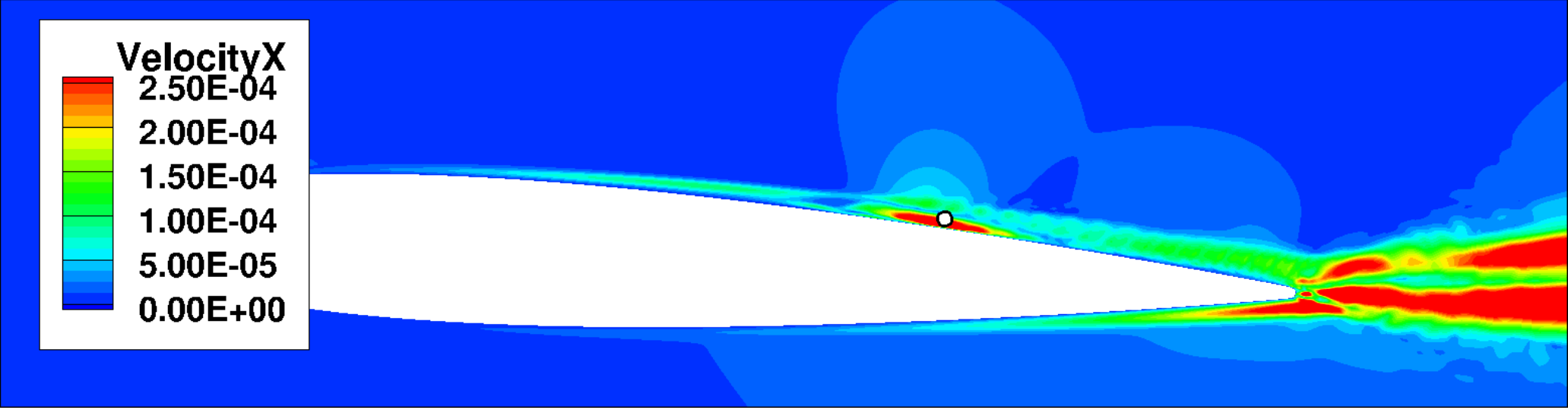}}
		\subfigure[U-velocity, $f= 0.47$]
		{\includegraphics[width=0.55\textwidth,trim={1mm 1mm 1mm 1mm},clip]
			{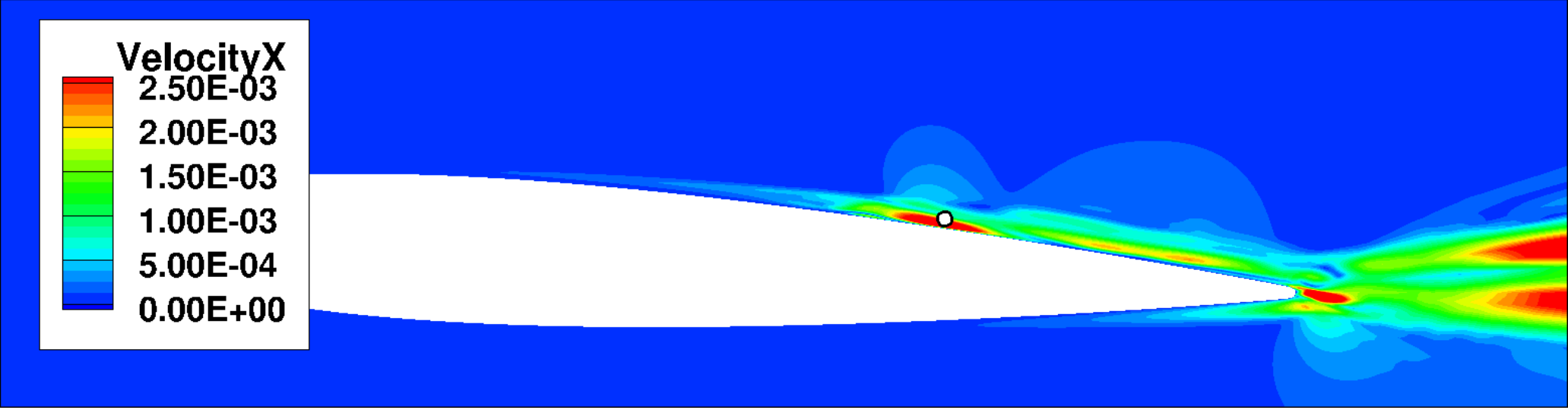}}
		\subfigure[U-velocity, $f= 3.78$]
		{\includegraphics[width=0.55\textwidth,trim={1mm 1mm 1mm 1mm},clip]
			{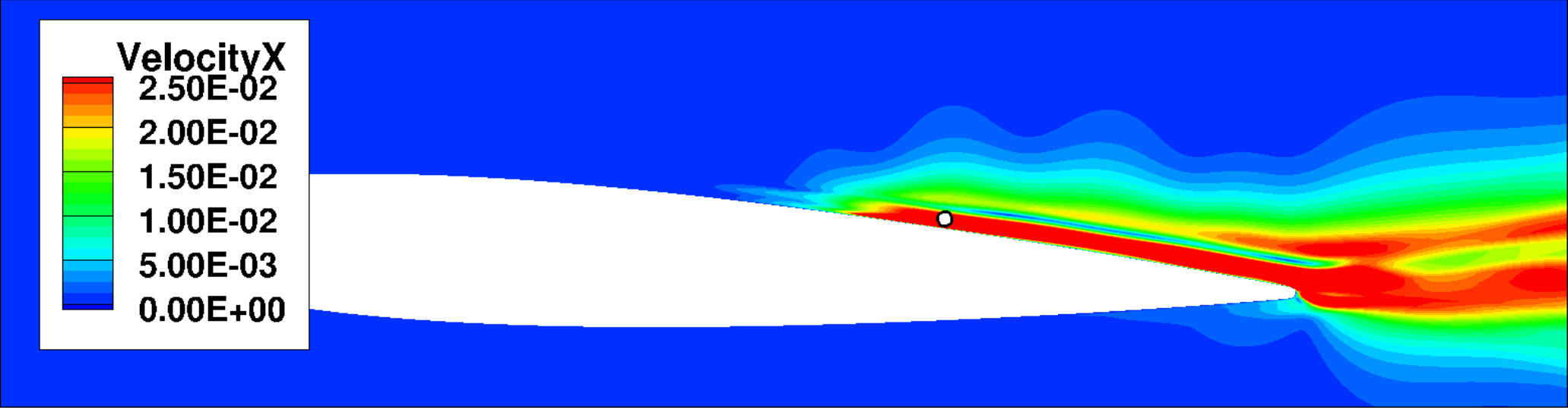}}
	\caption{Contours of Fourier mode amplitudes of longer temporal signal.}
	\label{fig_fourier_M03_3deg}
\end{figure}

It is possible to associate the previous high and low frequency tones to specific flow features. Figure \ref{fig_fourier_M03_3deg} presents contours of Fourier mode amplitudes for u-velocity obtained from the previous temporal signals. One can see that the lower frequencies are associated with fluctuations around $x=0.68$ (marked by the black and white dot in the figures). This region marks the end of the separation bubble that may have a flapping effect similar to the previous case with no incidence. The low frequency tone at $f=0.23$ is also related to fluctuations on the airfoil pressure side, near the trailing edge. On the other hand, for the frequency of the main tone, velocity fluctuations are associated with a region downstream the bubble, including the trailing-edge region. These fluctuations are related to flow instabilities that lead to development of vortical structures shed by the trailing edge. It is important to highlight that maximum amplitudes of Fourier modes computed for $f=3.78$ are 100 and 10 times higher than those obtained for $f=0.23$ and $0.47$. This indicates that the shedding of vortical structures is a more energetic process than the bubble motion for the current case with angle of incidence, differently from the previous case with zero incidence, what can be inferred from a direct comparison between the FFTs of u-velocity for both cases.
%

\begin{figure}[H]
	\subfigure[]
	{\includegraphics[width=0.33\textwidth,trim={2mm 2mm 2mm 2mm },clip]{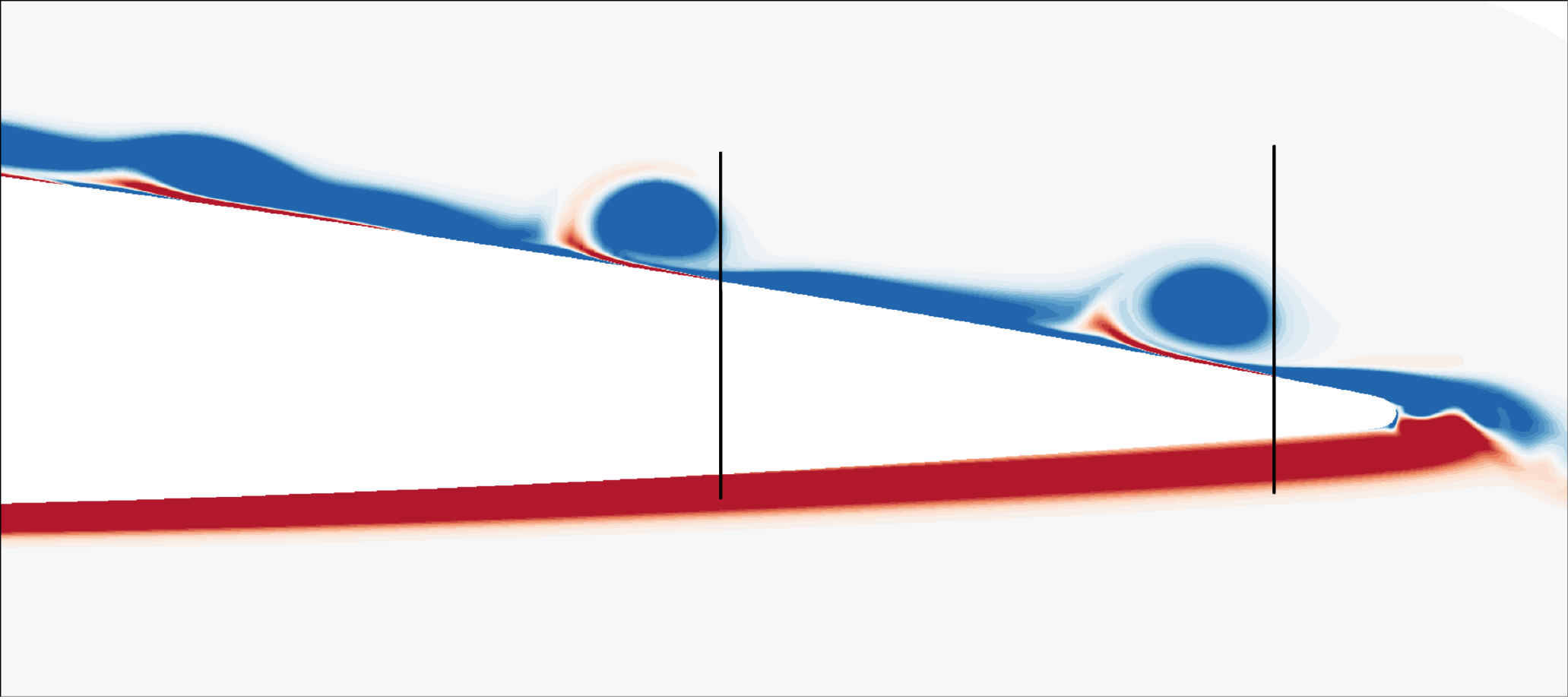}}
	\subfigure[]
	{\includegraphics[width=0.33\textwidth,trim={2mm 2mm 2mm 2mm },clip]{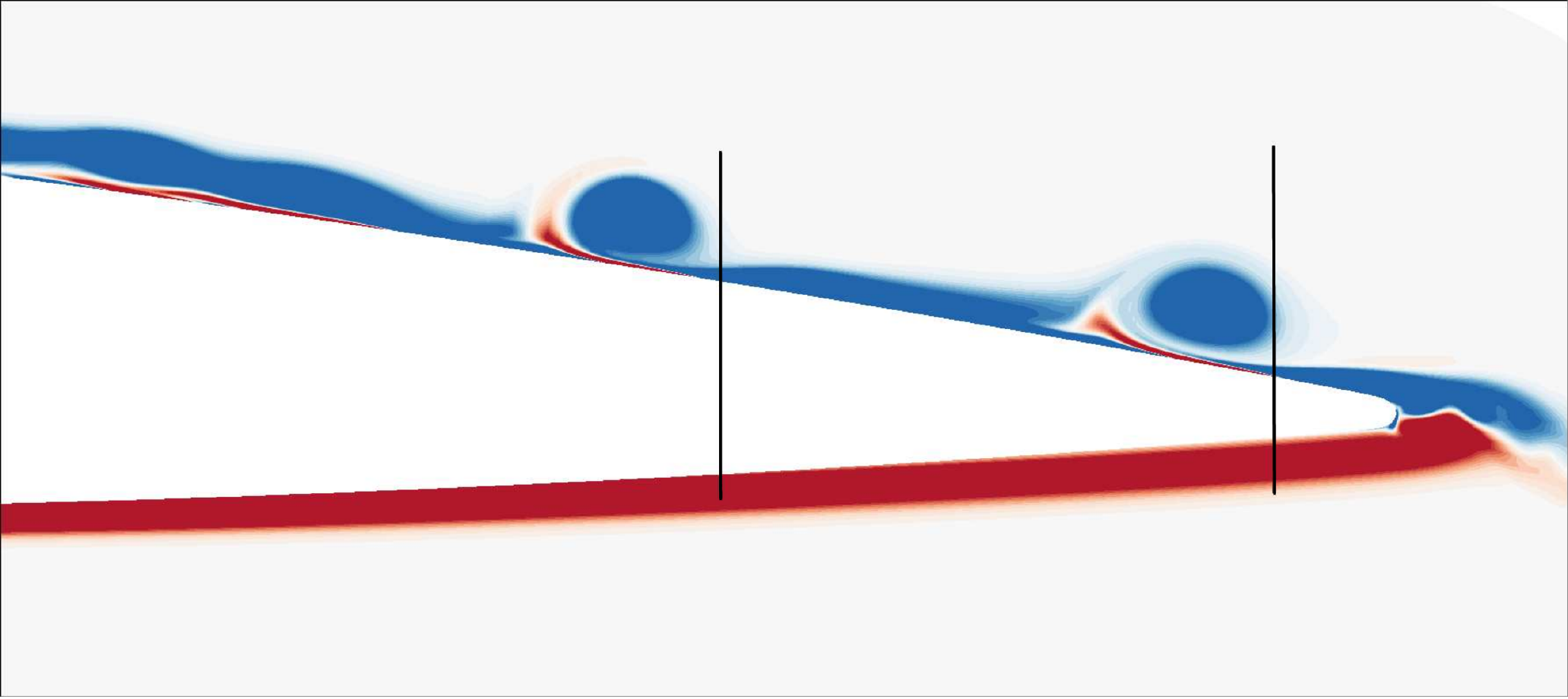}}
	\subfigure[]
	{\includegraphics[width=0.33\textwidth,trim={2mm 2mm 2mm 2mm },clip]{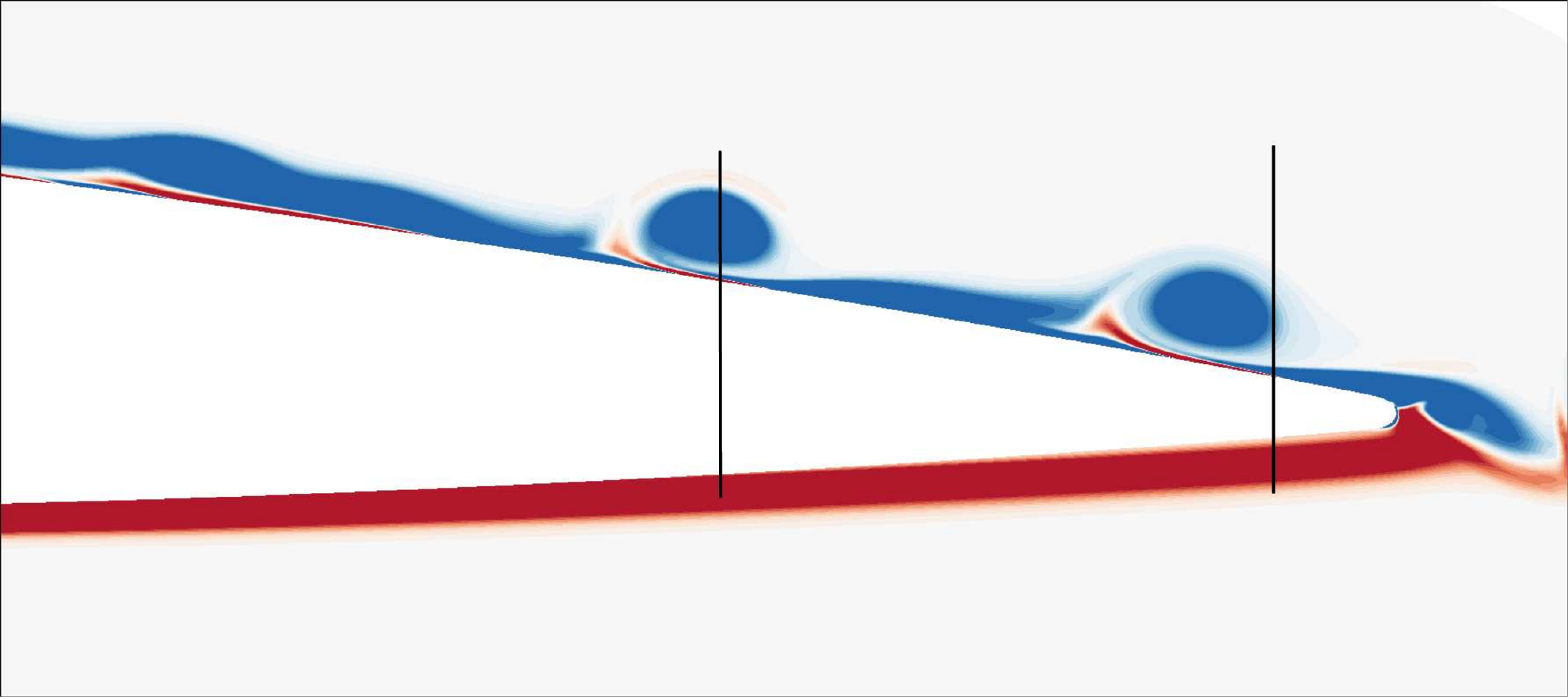}}
	\caption{Flow snapshots of vorticity contours indicating delay and advancement of vortical strutures.}
	\label{fig_flow_M03_3deg}
\end{figure}

\begin{figure}[H]
	\begin{subfigmatrix}{2}
		\subfigure[U-velocity]
		{\includegraphics[width=0.495\textwidth,trim={2mm 2mm 20mm 10mm },clip]{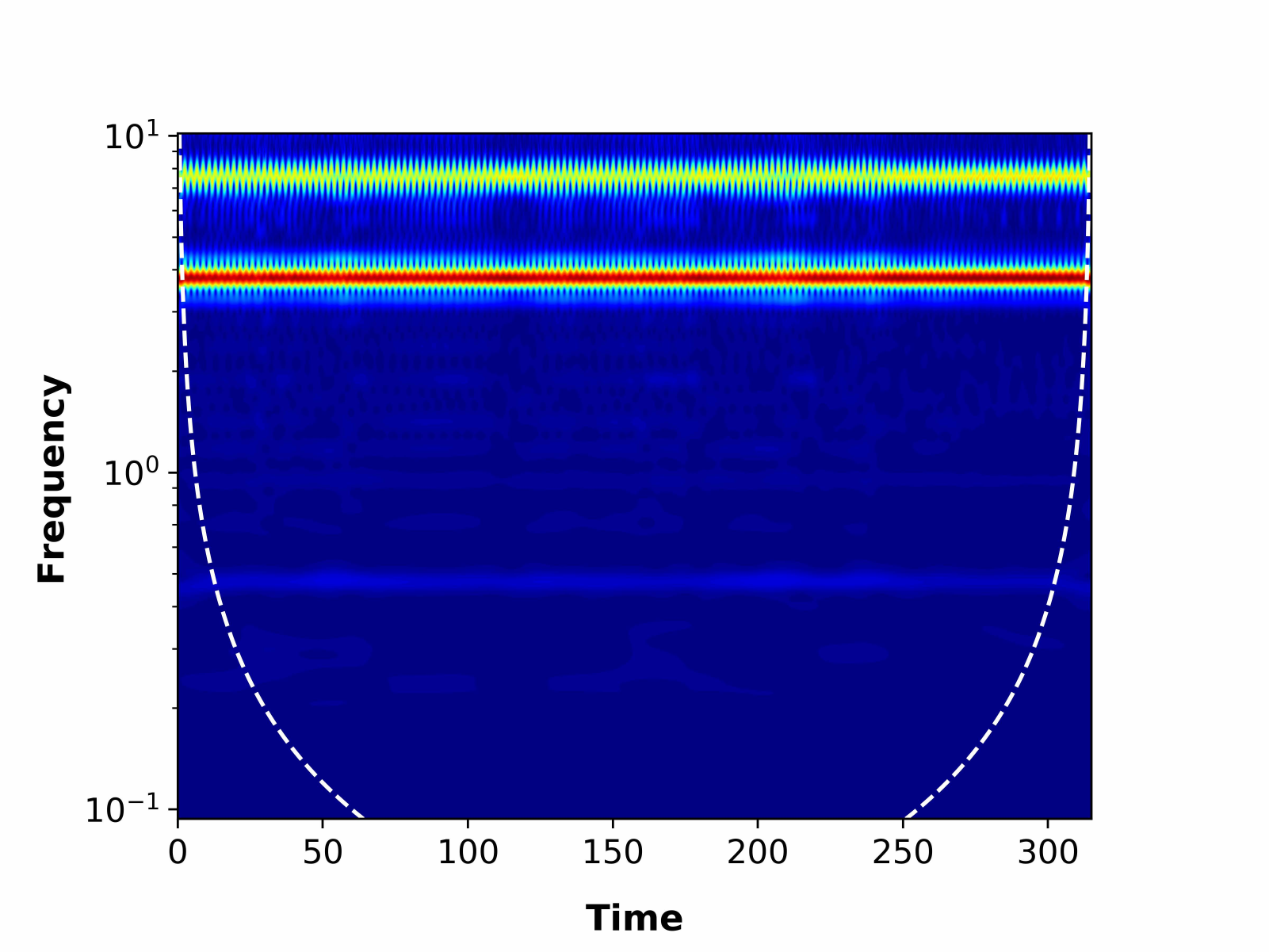}}
		\subfigure[Pressure]
		{\includegraphics[width=0.495\textwidth,trim={2mm 2mm 20mm 10mm },clip]{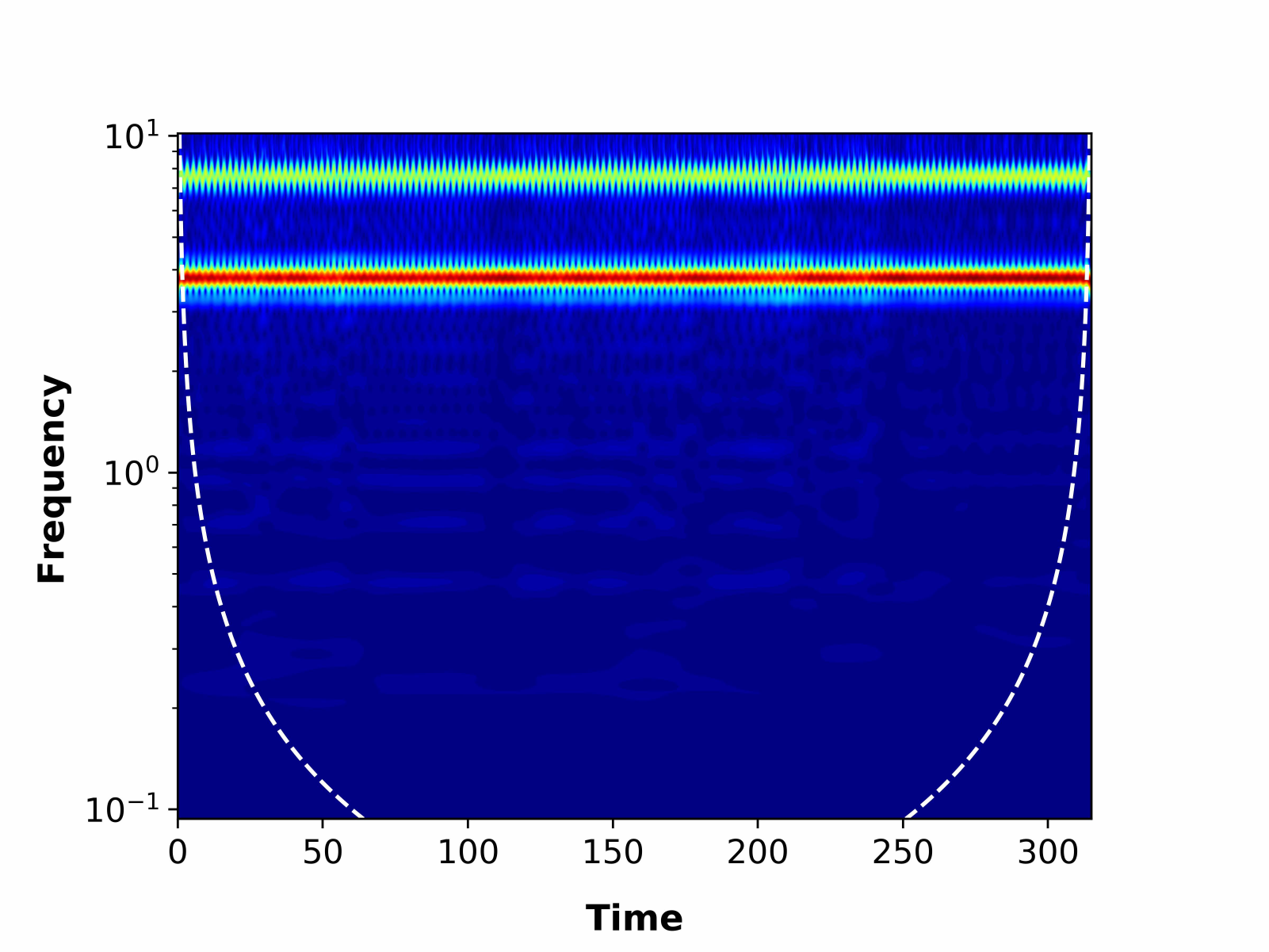}}
	\end{subfigmatrix}
	\caption{Spectrogram of temporal signal.}
	\label{fig_wavelet_M03_3deg}
\end{figure}

For the current case, the flow dynamics around the trailing edge is dictated by advection of vortical structures that develop along the suction side boundary layer. The process is not as complex as that shown in Fig. \ref{fig_fourier_M03} for zero deg. For the 3 deg. angle of attack configuration, snapshots of vorticity contours are presented for different time instants in Fig. \ref{fig_flow_M03_3deg}. Flow structures are similar, however, their generation are impacted by a frequency modulation from the separation bubble. This effect can be observed by the vertical black lines which are positioned in the same location for all figures. Comparing Figs. \ref{fig_flow_M03_3deg}(b) and (c) with Fig. \ref{fig_flow_M03_3deg}(a), it is possible to observe a slight delay and advancement of vortical strutures, respectively. Such effects can be observed as a frequency modulation in the u-velocity and pressure spectrograms of Fig. \ref{fig_wavelet_M03_3deg}. For this case, the wavelet transform shows a more regular flow behavior with the main tone at $f=3.78$ and its first harmonic, besides their frequency modulations. The lower frequency $f=0.47$ can be seen in the u-velocity spectrogram as a lighter blue line.
%
%

\subsection{A Model for Secondary Tones}

Secondary tones represent a predominant feature of airfoil noise at moderate Reynolds numbers. As previously discussed, flow separation and instabilities have an impact on secondary tones through amplitude and frequency modulation of near-field velocity and pressure signals and, therefore, on acoustic emission. In this section, we employ artificial model signals that present similarities with those found in the current flows for pressure and velocity fluctuations. With these model signals it is possible to assess the differences between amplitude and frequency modulations on spectra. Here, it is shown that frequency modulation is more prominent in the current problem of secondary tones.

\begin{figure}[H]
	\begin{subfigmatrix}{2}
		\subfigure[Window function: composition of sine waves]
		{\includegraphics[width=0.495\textwidth,trim={2mm 2mm 2mm 2mm },clip]
			{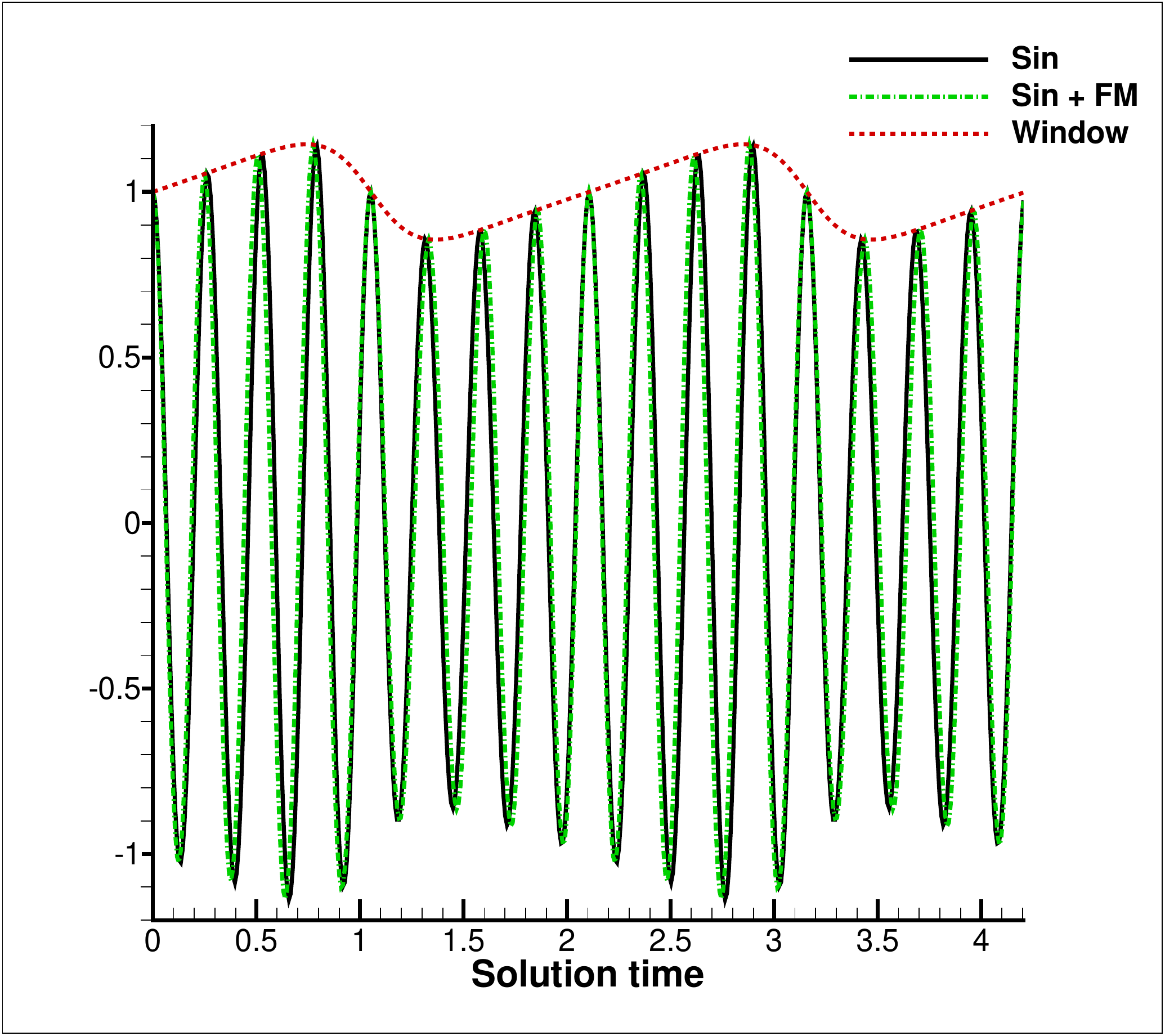}}
		\subfigure[Window function: Tukey]
		{\includegraphics[width=0.495\textwidth,trim={2mm 2mm 2mm 2mm },clip]
			{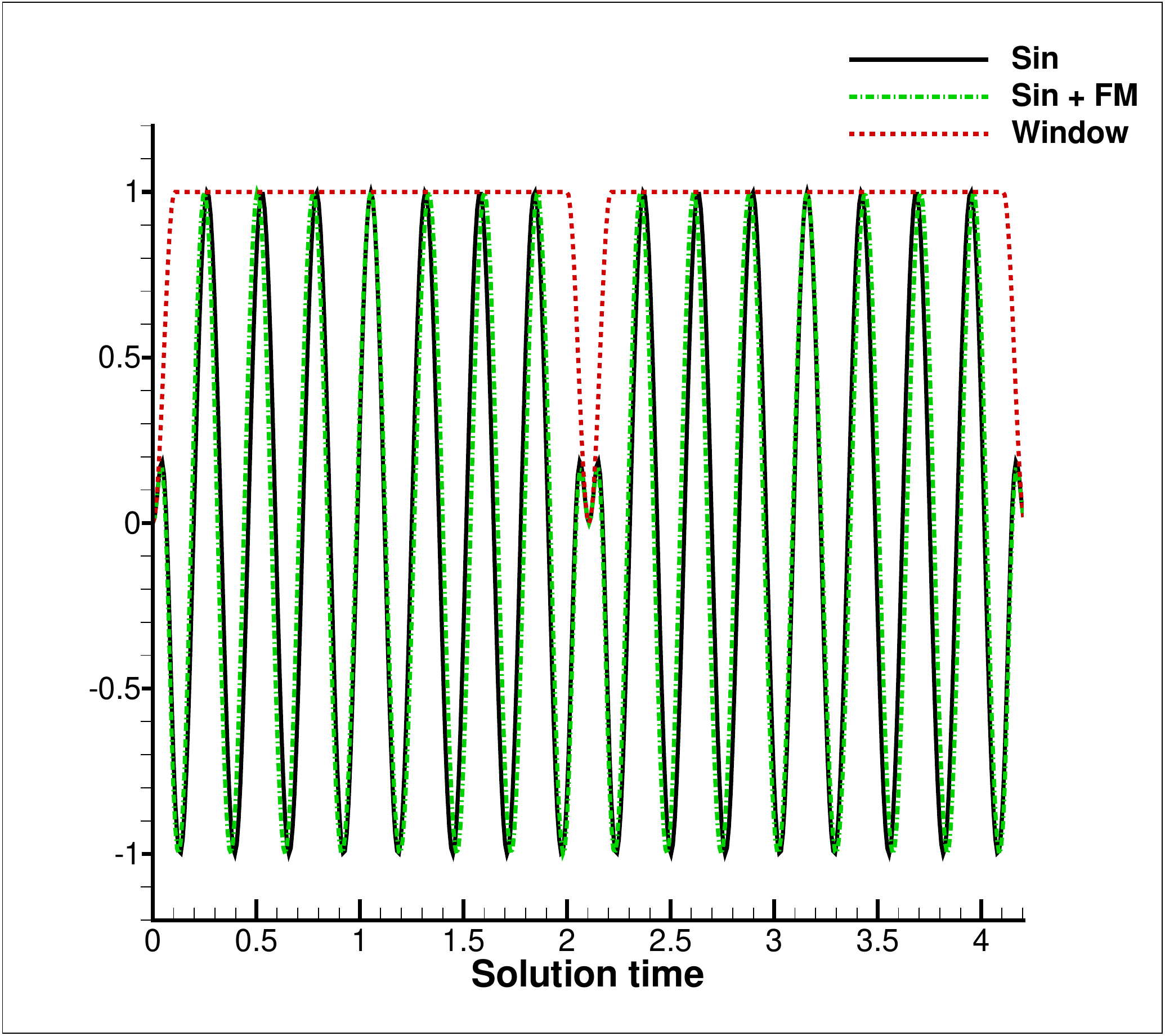}}
	\end{subfigmatrix}
	\caption{Sinusoid with amplitude and frequency modulations.}
	\label{fig_model_temporal}
\end{figure}

Figure \ref{fig_model_temporal} shows two artificial signals that share similarities with those from Figs. \ref{fig_time_M03} and \ref{fig_time_M03_3deg}. These signals are composed of a windowed sine wave with frequency $f_{c}$ and amplitude $A$ depicted by solid black lines, represented by Eq. \ref{janelamento}.
\begin{equation}
\label{janelamento}
y(t) = A \sin \left[2 \pi f_{c} t\right] \, w(t) \mbox{ .}
\end{equation}
The frequency modulation of the signal is shown by dashed green lines using Eq. \ref{freq_mod}
\begin{equation}
\label{freq_mod}
y(t) = A \sin \left[ 2 \pi f_{c} t + \sin(2 \pi f_{m} t) \right] \, w(t) \mbox{ ,}
\end{equation}
%
Two different window functions are shown in red dashed lines in Figs. \ref{fig_model_temporal}(a) and (b). The first window, a sinusoidal modulation, is employed to approximate the behavior of the signal in Fig. \ref{fig_time_M03}(a). It is given by a combination of sine waves as
\begin{equation}
w(t) = \sum\limits_{i=0}^{13} A_i \sin{ \left[2.0 \pi f_{w} t + mod(i,2) \, \pi\right]} \mbox{ ,}
\end{equation}
where the magnitude of each sinusoidal is given by $A_{i} = 0.125/3.0^i$. A phase correction is necessary for the odd terms and is enforced by the modulus equation $mod(i,2) \, \pi$. In order to mimic the amplitude modulation observed in Fig. \ref{fig_time_M03}(b), which is represented by regions with nearly constant values followed by a fast reduction in the magnitude, we employ a Tukey window function, given by
\begin{equation}
w(t) = \begin{cases}
\vspace{2mm} \hspace{2mm} \frac{1}{2} \left\{ 1 + \cos \left[ \pi \left( \frac{2n}{\alpha N} -1 \right) \right] \right\}, & 0 \le n < \frac{\alpha N}{2}  \\
\vspace{2mm} \hspace{2mm} 1, & \frac{\alpha N}{2} \le n <  N \left( 1 - \frac{\alpha}{2} \right)  \\
\vspace{2mm} \hspace{2mm} \frac{1}{2} \left\{ 1 + \cos \left[ \pi \left( \frac{2n}{\alpha N} - \frac{2}{\alpha} + 1 \right)  \right] \right\}, & 1 - \frac{\alpha}{2} \le n < N \mbox{ ,}
\end{cases}
\end{equation}
where the $\alpha$ parameter controls the extent of the plateau region and is used as 0.1 here for a sharper modulation. The $n$ index is related to time $t$ by $t=n \Delta t$ and the window size is controlled by the parameter $N$.

\begin{figure}[]
	\begin{subfigmatrix}{3}
		\subfigure[Sine wave (no window)]
		{\includegraphics[width=0.32\textwidth,trim={15mm 2mm 20mm 20mm },clip]
			{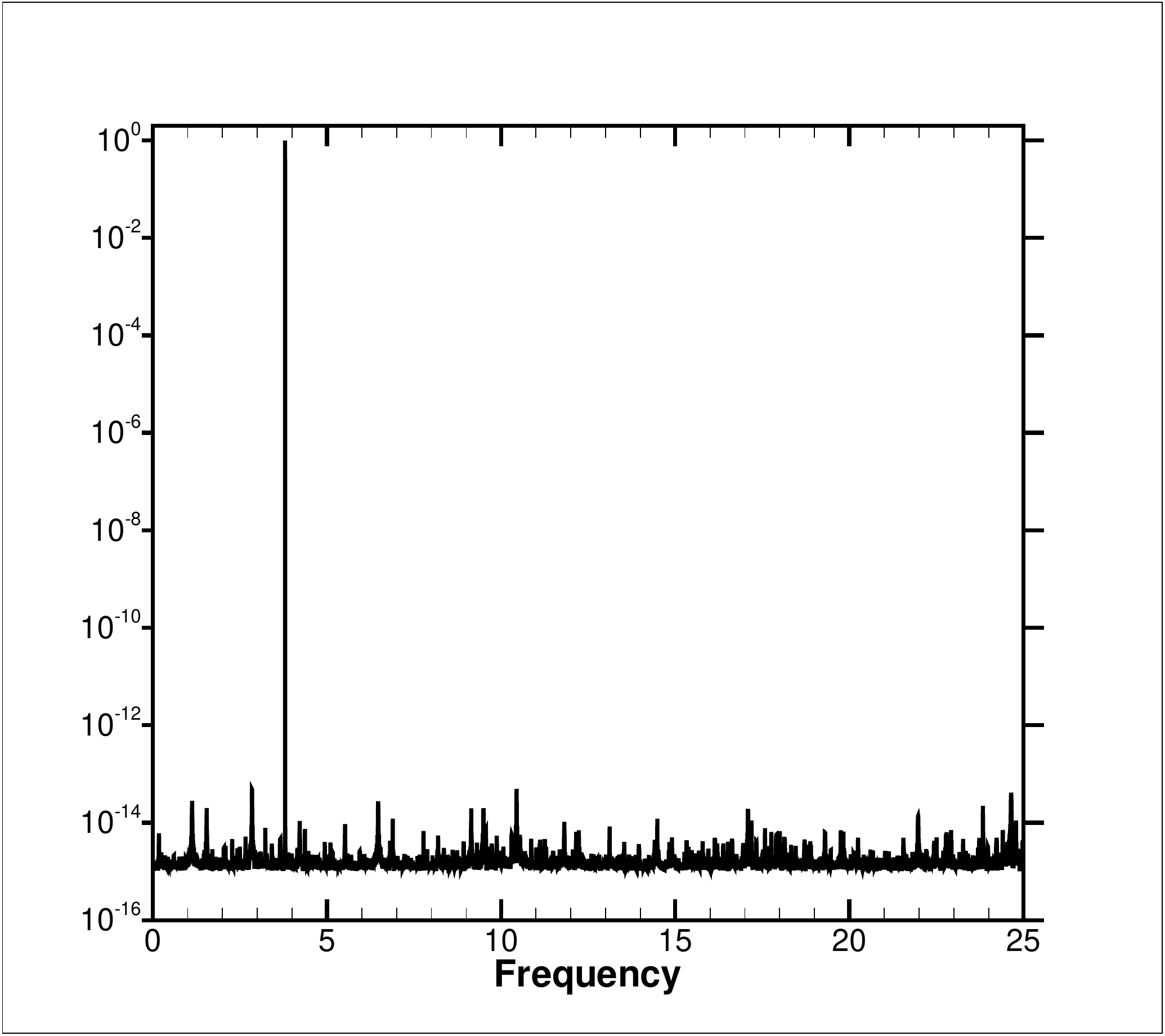}}
		\subfigure[AM with sinusoidal window]
		{\includegraphics[width=0.32\textwidth,trim={15mm 2mm 20mm 20mm },clip]
			{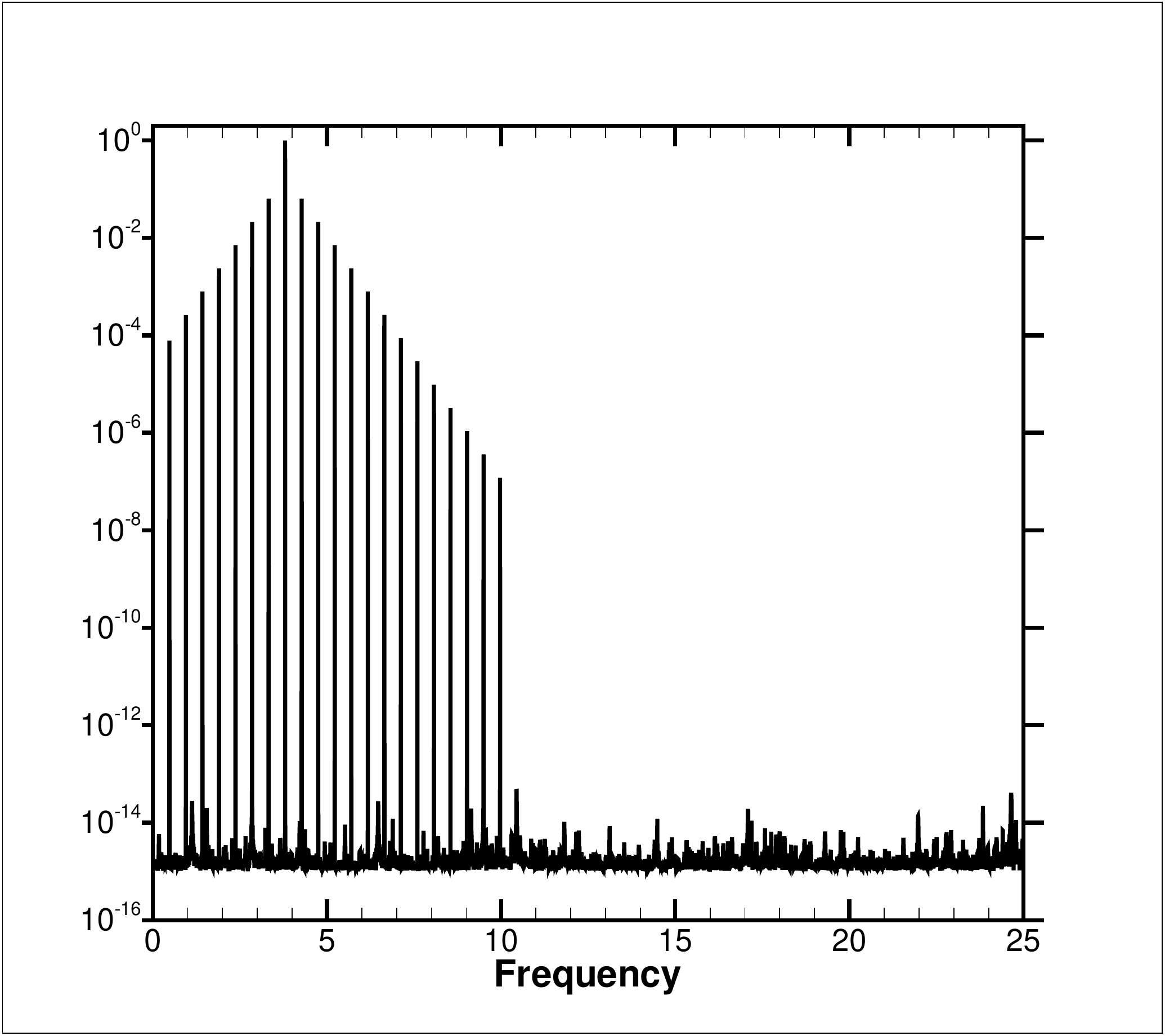}}
		\subfigure[AM with Tukey window]
		{\includegraphics[width=0.32\textwidth,trim={15mm 2mm 20mm 20mm },clip]
			{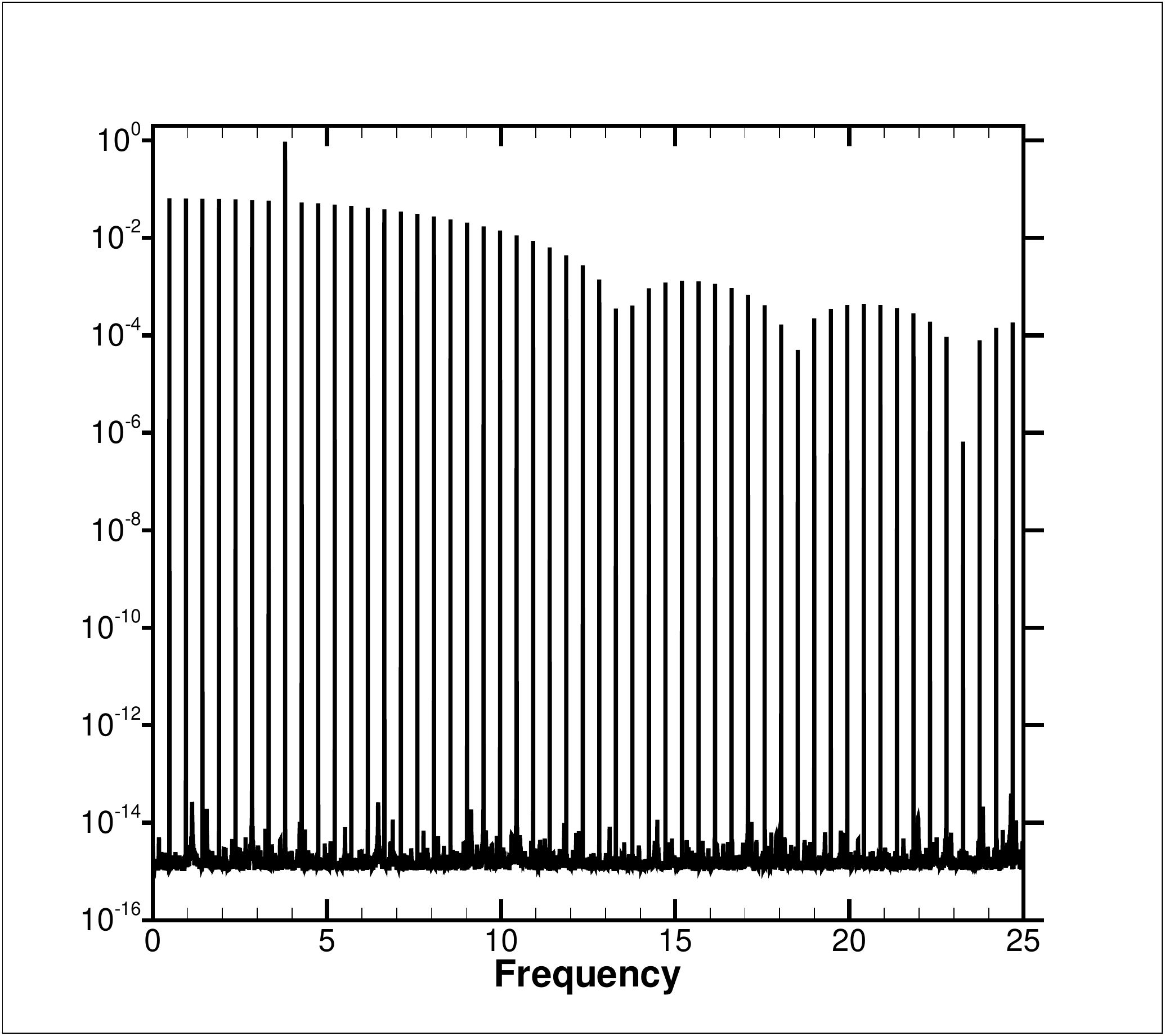}}
		\subfigure[FM of sine wave (no window)]
		{\includegraphics[width=0.32\textwidth,trim={15mm 2mm 20mm 20mm },clip]
			{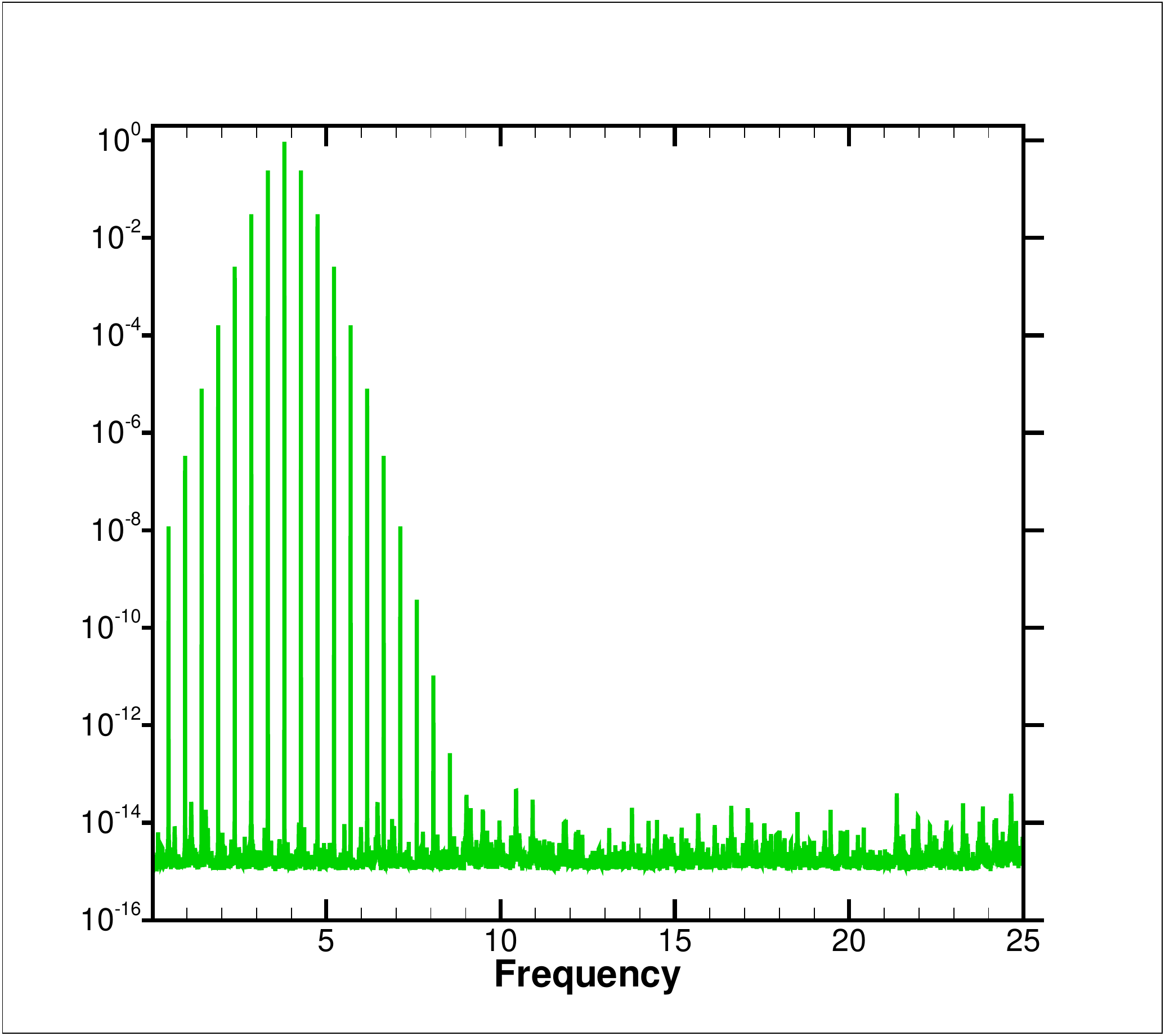}}
		\subfigure[FM + AM with sinusoidal window]
		{\includegraphics[width=0.32\textwidth,trim={15mm 2mm 20mm 20mm },clip]
			{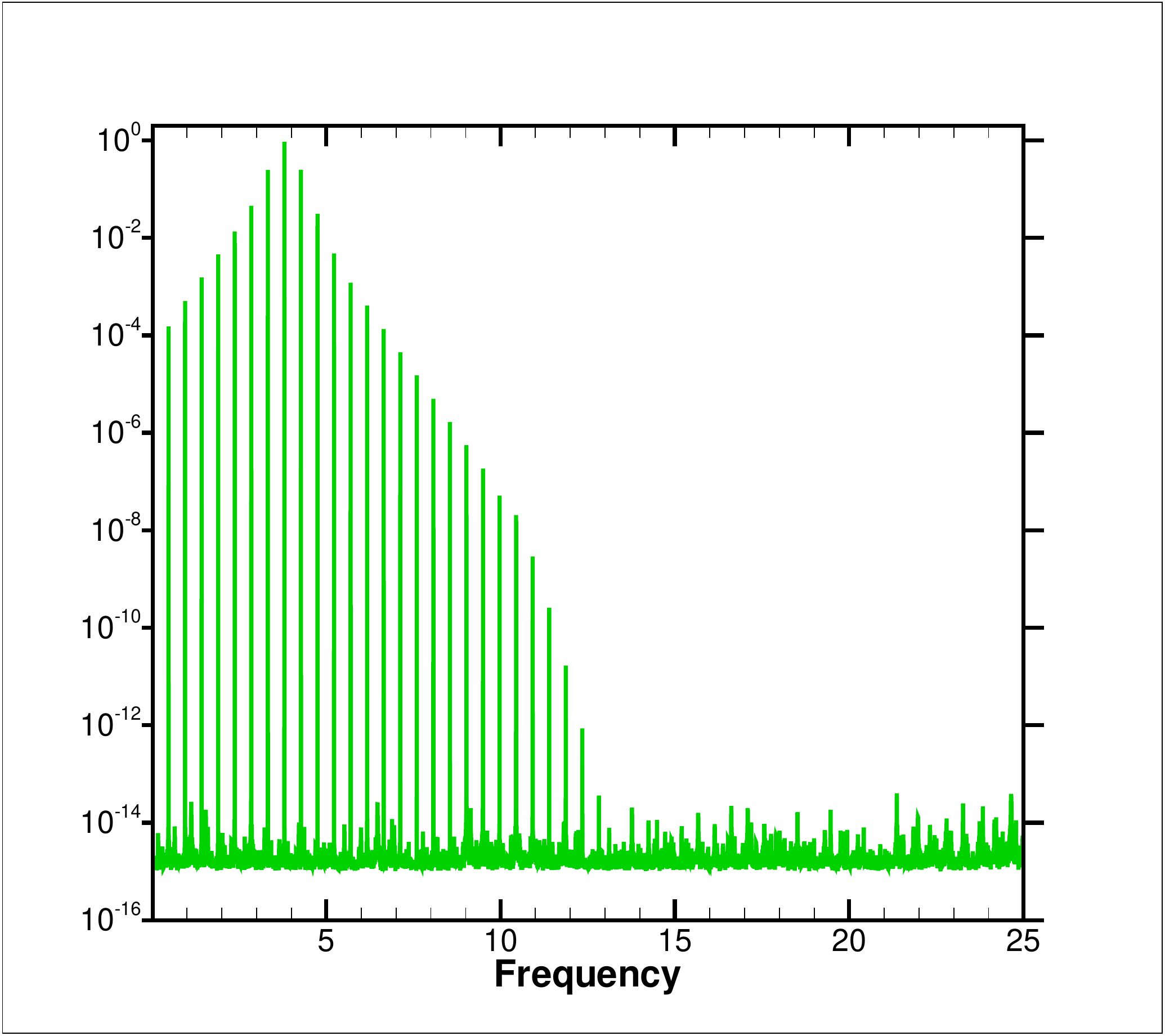}}
		\subfigure[FM + AM with Tukey window]
		{\includegraphics[width=0.32\textwidth,trim={15mm 2mm 20mm 20mm },clip]
			{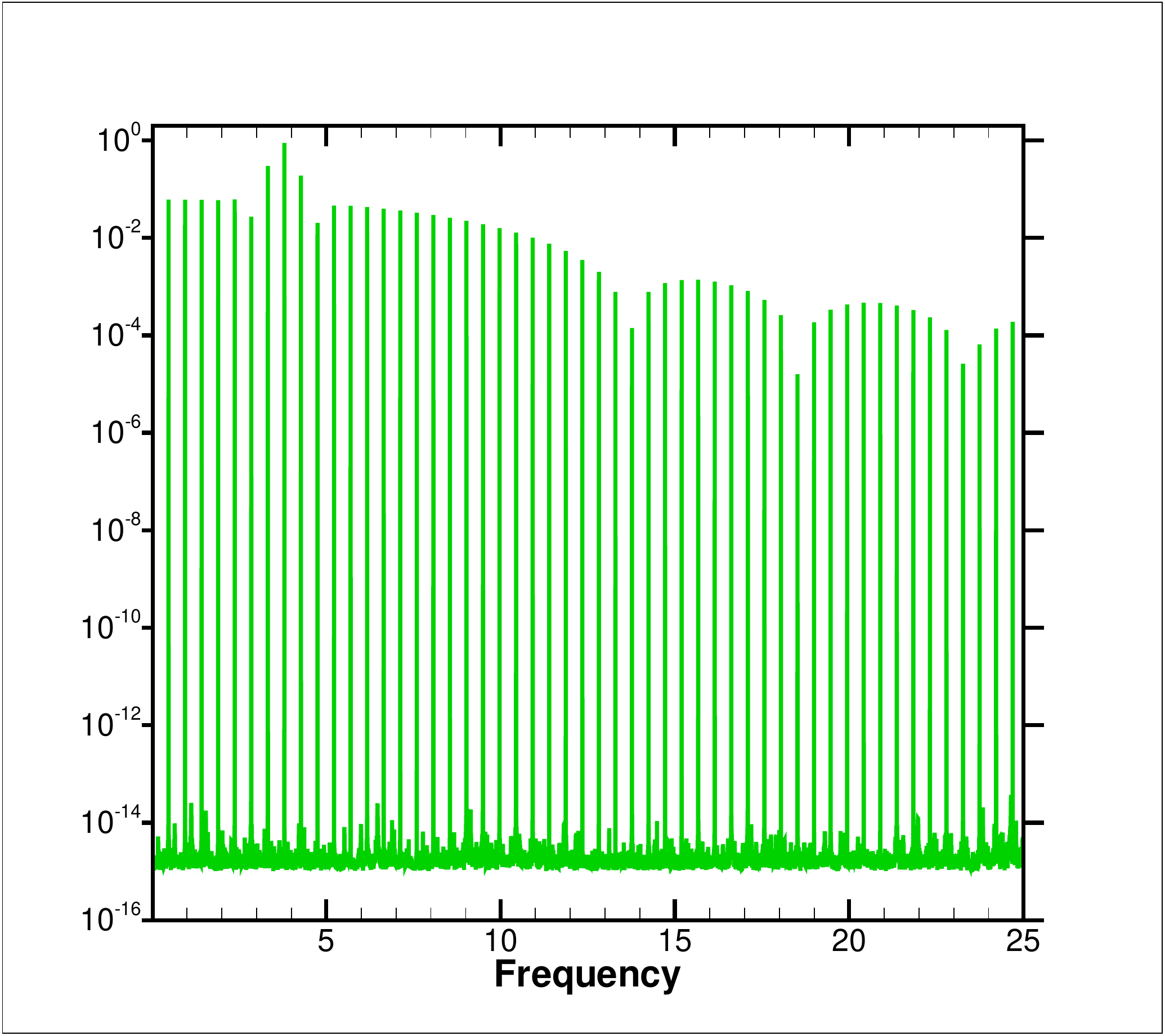}}
	\end{subfigmatrix}
	\caption{Power spectral densities of model signals with amplitude modulation (AM) and frequency modulation (FM).}
	\label{fig_model_spectra}
\end{figure}

The pure amplitude modulation of Eq. \ref{janelamento} is the product of the window function and the original sine curve and it only modifies the magnitude of the signal. Hence, in the frequency domain it is represented by a convolution of the window response with the spectral content of the signal. In this sense, in Fig. \ref{fig_model_spectra} one can observe PSDs of the different model signals obtained by the combination of a sine wave of center frequency $f_{c} = 3.78$ and window functions of frequency $f_{w} =0.47$. These values are imposed to model the behavior observed for the flow at 3 deg. incidence. Figure \ref{fig_model_spectra}(a) presents a simple periodical sinusoid (no windowing) where a single tone is observed. The effects of amplitude modulation via different window functions are presented in Figs. \ref{fig_model_spectra}(b) and (c) for sinusoidal and Tukey functions and multiple secondary tones (sidelobes) can be observed in the spectra besides the main tone.

The frequency modulation is characterized by the compression and dilatation of the signal, without modification of its amplitude. The spectral response is also composed of a center frequency and multiple tones equidistant by the modulation frequency, as shown in Fig. \ref{fig_model_spectra}(d). In the model problem proposed here, this frequency is $f_{m} = 0.47$.
For the present airfoil flow, the amplitude modulation effect is caused by flapping of the separation bubble that modifies the shedding mechanism of vortical structures as shown in Fig. \ref{fig_flow_M03}. This shedding leads to a natural amplitude modulation of the pressure and velocity signals that affect the subsequent noise emission. Although it is clear from literature that amplitude modulation impacts the overall noise generation process, here we show how the signal modulation modifies the spectral response. Motion of the separation bubble also leads to a frequency modulation effect on the pressure and velocity signals. 
When such signals are further modulated in amplitude by the sinusoidal and Tukey window functions, we obtain the signals depicted in Figs. \ref{fig_model_spectra}(e) and (f), respectively.

\begin{figure}[]
	\begin{subfigmatrix}{2}
		\subfigure[Sine wave (no window)]
		{\includegraphics[width=0.495\textwidth,trim={2mm 2mm 20mm 10mm },clip]
			{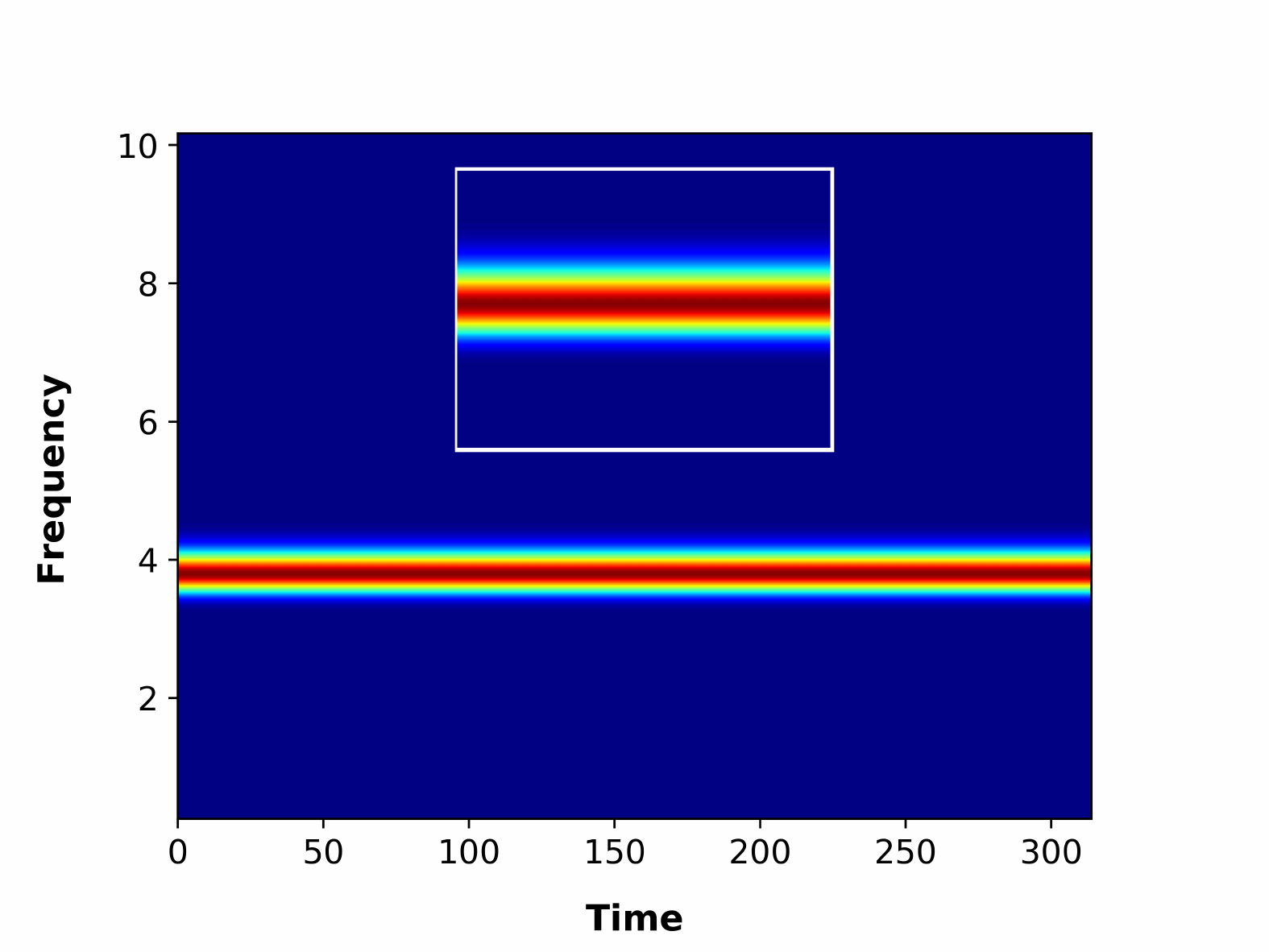}}
		\subfigure[AM with sinusoidal window]
		{\includegraphics[width=0.495\textwidth,trim={2mm 2mm 20mm 10mm },clip]
			{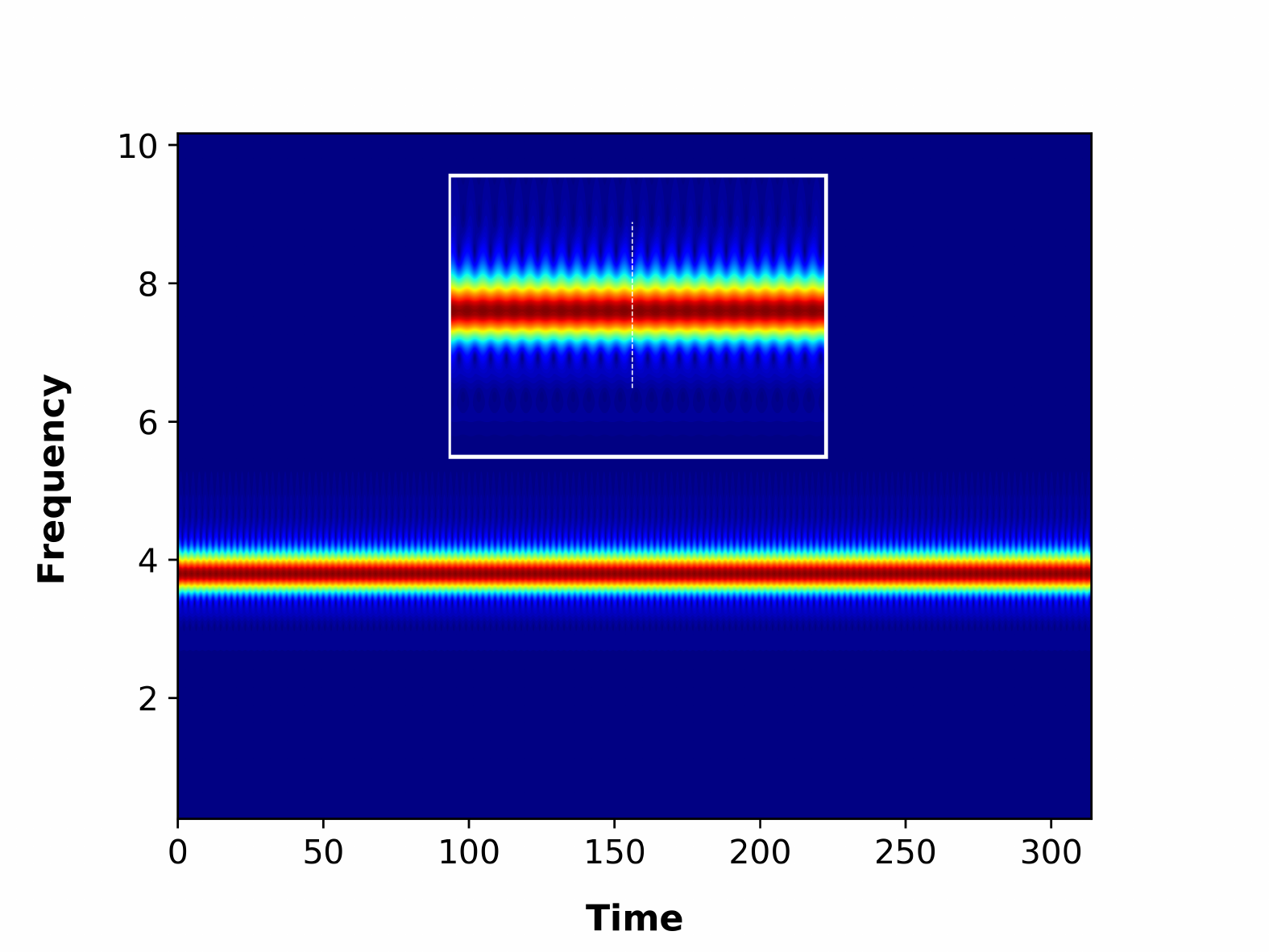}}
		\subfigure[FM + AM with sinusoidal window]
		{\includegraphics[width=0.495\textwidth,trim={2mm 2mm 20mm 10mm },clip]
			{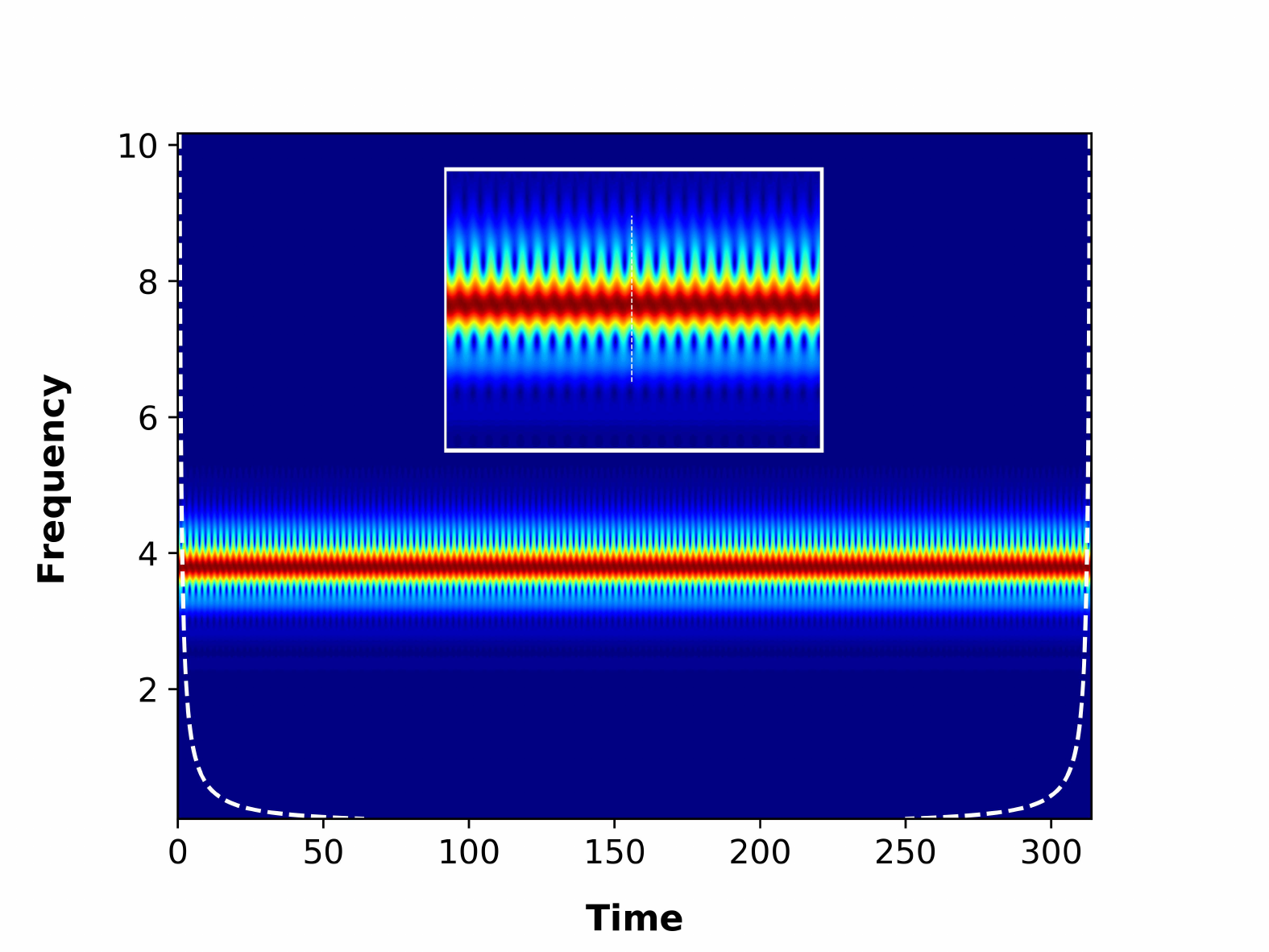}}
		\subfigure[Pressure signal of Fig. \ref{fig_wavelet_M03_3deg}(b)]
		{\includegraphics[width=0.495\textwidth,trim={2mm 2mm 20mm 10mm },clip]
			{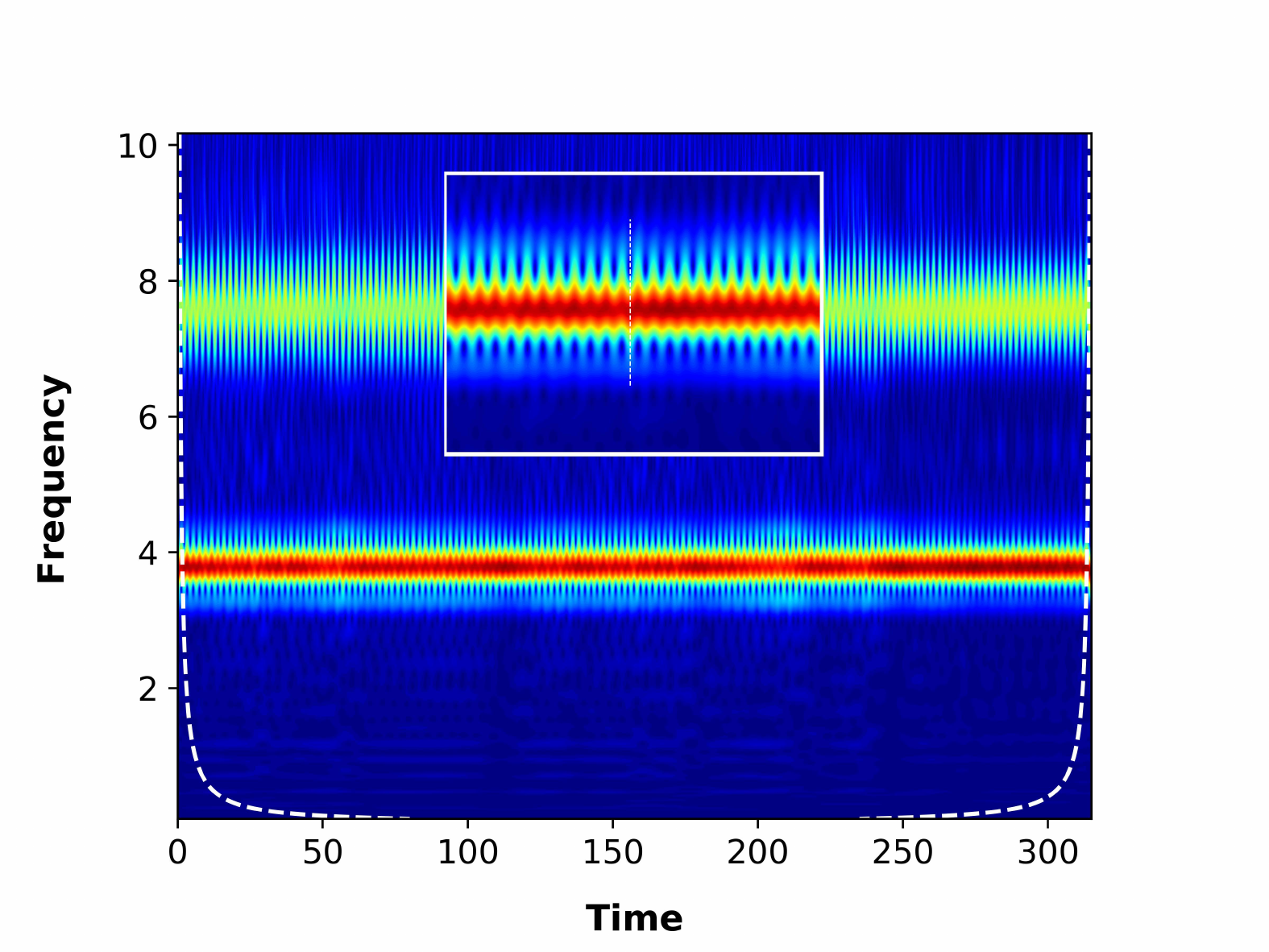}}
	\end{subfigmatrix}
	\caption{Spectrogram of model signals with detail view of center frequency.}
	\label{fig_model_wavelet}
\end{figure}

Continuous wavelet transforms are applied to understand the time-frequency behavior of the model signals. The benefit of this transformation is the time-frequency decomposition, differently of the Fourier transform. The spectrograms of the artificial signals from Figs. \ref{fig_model_spectra}(a), (b) and (e) are presented in Fig. \ref{fig_model_wavelet}. As a baseline for comparison, we present in Fig. \ref{fig_model_wavelet}(a) the wavelet transform of the pure-sinusoidal wave, where a constant magnitude response is obtained. When applying a modulation in terms of amplitude, Fig. \ref{fig_model_wavelet}(b), the magnitude of the coefficients are modified and secondary lobes appear on both sides of the dominant tone. On the other hand, the frequency modulation shown in Fig. \ref{fig_model_wavelet}(c) results in the alternating position of the secondary tones that switch from either side of the center frequency. Comparing these results to the original flow pressure signal from Fig. \ref{fig_model_temporal}(b) and reproduced in Fig. \ref{fig_model_wavelet}(d), it is possible to see the same behavior of frequency modulation. Hence, the model signals show that frequency modulation plays a very important role in the dynamics of vortical structures on the airfoil suction side in the flow at 3 deg. angle of incidence. For the  configuration with zero incidence, intermittency, amplitude and frequency modulations drive the dynamics of the more complex flow.
%

\section{Conclusions}

A study of tonal trailing-edge noise is performed for moderate Reynolds number flows past a NACA 0012 airfoil. Direct numerical simulations are conducted for Mach numbers 0.1, 0.2 and 0.3 at 0 deg. angle of attack, and for Mach number 0.3 at 3 deg. angle of attack. For all cases the Reynolds number is fixed at $Re_{c}=10^5$.  The role of amplitude and frequency modulations on the presence of secondary tones is analyzed by investigation of artificial signals that present similarities with those obtained by DNS of the current flows.

Despite the geometric symmetry for the cases with zero incidence, the flows become non-symmetric independently of the Mach number. In these cases, a separation bubble forms only on one side of the airfoil and has an important effect on flow instabilities developed along the boundary layer. The velocity and pressure temporal signals measured near the trailing edge exhibit changing patterns due to a low frequency motion of the bubble. Power spectral densities of these signals display a main tone with equidistant secondary tones as discussed in literature. However, depending on the specific flow period analyzed, the main tones are shifted by the bubble frequency. Temporal evolution of flow structures is visualized by snapshots of vorticity contours and a correlation of vortex shedding with near-field pressure fluctuations is presented. It is shown that vortex interaction occurs through pairing and merging dominated by flapping of the separation bubble. This leads to intense amplitude and frequency modulations, besides intermittency, that are observed in a spectrogram of the temporal signals.

Boundary layer instabilities shed due to a separation bubble develop along the airfoil suction side for the case of 3 deg. angle of incidence. Again, noise spectra display secondary tones but velocity and pressure temporal signals do not show strong pattern variations as those observed for the zero incidence flow configuration at the same Mach number. However, similarly to the case with zero incidence, low frequency motion of the bubble is responsible for frequency modulation of near-field flow properties. This effect is visualized by snapshots displaying vortical structures and a spectrogram of pressure and velocity signals.

One can conclude that the low-frequency motion of the separation bubble induces a frequency modulation of the flow instabilities developing along the airfoil boundary layer. For the zero deg. angle of attack case this effect is more prominent and leads to complex interaction of vortical structures. Hence, a strong amplitude modulation also occurs acting as a window function to the near-field pressure and velocity signals which, subsequently, affect the trailing-edge noise scattering mechanism. For the case of 3 deg. angle of attack, bubble motion is mostly responsible for frequency modulation as shown by a comparison between the original DNS pressure signal computed near the trailing edge and a sinusoid with the main frequency of the original pressure signal modulated by the frequency of the separation bubble.

\section*{Acknowledgements}

The authors of this work would like to acknowledge Fun\-da\-\c{c}\~{a}o de Amparo \`{a} Pesquisa do Estado de S\~{a}o Paulo, FAPESP, for supporting the present work under research grants No.\ 2013/03413-4, 2013/08293-7, 2018/11835-0 \@. The authors thank Centro Nacional de Processamento de Alto Desempenho em S\~ao Paulo, CENAPAD-SP, for providing the computer resources for the numerical simulations under project 551.


\bibliographystyle{model1-num-names}
\bibliography{bibfile}

\end{document}